\documentclass[aps,prb,notitlepage,citeautoscript,superscriptaddress,eqsecnum,reprint, longbibliography]{revtex4-2}

\usepackage{xr}
\usepackage{amsmath}
\usepackage{amsfonts, amssymb,amsxtra}
\usepackage[]{graphicx}
\usepackage{grffile}
\usepackage{amsfonts}
\usepackage{framed}
\usepackage{bm}
\usepackage{braket}
\usepackage{xcolor}
\usepackage{chngcntr}
\usepackage{comment}
\usepackage[normalem]{ulem}

\usepackage{algorithm}
\usepackage{algpseudocode}

\counterwithout{equation}{section} 

\usepackage{bm}
\usepackage{epsfig}
\usepackage{blindtext}
\usepackage[most]{tcolorbox}
\usepackage{verbatim}
\usepackage{color}
\usepackage{diagbox}
\usepackage{outlines}
\usepackage{cancel}
\usepackage{simpler-wick}

\usepackage[colorlinks=true]{hyperref}
\hypersetup{
    unicode=false,          
    pdftoolbar=true,        
    pdfmenubar=true,        
    pdffitwindow=false,     
    pdfstartview={FitH},    
    pdftitle={},    
    pdfauthor={},     
    pdfsubject={},   
    pdfcreator={},   
    pdfproducer={}, 
    pdfkeywords={} {} {}, 
    pdfnewwindow=true,      
    colorlinks=true,       
    linkcolor=magenta, 
    citecolor=blue,        
    filecolor=magenta,      
    urlcolor=blue           
}

\newcommand{\avg}[1]{\langle #1 \rangle}

\begin{document}
\title{Simulating the Transverse Field Ising Model on the Kagome Lattice using a Programmable Quantum Annealer}
\author{Pratyankara Narasimhan}
\affiliation{Center for Materials Theory, Department of Physics and Astronomy, Rutgers University, Piscataway, NJ 08854, USA}
\author{Stephan Humeniuk}
\email{stephan.humeniuk@gmail.com}
\affiliation{Center for Materials Theory, Department of Physics and Astronomy, Rutgers University, Piscataway, NJ 08854, USA}
\author{Ananda Roy}
\affiliation{Center for Materials Theory, Department of Physics and Astronomy, Rutgers University, Piscataway, NJ 08854, USA}
\author{Victor Drouin-Touchette}
\email{victor.drouin-touchette@usherbrooke.ca}
\affiliation{Center for Materials Theory, Department of Physics and Astronomy, Rutgers University, Piscataway, NJ 08854, USA}
\affiliation{Institut Quantique and Département de Génie Électrique et de Génie Informatique, Faculté de Génie,
Université de Sherbrooke, Sherbrooke, Québec, J1K 2R1, Canada}
\date{\today}

\begin{abstract}
The presence of competing interactions due to geometry leads to frustration in quantum spin models. As a consequence, the ground state of such systems often displays a large degeneracy that can be lifted due to thermal or quantum effects. One such example is the antiferromagnetic Ising model on the Kagome lattice. It was shown that while the same model on the triangular lattice is ordered at zero temperature for small transverse field due to an order by disorder mechanism, the Kagome lattice resists any such effects and exhibits only short range spin correlations and a trivial paramagnetic phase. We embed this model on the latest architecture of D-Wave’s quantum annealer, the Advantage2 prototype, which uses the highly connected Zephyr graph. Using advanced embedding and calibration techniques, we are able to embed a Kagome lattice with mixed open and periodic boundary conditions of 231 sites on the full graph of the currently available prototype.
Through forward annealing experiments, we show that under a finite longitudinal field the system exhibits a one-third magnetization plateau, consistent with a classical spin liquid state of reduced entropy. An anneal-pause-quench protocol is then used to extract an experimental ensemble of states resulting from the equilibration of the model at finite transverse and longitudinal field. This allows us to construct a partial phase diagram and confirm that the system exits the constrained Hilbert space of the classical spin liquid when subjected to a transverse field. We connect our results to previous theoretical results and quantum Monte Carlo simulation, which helps us confirm the validity of the quantum simulation realized here. With these results, we are able to provide new understanding into the nature of the phase diagram of this model while  extracting insight into the performance of the D-Wave quantum annealer to simulate non-trivial quantum systems in equilibrium.
\end{abstract}

\maketitle

\begin{figure*}[t]
\includegraphics[width=0.8\textwidth]{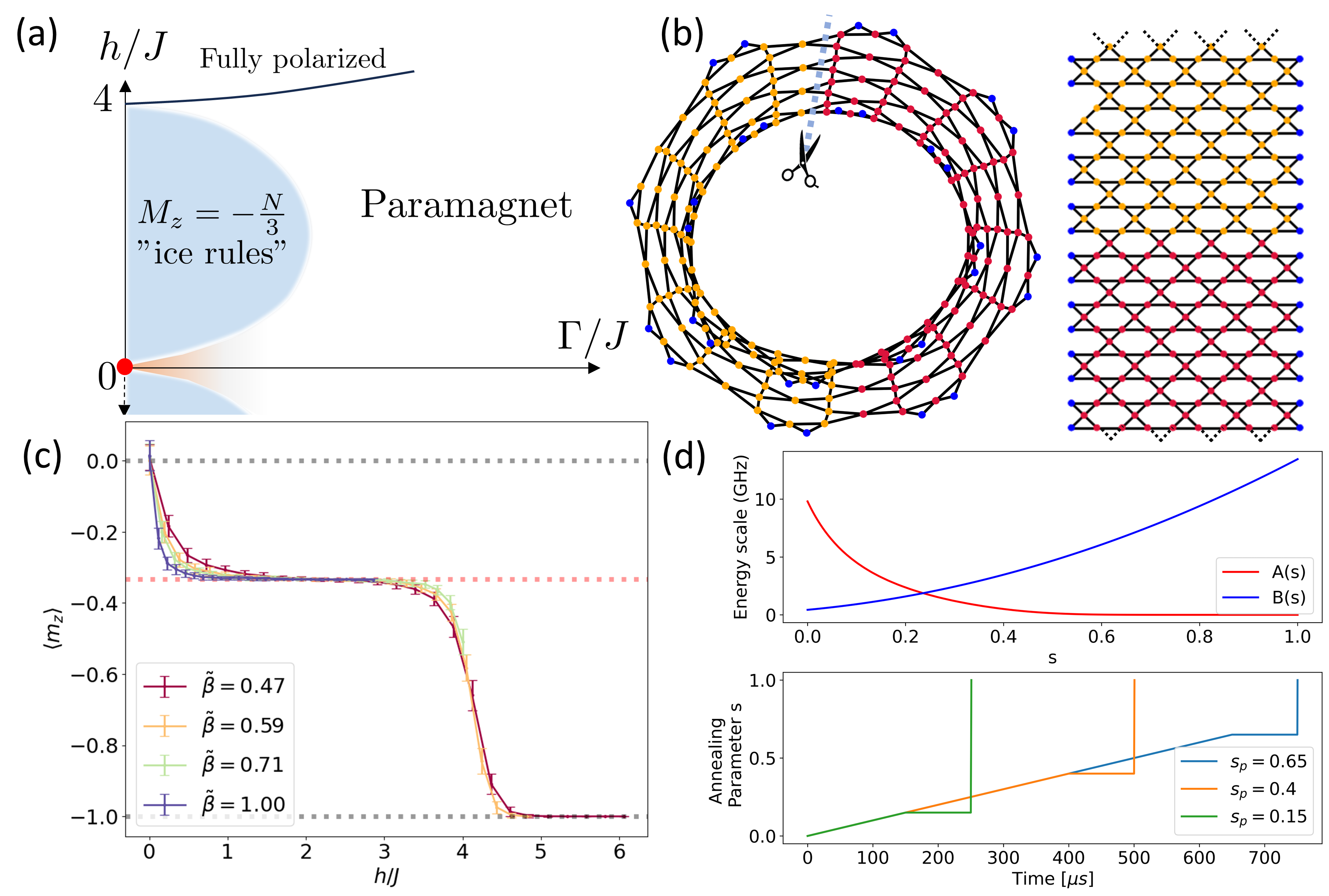} 
\caption{(a) Theoretically proposed phase diagram \cite{moessner2000two} of the transverse field Ising model on the Kagome lattice with antiferromagnetic coupling. The blue region corresponds to the extent of the stable "ice rules" manifold, with a one-third magnetization plateau. This region is predicted to host a long-range dimer ordered phase through a quantum order-by-disorder mechanism. We also show in slight orange that the quantum paramagnet continues for all $\Gamma/J$ at $h=0$. (b) Our experimentally realized embedding of a 231 site Kagome lattice with periodic boundary conditions in the $y$ direction. The tube-like structure as well as a flattened version are shown. Sites in blue at the edges are represented through ferromagnetic 2-chains, while red and orange sites are the two "layers" that can be simultaneously embedded in the Zephyr graph. The full embedding on the Zephyr graph in shown in Fig.~\ref{fig:embedding}. (c) The bulk magnetization per site as a function of the longitudinal field $h/J$ for various coupling constants $\tilde{\beta} J$ at zero transverse field (classical limit). Lowering $\tilde{\beta}$ acts like increasing the temperature, as can be seen by the decreased sharpness of the transition to the magnetization plateau (red dashed line). (d) Top: the prescribed annealing energy scales $A(s)$ and $B(s)$ on D-Wave's $\rm Advantage2\_prototype1.1$ device as a function of the annealing fraction $s$. Bottom: the annealing-pause-quench schedule $s(t)$ we use to simulate the model at a finite transverse field $\Gamma/J \sim A(s_p)/B(s_p)$.}
\label{fig:coumpound1}
\end{figure*}

\section{Introduction} 
\label{sec:introduction}

Frustration occurs in antiferromagnetic spin models where the interactions, either by geometry or other paths, cannot all be fully satisfied. An elementary exercise of frustration consists of anti-aligning Ising spins on a single triangle. After easily satisfying the first bond, the other two cannot be mutually anti-aligned no matter the choice of the third spin. Therefore, the ground state is comprised of the six degenerate states that have two bonds fulfilled and one bond unsatisfied. Examples of magnetic models exhibiting frustration are the antiferromagnetic Ising model on the triangular lattice \cite{moessner2000two}, the $J_1 - J_2$ model on the square lattice \cite{chandra1990ising}, the Shastry-Sutherland model \cite{shastry1981exact} and the Kitaev model on the honeycomb lattice \cite{kitaev2006anyons}. At first sight, the frustration in these spin systems leads to ground states with extensive degeneracy. However, in many situations, external or internal perturbative terms lift this degeneracy and lead to ordered states, as is the case in the triangular lattice Ising antiferromagnet \cite{moessner2000two}. If this degeneracy is not lifted in the presence of a perturbation, the disordered state persists down to $T=0$, resulting in a quantum paramagnet. The quantum paramagnet is a cousin of the more famous quantum spin liquids (QSLs) \cite{anderson1973resonating, balents2010spin}; both remain disordered down to zero temperatures, but the QSLs present non-trivial entanglement entropy features. Although the absence or presence of an ordered ground state has been settled for some specific models such as the Kitaev model on the honeycomb lattice \cite{kitaev2006anyons}, in other cases, computational analysis remains too complex to carry out to large enough lattices so that the exact nature of the ground state remains unknown. One such example is the antiferromagnetic Heisenberg model on the Kagome lattice \cite{Iqbal_2011, Kolley_2015, He_2017}, where debate still rages on whether the ground state is a $\mathbb{Z}_2$ gapped QSL and a gapless $U(1)$ liquid. These questions are important beyond the academic interest; many proposed theoretical models for QSLs lead to non-trivial topological properties \cite{ebadi2021quantum, semeghini2021probing, kattemolle2022variational, iqbal2023creation, google2023non}, which, if prepared and manipulated correctly, could lead to efficient platforms for topological quantum computing by the error-protected manipulation of non-local degrees of freedom. In an era of accessible quantum computers and quantum simulators, the question of efficiently and faithfully preparing QSLs on quantum devices, or even the study of whether such states of matter are the ground state of the system studied, motivates the study presented in this paper. 

Chief among this group of models of magnetic frustration rests the antiferromagnetic Ising model on the Kagome lattice (AFI-K). In this paper, we investigate the different quantum phases of this model in the presence of local fields. Although the model has been extensively studied in the past using quantum Monte-Carlo (QMC), recent theoretical work has cast doubt on some previously settled aspects of the phase diagram. Our goal is to explore this phase diagram through novel tools, such as a quantum annealer as a quantum simulator, as well as QMC studies. The Hamiltonian of the AFI-K model is

\begin{equation}
    \mathcal{H}_{\rm Kag} = J \sum_{\avg{i,j}} \sigma^z_i \sigma^z_j + \Gamma \sum_i \sigma_i^x  + h \sum_i \sigma^z_i \;,
    \label{eq:kagham}
\end{equation}

\noindent where the edges $\langle i,j \rangle$ are those of the two-dimensional Kagome lattice formed of corner-sharing triangles, as shown in Fig.~\ref{fig:coumpound1} (b), and both a longitudinal field $h$ and a transverse field $\Gamma$ can be applied. It is known that the AFI-K model with $h = \Gamma = 0$ is a quantum paramagnet down to $T=0$ \cite{moessner2000two}, owing to an extensive number of degenerate states due to the interplay of the frustration and the nearly disconnected geometry of the lattice \cite{powalski2013disorder}. For the triangular lattice version of this model, a phenomena known as order by disorder \cite{villain1980order} happens in the presence of a transverse field \cite{moessner2001ising, moessner2001phase, moessner2001resonating, moessner2001short}.  This occurs when the large degeneracy of a set of states is broken as the system chooses the subset of states with the "softest" fluctuations and thus maximizes the entropy. In the triangular lattice, this means that the quantum paramagnet at $h = \Gamma = 0$ does not survive finite $\Gamma$. This was recently observed using a quantum annealer in ~\cite{king2018observation}, which was able to track the successive Kosterlitz-Thouless phase transitions \cite{isakov2003interplay} as microscopic defects within the six-fold degenerate manifold, formed bound pairs leading to quasi-long-range order. The final onset of a six-state clock long-range ordered phase was not observed on the annealer due to insufficiently low device temperature, though it has been seen in quantum Monte-Carlo studies \cite{wang2017caution}. Nevertheless, that study remains an important advancement in the benchmarking of quantum annealers as quantum simulators of spin systems, and since then multiple studies on these devices have been able to provide new insights on spin models \cite{lopez2023field,lopez2024quantum}.

On the other hand, studies on the Kagome lattice model revealed an interesting setting of \textit{disorder by disorder} \cite{fazekas1974ground}, where quantum effects are not able to lift the large degeneracy of a classically disordered state. Indeed, Moessner, Sondhi and Chandra \cite{moessner2000two} proposed that the system would remain a quantum paramagnet, for all values of $\Gamma$. Furthermore, the application of a longitudinal field $h$ was found to partially lift the degeneracy of the ground state to the Kagome spin ice state with magnetization $M/N = \pm \frac{1}{3}$ and constrained dynamics of defects due to strong constraints on each triangle. These magnetization plateaus and related dynamics were previously observed for $\Gamma=0$ (no transverse field, thus a classical model called Kagome spin ice) using D-Wave's Advantage quantum annealer \cite{lopez2023kagome}. A schematic of the theoretically proposed phase diagram \cite{moessner2000two, moessner2001ising} is shown in Fig.~\ref{fig:coumpound1} (a). On the $\Gamma = 0$, $h/J \neq 0$ line, the system is in the Kagome spin ice state, where each triangle consists of two spins aligning with the field $h$ and one spin anti-aligned. These unusual states have a net nonzero moment but still have a high degeneracy such that the entropy is lowered from the $h=0$ point but remains extensive. The authors of Ref.~\cite{moessner2000two} suggested that applying an infinitesimal $\Gamma$ leads to a long-range dimer star ordered phase, by virtue of a mapping to a hexagonal lattice dimer model \cite{moessner2001phase}, while other authors discuss alternative scenarios \cite{nikolic2005theory, nikolic2005disordered}. This ordered manifold is thought to correspond to states with maximally flippable hexagons (hence the stars), which can coherently flip their spins with an amplitude of the order of $(\Gamma/h)^6$. This proposed theoretical state would then vanish for higher $\Gamma$ into a trivial quantum paramagnet connected to the $\Gamma \rightarrow \infty$ limit. Finally, for $h/J > 4.0$ \footnote{This is obtained by comparing the energy of a polarized configuration, where each bond carries an energy of $J$ while each site an energy of $-h$. There are two bonds per site, resulting in an energy per site of $2J - h$. Comparing this with the energy of a $M/N = \pm \frac{1}{3}$ state, where the energy per site is, on average, $-\frac{2J}{3} - \frac{h}{3}$, one obtains the value of $h/J = 4.0$ as the field beyond which the fully polarized state is favored.}, it is no longer energetically favorable for any of the individual spins to be anti-aligned to the field $h$, and thus the system is fully polarized. We note however that recent numerical studies in Ref.~\cite{wu2019tunneling} shows that this maximally flippable state, which can also be interpreted as a valence-bond solid, is not gapped from the other dimer states by the transverse field. A new scenario was thus advanced, where the increase in the transverse field simply brings the system through a crossover from the restrained manifold of states with lowered entropy and dimer constraints to a trivial paramagnet. 

Here, we use a programmable quantum annealer, namely D-Wave's latest prototype, the $\rm Advantage2\_prototype1.1$ device, to simulate the AFI-K model for a wide range of longitudinal fields $h/J$ and transverse fields $\Gamma/J$. In fact, the highly connected lattice of the latest D-Wave prototype allows for a rather straightforward embedding of the model onto the lattice of connected flux qubits. We then use an annealing-pause-quench (APQ) schedule, which has been proposed as a tool to extract an ensemble of configurations for a given transverse field, which itself is extracted from the exact pause point $s_p$ that one chooses after the anneal. These details are presented in section \ref{sec:methods}.  In order to provide further context to some of the puzzling features of our results, we use a quantum Monte-Carlo (QMC) algorithm to simulate the system classically for comparable system sizes. This approach has two benefits. Firstly, by connecting to the known features of the phase diagram, we can benchmark the performance of the quantum annealer as a quantum simulator. Secondly, our combined APQ and QMC results bring new light on the phase diagram after the recent proposals of Ref.~\cite{wu2019tunneling} and we comment on the observed absence of a long-range dimer ordered state in our results, in spite of previous predictions.

In section~\ref{sec:results}, we show our main results, which can be summarized as follows. First, we confirm the disorder by disorder mechanism at $h=0$ and the quantum paramagnetic state which persists for all $\Gamma \neq 0$. Then, we explore the order by disorder mechanism at $h/J < 4.0$, and are able to see a crossover between the Kagome spin ice regime, with an extensive degeneracy and obeying the "ice rules", and the disordered regime. This investigation is helped by an analysis of the spin structure factor, which is presented in subsection~\ref{subsec:fft}, and an in-depth analysis of the histogram of the phase angle of the complex order parameter in subsection \ref{sec:dimer}. Both of these sections help us conclude that our quantum simulation of the AFI-K model does not lead to the observation of the long-range dimer star ordered state, which is also consistent with our QMC results. Instead, we present a reconciliation of the order-by-disorder scenario of Moessner et al. with the concept of magnetic moment fragmentation in Kagome spin ice \cite{museur2023neutron}. This helps us construct a partial phase diagram, which is presented in subsection~\ref{sec:crossover}. We find that a conspiracy of device and schedule specific effects hamper our ability to go to low enough temperature, where a true valence bond solid in the dimers may lie. Finally, we present our conclusions and outlook for future work in section \ref{sec:conclusion}. Details on the embedding methods, device calibration and the QMC simulations are presented in the appendices.


\section{Annealing Methods} 
\label{sec:methods}

Quantum annealing (QA) is a heuristic generalization of the more abstract quantum adiabatic algorithm (QAA) \cite{farhi2001quantum}, which was designed at first to tackle NP-hard optimization problems, such as obtaining the ground state of spin glass systems. In both of these protocols, one implements a time dependent Hamiltonian $H(t) = (1-t/T_a)H_{m} + (t/T_a) H_c$ where the annealing takes a time $T_a$, $H_m$ is a mixing term (such as a transverse field term) for which a ground state is easily obtained and is orthogonal to the $Z$ basis, and $H_c$ is the cost function or classical term that is diagonal in the $Z$ basis. For more information on these techniques, see the following reviews \cite{albash2015reexamining, das2008colloquium, hauke2020perspectives}. While QAA requires in the best case a total annealing time $T_a \sim \Delta_{\rm min}^{-2}$ scaling with the minimum gap to obtain the ground state of $H_c$, that requirement is void in QA, where the computation time $T$ is kept fixed, at a cost of many shots. This heuristic approach stems from early work \cite{brooke1999quantum,kadowaki1998quantum} where the general ideas of thermal annealing were ported to quantum magnets through the use of the transverse field $H_M = \Gamma \sum_i \sigma_i^x$. The promise that implementing such QA algorithms could lead to a faster heuristic approach to NP-hard optimization problems led to the development of quantum annealing devices. Quantum processing units (QPUs) implementing the QA algorithm and developed by D-Wave can be accessed from the cloud. The programmable quantum annealer is formed of a highly connected graph of superconducting flux qubits with flux-tunable Josephson junctions, with up to thousands of qubits available. Since then, new platforms have emerged using tunable Rydberg atoms manipulated using optical tweezers, as offered by companies QuEra and Pasqal \cite{henriet2020quantum}. In all of these companies, qubits are controlled globally and variants of QA are implemented to solve optimization problems or perform quantum simulations. 

The specific quantum annealer used in this study is the $\rm Advantage2\_prototype1.1$ prototype device, which implements the following tunable transverse field Ising model (TFIM):

\begin{equation}
\begin{split}
        &\mathcal{H}_{\rm DWave} (s) = - \frac{A(s)}{2} \Big[ \sum_i \sigma_i^x \Big] \\
        &\qquad + \frac{B(s)}{2} \Big[ \sum_{\avg{i,j}} J_{ij, \rm phys} \sigma^z_i \sigma^z_j + \sum_i h_{i,\rm phys} \sigma^z_i \Big] \;,
\end{split}
\end{equation}

\noindent where $s$ is the annealing fraction, which can be tuned continuously. $A(s)$ controls the strength of the transverse field whereas $B(s)$ controls the strength of the classical energy in the $Z$ basis. The functions $A$ and $B$ as a function of $s$ are fixed for a given device, and shown in Fig.~\ref{fig:coumpound1} (d). Measurements are only done at $s=1$ where the transverse field is absent. The user can specify the function $s(t)$, which can lead to fast or slow anneals and to pauses and quenches. Three constraints are put on $s(t)$: $s(0) = 0$ for forward annealings, $s(t=T) = 1$ at time $T$ such that $1 \mu \text{s} < T < 2000 \mu$s, and $ds/dt \leq 1 \mu \text{s}^{-1}$. The last one constrains the speed of quenches, so that fast anneals are unavailable to the cloud user. Finally, the programmability of this device comes fully to fruition with the user-specified $J_{ij, \rm phys}$ and $h_{i, \rm phys}$. The "phys" label here applies to the microscopic terms of $\mathcal{H}_{\rm DWave}$, in order to differentiate them from those of the AFI-K Hamiltonian of Eq.~\ref{eq:kagham}. There are 563 individually addressable $h_{i, \rm phys} \in [-4,4]$ and 4790 couplers $J_{ij, \rm phys} \in [-2, 1]$ in the $\rm Advantage2\_prototype1.1$ prototype.

The availability of the D-Wave quantum annealer has led to many studies on its use to study the properties of quantum spin systems, in particular on variants of the TFIM model due to its natural implementation in D-Wave's devices. Among such studies are the field-induced magnetic phases in quasicrystals \cite{lopez2023field}, the quantum critical dynamics in spin chains \cite{king2022quantum, king2022coherent}, the out-of-equilibrium behavior of spin chains \cite{king2021quantum}, the frustrated $J_1-J_2$ model \cite{park2021phase}, the frustrated Ising model on bathroom tile lattice \cite{hearth2022quantum}, the implementation of a $\mathbb{Z}_2$ spin liquid \cite{zhou2021experimental, zhou2022probing}, the study of the multiple magnetic phases of classical Kagome ice \cite{lopez2023kagome} and of classical spin ice \cite{king2021qubit}, the spin glassy behavior in a random 3D magnet \cite{harris2018phase}, the order-by-disorder phenomena in the triangular lattice antiferromagnetic Ising \cite{king2018observation, king2021scaling}, the study of Griffiths-McCoy singularity on random graphs \cite{nishimura2020griffiths} and the magnetization plateaus in the Shastry-Sutherland model \cite{kairys2020simulating}. Although much of these works are done by or in collaboration with D-Wave itself, the cloud access to the QPUs has resulted in more independent teams performing novel research in recent years, which has showed that more and more quantum spin systems can be simulated. This, in turn, helps benchmark the latest devices, as many behaviors of these systems are either well known through exact methods or can be efficiently simulated on classical computers through Monte-Carlo methods.

\subsection{Embedding}

\begin{figure}[t]
\includegraphics[width=\columnwidth]{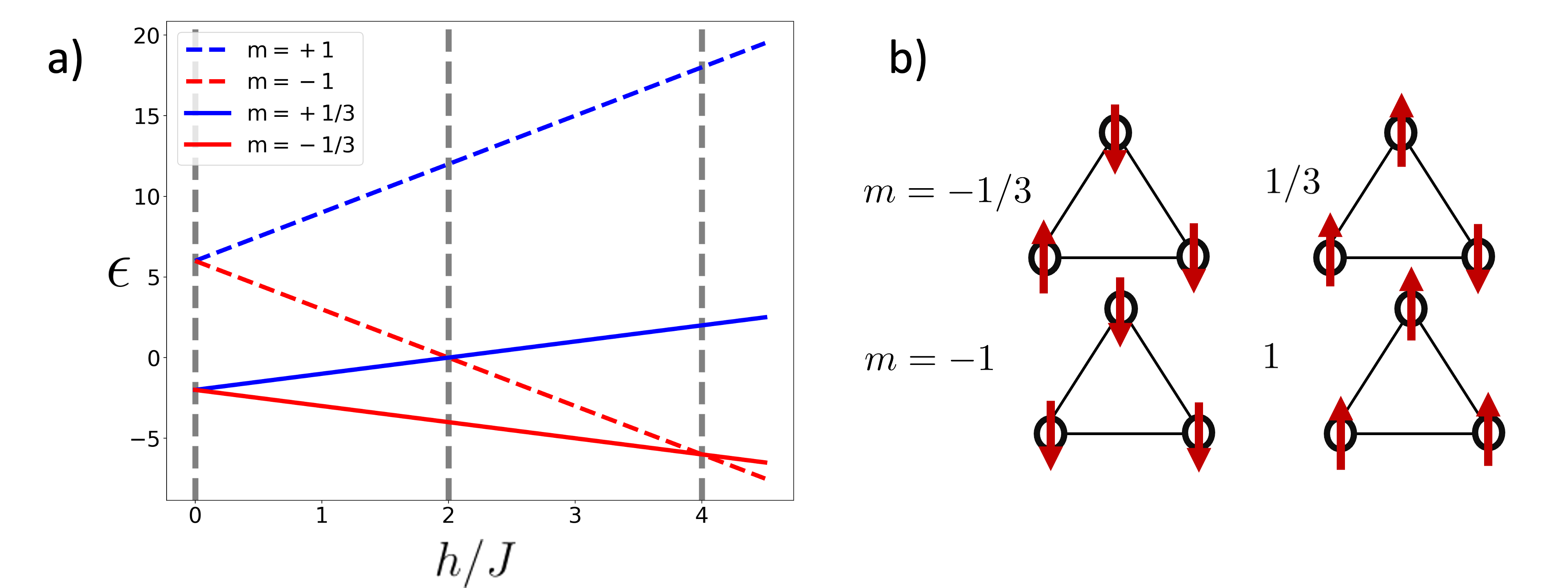} 
\caption{(a) Classical energy $\epsilon$ of different Ising configurations $\{z_i\}$ on an antiferromagnetic triangle with longitudinal field, with $\epsilon = 2J \sum_{ij} z_i z_j + h \sum_i z_i$, where the nearest neighbor interaction is set at $2J$ to reflect the local environment of the Kagome lattice with its four neighbors per site. We see that, for $0 < h/J < 4.0$, the $m = 1/3$ state which aligns with the field is lower in energy, whereas for $h/J > 4.0$ the fully polarized state is lower. (b) The different local Ising configurations corresponding to the different branches of (a). Note that, for $h/J = 2.0$, the gap to the excited state is maximal, and that the nature of the first excited state changes there between high and low longitudinal fields.}
\label{fig:energetics}
\end{figure}

The $\rm Advantage2\_prototype1.1$ device realizes the highly connected Zephyr graph \cite{boothby2020next,boothby2021architectural, boothby2021zephyr}. Each qubit can have up to 20 active couplers $J_{ij, \rm phys}$ to other qubits. This increased connectivity compared to the previous Chimera and Pegasus graphs means that the frustrated unit of the Kagome lattice, an antiferromagnetically coupled triangle of Ising spins, as displayed in Fig.~\ref{fig:energetics} (b), is natively implemented on the device without having to resort to the formation of large chains of ferromagnetically coupled qubits. After creating a unit cell which we can tile on the full $Z_4$ Zephyr graph, we realized that this embedding was sparse enough to allow us to embed another sheet of the Kagome lattice using an alternate version of our unit-cell embedding. Finally, both lattices could be connected at the edges ("stitched"), forming a periodic cylinder, as shown in Fig.~\ref{fig:coumpound1} (b), with the red and orange sites corresponding to the two lattices that we have connected. The full embedding is shown in Fig.~\ref{fig:embedding}, in the appendix. 

Two issues remained after doing this. Firstly, some qubits and couplers are absent in the actual machine, showing up as defects from the ideal $Z_4$ Zephyr graph. We then created alternate local embeddings for the unit cells that used different qubits/couplers in order to bypass missing ones. Further details on this can be found in section \ref{appendix1}. Secondly, the lattice one can embed with such procedure is rather small, and some sites are not part of an elementary triangle of the Kagome lattice (our unit cell relied on tiling of hexagons, whereas triangles are the natural unit cell structure of the Kagome lattice). In order to create a uniform cylindrical lattice with easy to tackle boundary conditions, we took advantage of the large amount of unused qubits in the graph at the open edges of the Zephyr graph and inserted manually boundary sites that are represented by 2-chains which are pairs of strongly ferromagnetically coupled qubits representing a single site on the lattice. These sites are shown in blue in Fig.~\ref{fig:coumpound1} (b). 

We then have that our model is formed of sites that have different effective transverse fields, with 2-chains having $\Gamma_2^{\rm eff} \sim \frac{A(s)^2}{B(s)}$, due to the use of ferromagnetic chains, while sites represented by a single qubit have a transverse field $\Gamma = A(s)/2$. We use anneal offsets, a tuning knob provided by D-Wave, where an individual qubit's annealing can be delayed/advanced compared to others, thus providing an ability to individually tune $\Gamma/J$ for individual qubits. This tool has been previously proposed as a means to enhance the fidelity of solving \textit{classical} QUBO problems \cite{yarkoni2019boosting}. In our case, we present for the first time the use of anneal offsets to calibrate the APQ schedule; previous work focused on fast anneals exclusively \cite{chern2023tutorial}. With the anneal offsets, we can make it so that the effective transverse field on 2-chains is the same as the bare transverse field: $\Gamma = \Gamma_2^{\rm eff}$, which is essential for faithful quantum simulation. The details of this procedure are explained in Sec.~\ref{appendix2}. Using this calibration has a few consequences; we need to adjust $h_{i,\rm phys}$ and $J_{ij, \rm phys}$ for these blue sites. Furthermore, we implemented manually a form of mean-field boundary condition, where we attempt to minimize the effect of this open cylindrical boundary on the bulk sites of the system (red and orange sites). Though a full self-consistent loop where $\avg{m_z}_{\rm bulk} = \avg{m_z}_{\rm edge}$ is satisfied would have been preferable, we lacked QPU access time to perform this fully. This led us to implement a more heuristic approach where a local field $h_i = 0.5 h$ is applied on the boundary sites (blue), with $h$ the longitudinal field applied to all other bulk sites. In all data shown in this paper, observables are only calculated on bulk sites, as there remains a substantial bias at the edges, though one that is less than without this adjustment. Together, these embedding methods permit the full embedding of our 231 site cylindrical Kagome lattice. 

\begin{figure*}[t]
\includegraphics[width=\textwidth]{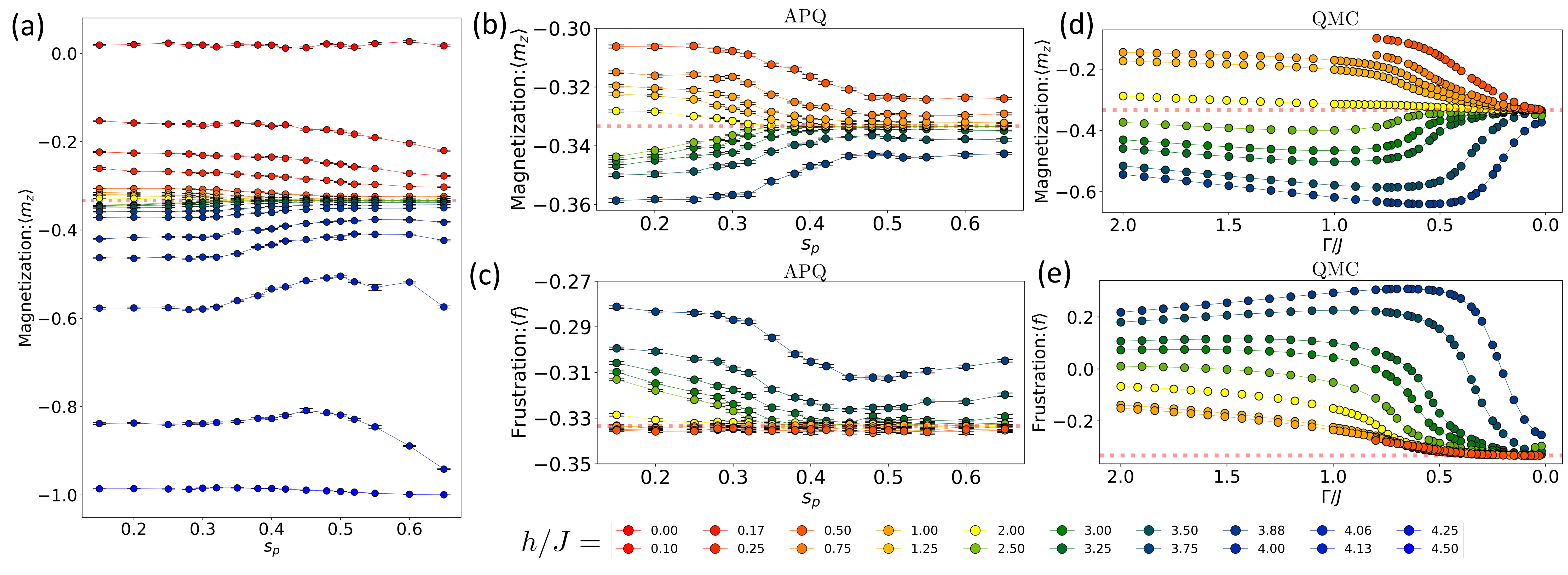} 
\caption{(a) Bulk magnetization $\avg{m_z}$ as a function of $s_p$ the pause point of the APQ schedule, for diverse values of $h/J$. (b) Closeup of the magnetization of (a) around the $\avg{m_z} = \frac13$ plateau. (c) The bond frustration $\avg{f}$  [see Eq.~\ref{eq:frustration}] for the bulk sites as a function of the pause point $s_p$ for the same $h/J$ values of the closeup in (b). In (d-e) we show quantum Monte-Carlo results (QMC) for diverse values of $\Gamma/J$ (notice the inverted horizontal axis), in order to qualitatively compare with the experimental APQ results. Simulations are performed for $\beta J = 10$ and a periodic system of size $L = 9$. (d) $\avg{m_z}$ magnetization as obtained from the QMC analysis. Considering that $s_p \rightarrow 0.65$ is $\Gamma \ll J$ and that $s_p \rightarrow 0.15$ is $\Gamma \gg J$, we see that the same trends are present in (a). (e) The bond frustration $\avg{f}$ obtained from QMC analysis, again showing qualitative agreement with (c).}
\label{fig:coumpound2}
\end{figure*}

Note that we also attempted to embed the Kagome lattice on the Chimera graph (2000Q device, now unavailable), which led to some sites being represented by FM chains of three qubits while others had FM chains of two qubits. This non-uniformity in chain length meant that some sites would have \textit{drastically} different dynamics since their effective transverse field $\Gamma^{\rm eff}$ would be different: $\Gamma^{\rm eff}_2 \neq \Gamma^{\rm eff}_3$. It is possible to fix this, but this process became quickly more involved than simply reformulating our problem for the Zephyr graph. In Ref.~\cite{lopez2023kagome}, the authors are able to embed the Kagome lattice on the Pegasus graph. Due to triangles being unavailable as minimal cycles in this graph, they had to minor-embed using chains. Thus, the sites of their Kagome lattice embedding were represented uniformly, through chains of three ferromagnetically ($J_{ij} = K = -1.8$) coupled qubits. The effective transverse field on each site was then $\Gamma_3^{\rm eff} \sim \Gamma^3/K^2 \sim \frac{A(s)^3}{B(s)^2}$. The authors then exposed classical spin configurations to a short $1 \mu$s exposure of this effective transverse field, thus simulating quasi-classical dynamics between degenerate classical ground states. In our study, the transverse field $\Gamma \sim A(s)$ is kept to its bare value since all sites are represented by an effective single qubit. This permits us to explore larger values of transverse field than if we had used $n$-chains.


\subsection{Anneal-pause-quench} \label{subsec:apq}

Our focus being on the effects of finite transverse field $\Gamma/J$ on the magnetic properties of the AFI-K model, we opted for an anneal schedule called the anneal-pause-quench (APQ) sequence, as previously implemented on the quantum annealer \cite{harris2018phase, king2018observation, king2021scaling, nishimura2020griffiths}. This protocol consists of a slow anneal from the starting state $\ket{\psi(s=0)} = \prod_i \ket{\rightarrow}_i$ at $s=0$ up to a specified pause point $s=s_{p}$. This anneal takes a time $t_1 = 1000 s_{p} \mu$s, which is then followed by a pause of time $t_2 = 100 \mu$s at that point. This is meant to help the system equilibrate to the Hamiltonian $H(s_p)$. Finally, a readout quench to $s=1$ is executed with the maximal slope available on the cloud access, which thus takes a time $t_3 = (1-s_{p})\mu$s. This is displayed in Fig.~\ref{fig:coumpound1} (d). The objective of this protocol is that the ensemble of states generated after the anneal and pause is close to the distribution one would get from sampling the equilibrated thermal equilibrium for $T \ll J$ and with a finite $\Gamma, h$. The readout quench should also be fast enough to simply collapse the state onto the $Z$ axis for measurement, without deforming it. Whereas values of $h/J$ are easily set by the user, $\Gamma/J$ comes from the ratio of the $A$ and $B$ functions at the pause point $s_p$. 

There remains debate as to whether this procedure accurately samples the canonical distribution of $H(s_{p})$ \cite{izquierdo2021testing, nishimura2020griffiths}, but we show in this paper that it does seem to recover the main signatures that are expected in the AFI-K model, with some caveats. As others have pointed out, the readout quench is not fast enough to produce qualitatively correct data at large $\Gamma$ (i.e. small $s_p$). In Refs. \cite{king2018observation, king2021quantum, izquierdo2021testing}, the canting in the $\sigma_x$ direction that should be observed at large transverse field is mostly absent, and thus $\avg{\sigma^z}$ is larger than otherwise expected. This is because the readout quench is simply not fast enough, such that the canting slowly relaxes through the "quench". Our results are consistent with this interpretation, though the model seems to be resilient enough to this canting that we can nevertheless infer some qualitative insight.

\section{Results} 
\label{sec:results}

In order to test the Kagome lattice embedding we present in this paper, we first performed simple forward annealing experiments, which are specific cases of the APQ schedule where $s_p = 1$, i.e. there is no pause until the classical point. This replicates the results of Ref.~\cite{lopez2023kagome}. These lead to an ensemble of configurations for the classical Kagome Ising model with a longitudinal field $h$. We do this for different values of $J = \tilde{\beta} J_{\rm max}$, where $J_{\rm max} = 11.16 \; \text{GHz}$ - increasing $\tilde{\beta}$ is equivalent to decreasing the effective temperature $T/J$ at which we simulate the classical model. We compute the bulk magnetization per site $\avg{m_z} = \avg{\frac{1}{N} \sum_i \sigma_i^z}$, where we have $N$ sites in the bulk and $\avg{\cdots}$ refers to the average over all obtained configurations, which is typically on the order of $10^4$ for results obtained with APQ while we only performed $500$ reads to forward annealing experiments shown in Fig.~\ref{fig:coumpound1} (c).

As we can see in Fig.~\ref{fig:coumpound1} (c), which displays the bulk magnetization per site $\avg{m_z}$, our results for $\tilde{\beta} = 1$ show a sharp transition from $\avg{m_z} = 0$ at $h/J = 0.0$ to a plateau of $|\avg{m_z}| = \frac13$ (dashed red line is added as a guide). This plateau is consistent with the results obtained in Ref.~\cite{lopez2023kagome} as well as the literature predictions \cite{moessner2001ising}. As $\tilde{\beta}$ is decreased, the transition and plateau become less sharp, consistent with the higher temperatures. Finally, another transition at $h/J = 4.0$ occurs to a fully aligned state along the magnetic field. The intermediate phase with a magnetization plateau is due to a compromise between the longitudinal field and the antiferromagnetic interactions between sites among a triangle, where each triangle has a local configuration among the set $\{ \ket{\uparrow \uparrow \downarrow}, \ket{\uparrow \downarrow \uparrow}, \ket{\downarrow \uparrow \uparrow} \}$ (shown in Fig.~\ref{fig:energetics}), called the "ice-rule" manifold. In this classical model, the $\frac13$-plateau, along with all triangles being in one of those three configurations, is a signature of a classical spin liquid. As temperature is reduced, the system goes through a crossover from a high temperature paramagnet to this classical spin liquid phase where the entropy is reduced but still extensive. This is related to the question on dimer coverings on the hexagonal lattice, as one can assign one of three dimers to each local triangle arrangement \cite{moessner2000two}. The ground state is thus formed of an extensive number of states satisfying the "ice-rule" constraints, i.e. that each local triangle has magnetization $\frac13$. The study of the stability of this ice-rule manifold to the effect of a transverse field is exactly the subject of our work.

The result of our experiments using the APQ protocol are presented in Fig.~\ref{fig:coumpound2}, as well as results from quantum Monte-Carlo (QMC). In both cases, the same measurements were taken, such as the bulk magnetization $\avg{m_z}$ as well as the frustration average $\avg{f}$, given by

\begin{equation}
    \avg{f} = \frac{1}{2N_B} \sum_{i\in \mathcal{B}} \sum_{j \in n.n.(i)} \avg{\sigma^z_i \sigma^z_j} \;, \label{eq:frustration}
\end{equation}
\noindent where $n.n.(i)$ refers to all nearest neighbors of site $i$ in the bulk $\mathcal{B}$ of the lattice, and thus the sum is over all nearest neighbor bonds on the lattice. We find that it is a complementary notion to the magnetization, and $\avg{f} = -1/3$ is a strong indication that all triangles are in the "ice-rule" manifold, meaning that one out of three bonds is frustrated. Errorbars for the experimental APQ data is obtained by binning the 20000 measurements into 20 bins of 1000 measurements each, and then we perform a jackknife estimate of the mean and the 95$\%$ confidence interval. Errorbars for the QMC data are smaller than the data points. We note that, in the presented data, large $s_p$ corresponds to small transverse fields, while small $s_p$ corresponds to large transverse fields, with $A(s^{\ast})/B(s^{\ast}) = 1$ as $s^{\ast} \simeq 0.231$, such that the $x$ axis for QMC data is generally reversed to compare with APQ.

\begin{figure*}[t]
\includegraphics[width=\textwidth]{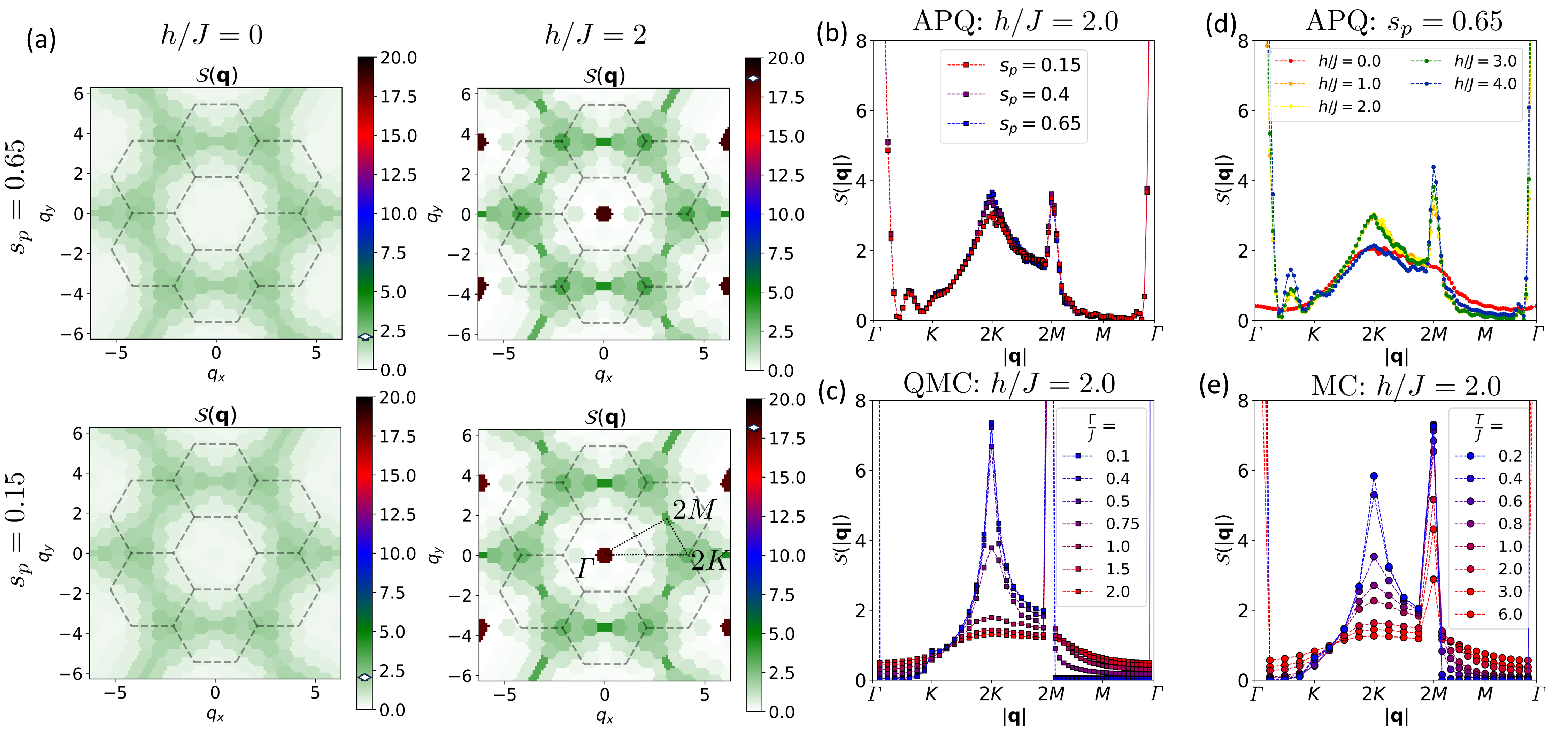} 
\caption{(a) Colorplots of the spin structure factor $\mathcal{S}(\mathbf{q})$ as a function of momenta $\mathbf{q}$, for a few $s_p$ and $h/J$ values. We note that, for $h/J =0.0$ (left column), the spin structure factor remains featureless as a function of $s_p$, signaling the absence of any transition into an ordered state as the transverse field in increased. For $h/J = 2.0$ (right column), the features at the $\mathbf{q} = 2K$ point decrease slightly with the increase of the transverse field ($s_p=0.15$). In (b-c), we compare a line cut through high symmetry points of the structure factor $\mathcal{S}(|\mathbf{q}|)$ (following the dashed path shown in (a)), at $h/J = 2.0$ for results obtained through the APQ schedule (b) and obtained through QMC (c). The QMC data is obtained for $\beta J = 10$ and a periodic system with $L = 24$. As seen in both APQ and QMC results, the broad peak at $\mathbf{q} = 2K$ stays constant over an extended range of $\Gamma/J$ before it drops as the transverse field $\Gamma/J$ is increased ($s_p$ is lowered) further. (d) Experimental APQ results of a line cut of the structure factor $\mathcal{S}(|\mathbf{q}|)$ for different values of $h/J$ at the low transverse field limit of $s_p = 0.65$, showing the continued presence of the broad red continuum. (e) Classical Monte-Carlo results ($\Gamma = 0$, finite $T/J$) of a line cut of the structure factor $\mathcal{S}(|\mathbf{q}|)$ for different temperatures $T/J$. Results are for a periodic system of size $L=12$. By comparing with (b), we see that our APQ results broadly correspond to $T_{\rm eff}/J \simeq 0.4$, much higher that the bare $T_{\rm device}/J$ ratio of the device, which is roughly $0.05$.}
\label{fig:coumpound3}
\end{figure*}

Let us now examine Fig.~\ref{fig:coumpound2} in detail. In (a), our APQ data shows first that the $h/J = 0$ data has nearly zero magnetization for all transverse fields (all pause points $s_p$), thus showing that disorder-by-disorder indeed occurs; quantum fluctuations cannot lift the degeneracy of the classical spin liquid at $h=\Gamma = 0$ \cite{moessner2000two,moessner2001ising,nikolic2005theory}. The finite value of the magnetization is due to stray fields in the physical device which confer small but non-negligible $\delta h$ to the flux qubits. This effect was minimized through flux bias offset calibration but remained in our final data, perhaps due to the extreme sensitivity of the state to polarizing fields. In addition to this absence of a bulk magnetization, we measured the spin structure factor $\mathcal{S}(\bm{q})$ for various $s_p$ at $h=\Gamma = 0$ (see Fig.~\ref{fig:coumpound3} a), for which we can see broad features indicative of the constrained dynamics of individual triangles being in the $m_{\triangle} = \pm 1/3$ local configurations. These features do not evolve as the transverse field $\Gamma$ is increased. We then see that increasing $h/J$ sharply leads to the ice rule state, where $\langle m_z \rangle = -\frac{1}{3}$. Note that the classical limit is for large $s_p$, i.e. on the right of the graph. Further increase of $h/J$ leads to the fully polarized phase as shown in Fig.~\ref{fig:coumpound1} (c). What is striking from our data is that the use of APQ for small values of $s_p$ (i.e. larger $\Gamma/J$) leads to either a decrease or an increase of the magnetization, depending on whether the longitudinal field is larger or smaller than $h/J = 2.0$. For clarity, we have shown in Fig.~\ref{fig:coumpound2} (b) a close-up around the magnetization plateau. We see that, for $s_p >0.3$, the magnetization is fixed to the plateau, while it starts to diverge from it as the transverse field is increased. This is consistent with the QMC data presented in (d), with the magnetization at $h/J = 2.0$ being especially resilient to the effect of the transverse field. We note that while the trends between the two methods are in agreement, there is significant quantitative difference between the two, especially at high $\Gamma$ (low $s_p$). This is a recurring theme, and is due to the forced canting of the spins during the readout quench which is not implemented fast enough to be a true quench \cite{king2018observation, king2021quantum, izquierdo2021testing}, as mentioned in Sec. \ref{subsec:apq}. 

In Fig.~\ref{fig:coumpound2} (c), we show the frustration parameter $\avg{f}$, which also presents some striking features. We plot here the same set of $h/J$ data as in (b). We note that, in the classical limit (large $s_p$), the $\avg{m_z} = -1/3$ plateau is also associated with a plateau in $\avg{f} = -1/3$, showing that the obtained magnetic states are those of the Kagome spin ice with its constrained dynamics due to the uniform presence of frustrated bonds. We see that for $h/J < 2.0$, the addition of transverse field does not meaningfully alter the frustration parameter, even though the magnetization changes. On the other hand, for $h/J > 2.0$, the frustration quickly departs from $\avg{f} = -1/3$ around $s_p \simeq 0.3$. These results are consistent with QMC (e), although with the same caveat that the readout process is too slow. In this particular case, the QMC data shows a continuous evolution of $\avg{f}$ with respect to $\Gamma$ for $h/J < 2.0$, although all curves seem to fall on top of each other. The APQ process is simply unable to reproduce the continuous background, which seems to be erased as spins cant back into the favorable ice-rule manifold during the readout process, thus fixing $\avg{f} = -1/3$. For $h/J > 2.0$, our agreement is better, and we think that this is due to the nature of the created excitations from the ice-rule manifold \cite{lopez2023kagome}, which seems to change between low and high longitudinal field. In Fig.~\ref{fig:energetics}, we show the classical energetics of different configurations of Ising spins on an antiferromagnetic triangle with a longitudinal field $h$. We note that while the gap between the lowest energy state is maximal at $h/J = 2.0$, the nature of the first excited state is different for low and high fields. We hypothesize that the application of the transverse field leads to the proliferation of different types of defects, which influences both the behavior of $\avg{m_z}$ and $\avg{f}$. For $h/J < 2.0$, defects of $m = + 1/3$ are created among the configuration of $\avg{m_z} \sim -1/3$, which lowers the absolute magnetization (upward trend of orange points in Fig.~\ref{fig:coumpound2} (b)) and since these defects maintain the ice-rule, they do not alter the frustration $\avg{f}$. For $h/J > 2.0$, defects of $m = -1$ are created, which increases the absolute magnetization (green and blue points in Fig.~\ref{fig:coumpound2} (b)) and alters the frustration $\avg{f}$ since these states violate the ice-rule. 

\begin{figure}[t]
\includegraphics[width=\columnwidth]{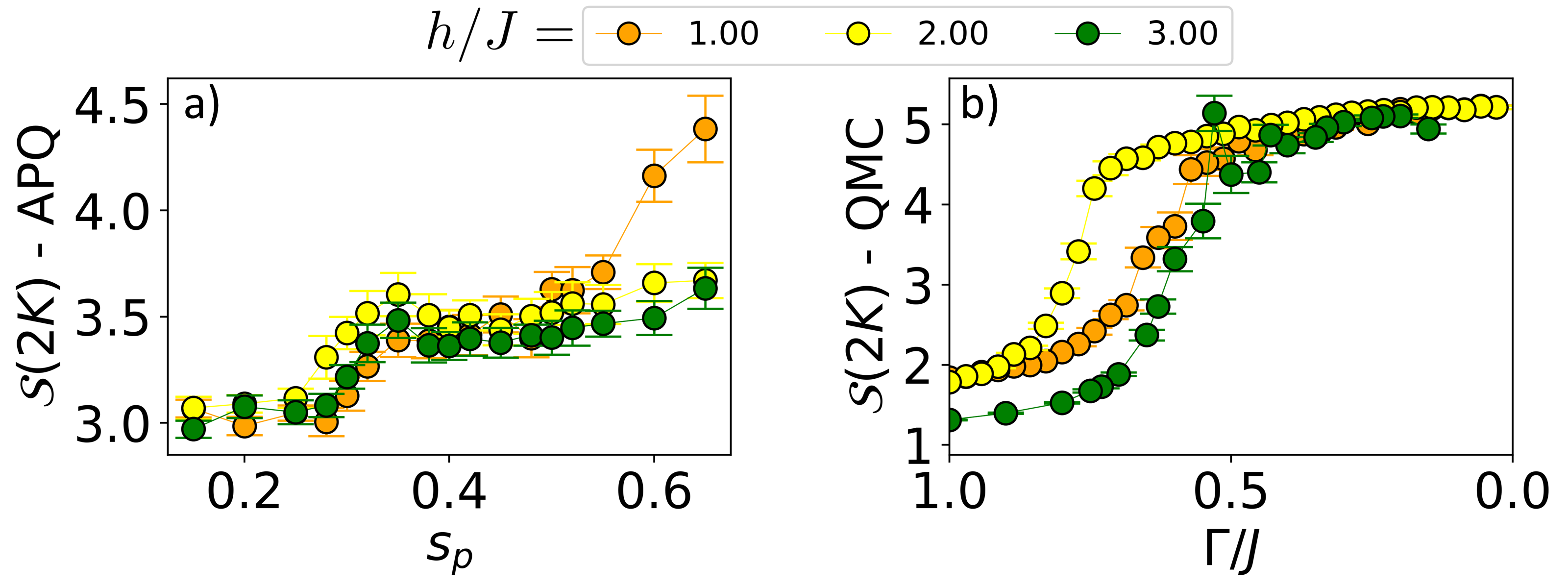} 
\caption{Comparison of the spin structure factor $\mathcal{S}(\mathbf{q})$ at the high symmetry point $\mathbf{q} = 2K$ for (a) the anneal-pause-quench (APQ) and our quantum Monte Carlo (QMC) simulations, for various values of $h/J$ deep in the blue lobe of Fig.~\ref{fig:coumpound1} (a). The $\Gamma/J$ horizontal axis of (b) is inverted to further the comparison with the $s_p$ pause point of APQ in (a). In both data, we see a sharp drop in $\mathcal{S}(2K)$ at moderate transverse field strengths, indicating the system exiting the constrained manifold of the ice-rule states. The QMC simulations are performed for $L=24$ and $\beta J=10$.}
\label{fig:dimer}
\end{figure}

\subsection{Spin Structure Factor}
\label{subsec:fft}

In order to further our insight into the different magnetic configurations generated by the APQ schedule at low/high $s_p$, we compute the spin structure factor of our $Z$ basis measurements, and compare it to QMC data. The spin structure factor $\mathcal{S}(\bm{q})$ is given by the two-dimensional Fourier transform of the spin configurations:

\begin{align}
    \sigma_{\bm{q}}^z &= \frac{1}{\sqrt{N_b}} \sum_{i} e^{i \bm{q}\cdot \bm{r}_i} \sigma_i^z \;, \\
    \mathcal{S}(\bm{q}) &= \avg{|\sigma_{\bm{q}}^z \sigma_{-\bm{q}}^z |} \;,
\end{align}

\noindent where the $\avg{\cdots}$ refers to an average over all obtained configurations. In most cases, we only look at the structure factor $\mathcal{S}(\bm{q})$, although in some cases important insight will be found in the complex-valued order parameter $\sigma^z_{\bm{q}} = m_{\bm{q}} e^{i \theta_{\bm{q}}}$. The present model of Kagome spin ice is equivalent to pyrochlore spin ice in a moderate magnetic field in $[111]$ direction \cite{moessner2003theory,turrini2022tunable}. The enforcement of the ice rules in the Kagome planes leads to dipolar correlation functions in real space \cite{youngblood1980correlations,garanin1999classical,moessner2003theory} and correspondingly to features in momentum space called \emph{pinch points} and diffuse scattering peaks in the structure factor. Specifically for Kagome spin ice, the structure factor at zero transverse field was previously obtained using a programmable quantum annealer in Ref.~\cite{lopez2023kagome}.

In Fig.~\ref{fig:coumpound3} we show the two-dimensional spin structure factor for four important points in the phase diagram, namely $h/J = 0.0$ and $2.0$ at both $s_p = 0.65$ (small transverse field) and $s_p = 0.15$ (large transverse field). Limits of Brillouin zones are outlined as dashed lines to indicate the six-fold rotational symmetry, while in one of the 2D plots we show a path in $\bm{q}$ space along the high symmetry points, $\bm{q} \in \{ \mathit{\Gamma} \rightarrow K \rightarrow 2K \rightarrow 2M \rightarrow M \rightarrow \mathit{\Gamma} \}$, which is then used for the line cuts of (b-e). Note that we use $\mathit{\Gamma}$ to denote the  $\bm q =  (0,0)$ high-symmetry point so as to avoid confusion with the transverse field $\Gamma$.

There are several conspicuous features in $S({\bm q})$ that are borne out both in the APQ and the QMC data. There is a sharp Bragg peak at 
$\mathit{\Gamma} = (0,0)$, a smaller Bragg peak at $2M = (\frac{2\pi}{a}, \frac{2\pi}{\sqrt{3}a})$ and a rather broad diffuse scattering peak at $2K = (\frac{8 \pi}{3a}, 0)$, which is the main focus of the following analysis. Only visible with a sufficient number of ${\bm q}$-points, for small $\Gamma/J$ and low temperatures there is a kink-like feature at ${\bm K} = (\frac{4 \pi}{3a}, 0)$. Here, reciprocal lattice vectors are written in terms of the lattice constant $a$ of the triangular Bravais lattice underlying the Kagome lattice.

The Bragg peak at the $\mathit{\Gamma}$-point, which is related to the uniform magnetization per site of $\avg{m_z}=-\frac{1}{3}$ ($h/J>0$) is clearly present for $h/J = 2.0$ in Fig.~\ref{fig:coumpound3} (a), as well as in (b-d). If the Bragg peak at $\bm{q}=2M$ and equivalent reciprocal lattice vectors, which correspond to nematic ordering of triangles with average magnetization $-\frac{1}{3}$, is subtracted, a sharp feature at the $2M$ points remains. These so-called \emph{pinch points} at $\bm{q}=2M$ are the hallmark of the spin-ice states in the one-third magnetization plateau.
\emph{Pinch points} are singular points in momentum space which are at the center of bow-tie like features and characterized by a highly anisotropic angular dependence of the structure factor: while $S(\bm{q})$ is non-zero and increases along the path $2M \rightarrow 2K$, it drops to zero sharply in the transverse direction along $2M\rightarrow \mathit{\Gamma}$. This feature is clearly seen for small $\Gamma/J$ (large $s_p$) in Fig.~\ref{fig:coumpound3}, for both APQ (b) and QMC (c), and for the lowest temperature in Fig.~\ref{fig:coumpound3} (e).
The \emph{pinch points} are absent in the disordered phase at $h/J=0$, see Fig.~\ref{fig:coumpound3} (a) and (d). This shows that the data obtained at $h/J = 0.0$ represents a quantum paramagnet as it does not present these \emph{pinch points}. Rather, the spin structure factor in the quantum paramagnet only shows a broad continuum that stems from the presence of triangles in the three $\ket{\uparrow \uparrow \downarrow}$ configurations as well as in the three $\ket{\downarrow \downarrow \uparrow}$. This continuum is present for all obtained $\mathcal{S}(\bm{q})$, as seen in Fig.~\ref{fig:coumpound3} (d).

One of the core features that is seen in the spin structure factor is the influence of the transverse field on the peak at $\bm{q} = 2K$. For QMC data in Fig.~\ref{fig:coumpound3} (c), there is a sharp reduction of this peak as the transverse field in increased, signaling the entrance into a quantum paramagnetic state and the loss of correlations within the ice-rule manifold. This decrease in $\mathcal{S}(2K)$ is also seen for APQ in Fig.~\ref{fig:coumpound3} (b). We compared the data at this specific high-symmetry point in more detail in Fig.~\ref{fig:dimer}. Both present qualitatively similar features for all $h/J$ values, where $\mathcal{S}(2K)$ flattens at small $\Gamma/J$ (large $s_p$), and crosses over into a smaller value at large $\Gamma/J$ (small $s_p$). The quantitative discrepancy between the QMC and APQ values is again due to the readout quench and the canting effect that partially destroys the states created during the pause at finite $\Gamma/J$.

Finally, we have also performed classical Monte-Carlo simulations at finite temperature to compare the effect of temperature and transverse field on the spin structure factor. This is presented in Fig.~\ref{fig:coumpound3} (e). One can see that increasing the temperature leads to mostly the same effects as increasing $\Gamma/J$ (as seen in (c)), with the exception of the pinch points, which are more resilient to transverse field than they are to the effect of thermal broadening. Thus, the fact that the feature at $\bm{q} = 2M$ in the APQ data does not change appreciably as $s_p$ is tuned helps to confirm that the main tuning knob in the APQ data is the transverse field, not the temperature of the system, even though the temperature is clearly finite.

\subsection{Possibility of dimer ordering} \label{sec:dimer}

\begin{figure*}
 \includegraphics[width=\textwidth]{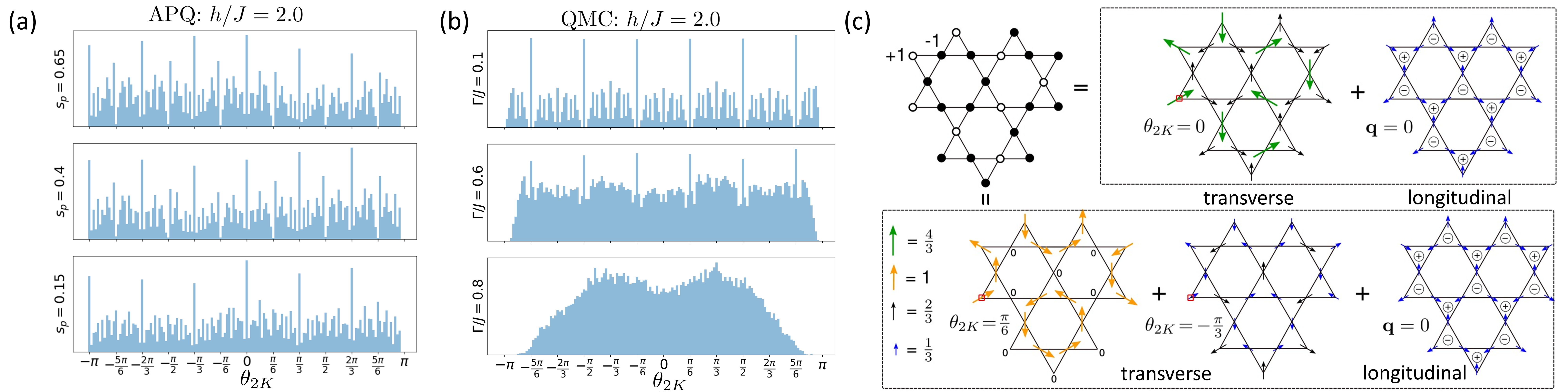}
 \caption{(a,b) Histogram of the phase angle $\theta_{2K}$ of the complex order parameter $\sigma^z_{\bm{q}}$ at momentum ${\bf q}=2K$ for the APQ schedule (a) and the QMC simulations (b) for increasing transverse field (decreasing pause point $s_p$). (c) A spin-ice configuration of up and down spins is mapped onto a planar vector field by associating an up (down) spin with an arrow of length 1 pointing into (out of) an upward-pointing triangle. The vector field resulting from a given $\sqrt{3} \times \sqrt{3}$ spin-ice configuration is decomposed into a divergence-free (transverse) fragment at ${\bf q}=2K$ and a divergence-full (longitudinal) fragment at ${\bf q}=0$ \cite{museur2023neutron}. The upper part of (c) shows the decomposition into one of the six dimer star ordered states (here, $\theta_{2K}=0$) and a longitudinal component carrying the average magnetization $\langle m_z \rangle = -\frac{1}{3}$. Other transverse configurations will correspond to symmetry-related $\theta_{2K}=2\pi n/6$.
 The lower part of (c) shows how the transverse fragment at ${\bf q}=2K$ can be further decomposed into a partially disordered state ($\theta_{2K}=\frac{\pi}{6}$) and another star dimer ordered state ($\theta_{2K}=-\frac{\pi}{3}$) of smaller
 amplitude. Equivalent decompositions lead to the other states from the two sets $\theta_{{\bf q}=2K} \in \{\pm \frac{5\pi}{6}, \pm \frac{3\pi}{6}, \pm \frac{\pi}{6}\}$ and $\theta_{{\bf q}=2K} \in \{0, \pm \frac{\pi}{3}, \pm\frac{2\pi}{3}, \pi \}$ in the transverse fragment. Thermal or quantum fluctuations act within the transverse fragment and enhance the amplitudes of these six-fold degenerate (clock-ordered \cite{isakov2003interplay}) states: As the phase angle histograms in panels (a) and (b) suggest, the different boundary conditions of the simulated Kagome lattices in APQ (mixed open-closed) and QMC (fully periodic) appear to lead to the selection of different sets of clock-ordered states, the partially disordered states in QMC and the dimer star ordered states in APQ. The parameters of the QMC simulation are $L=9$ and $\beta J = 10$.}
 \label{fig:histogram_phase}
\end{figure*}

Fig.~\ref{fig:histogram_phase} shows the histogram of the phase angle $\theta_{{\bf q}}$ of the complex order parameter $\sigma_{\bf q}^{z} = m_{{\bf q}} e^{i\theta_{\bf q}}$ at wavevector ${\bf q}=2K$, 
the ordering vector corresponding to a $\sqrt{3}\times \sqrt{3}$ unit cell. The results from the APQ simulations are shown in Fig.~\ref{fig:histogram_phase}(a) while the QMC results ared shown in Fig.~\ref{fig:histogram_phase}(b). In a longitudinal field there are three symmetry-related states with $\sqrt{3}\times \sqrt{3}$ unit cell which obey the ice rules and have average magnetization $m=-\frac{1}{3}$. The other set of three states has magnetization $m=\frac{1}{3}$. In the upper-left corner of Fig.~\ref{fig:histogram_phase}(c) one of these three states is depicted, which would be selected by quantum fluctuations from the manifold of the hexagonal lattice dimer model \cite{moessner2000two} at $\Gamma/J=0$ because the alternating spins on two out of three hexagons allow resonance processes at order $\left(\frac{\Gamma}{h}\right)^6$ ~\cite{moessner2000two}. However, it was already pointed out in Ref.~\cite{moessner2003theory} that in Kagome spin ice the structure factor at ${\bf q} = 2K$ does not exhibit a Bragg peak, but rather a diffuse scattering peak that diverges only logarithmically with system size, i.e. there is no long-range order at this wavevector. This is reminiscent to the broad peak observed in both APQ and QMC data in Fig.~\ref{fig:coumpound3}.

It turns out that the observation of an extended spin-ice phase
at finite-transverse field without long-range order and the mechanism of order-by-disorder
can be elegantly reconciled using the concept of magnetic moment fragmentation \cite{brooks-bartlett2014fragmentation, museur2023neutron, rougemaille2019cooperative}. First, spin configurations are expressed as a vector field in the plane
by associating an up-spin with an arrow pointing into an upward-pointing triangle (each site of the Kagome lattice can be identified to one and only one such up-triangle). 
According to the Helmholtz decomposition any vector field can be written as a sum of a transverse fragment, which is divergence-less,
and a longitudinal fragment, which is curl-less \cite{museur2023neutron}.
All states at ${\bf q}=2K=(\frac{8\pi}{3a}, 0)$ have vanishing average magnetization as is verified from the expression 
$\sum_{i=1}^{3}\sigma_i^{z} = m \sum_{i=1}^{3}\cos(\frac{8\pi}{3a} x_i + \theta_{2K}) = 0$
with the coordinates on a kagome triangle $x_i \in \{0, \frac{a}{4}, \frac{a}{2} \}$. Thus any spin-ice configuration consists of a divergence-free 
fragment of spin loops with zero average magnetization,
which contribute to the spatial dipolar correlations and the diffuse scattering \cite{moessner2003theory}, and a ${\bf q}=0$ fragment carrying the average magnetization of $-\frac{1}{3}$. The Helmholtz decomposition for a single spin ice configuration is shown schematically in Fig.~\ref{fig:histogram_phase}(c, upper row). 

A remnant of the order-by-disorder effect due to quantum and thermal fluctuations is apparent in the transverse fragment of the spin configuration \cite{museur2023neutron} at ${\bf q}=2K$ where two sets of six-fold degenerate (clock-ordered) states with $\sqrt{3} \times \sqrt{3}$ unit cell have relatively high amplitude compared to other $\sqrt{3} \times \sqrt{3}$ states. These are the \emph{partially disordered states} with phase angle of the Fourier transform of the spin configuration $\theta_{{\bf q}=2K} \in \{\pm \frac{5\pi}{6}, \pm \frac{3\pi}{6}, \pm \frac{\pi}{6}\}$ and the \emph{dimer star ordered states} \cite{museur2023neutron} with phase angle $\theta_{{\bf q}=2K} \in \{0, \pm \frac{\pi}{3}, \pm\frac{2\pi}{3}, \pi \}$. States from each set are depicted in Fig.~\ref{fig:histogram_phase}(c, lower row) where the transverse fragment is a dimer star ordered state with phase angle $\theta_{{\bf q}=2K}=0$, see Fig.~\ref{fig:histogram_phase}(c, top), which can be further decomposed into a partially disordered state plus smaller corrections, see Fig.~\ref{fig:histogram_phase}(c, bottom).

The key observation is that the $\sqrt{3} \times \sqrt{3}$ ordered state from the spin-ice manifold in the upper-left corner of Fig.~\ref{fig:histogram_phase}(c) is three-fold degenerate if the magnetization per triangle is enforced to be $-\frac{1}{3}$. On the other hand, the $\sqrt{3} \times \sqrt{3}$ spin configurations 
from the transverse (i.e. divergence-less) fragment are allowed to be six-fold degenerate as they have average magnetization per triangle equal to zero. This shows that the selection of clock-ordered $\sqrt{3} \times \sqrt{3}$ states by quantum or thermal fluctuations occurs in the transverse fragment of the magnetization. As the height of the diffuse peak at ${\bf q}=2K$ increases only logarithmically with the system size \cite{moessner2003theory}, it is evident that the slightly favoured $\sqrt{3} \times \sqrt{3}$ states do not develop long-range order. 

Whereas in the APQ results the dimer star ordered states are enhanced (see the large peaks at $\theta_{2K} = 0 \pm 2\pi n/6$ in Fig.~\ref{fig:histogram_phase}(a)), in the QMC simulations the preferred phase angles are those of the partially disordered states (see the dominant peaks at $\theta_{2K} = -\pi/6 \pm 2\pi n/6$ in Fig.~\ref{fig:histogram_phase}(b)). A contribution from the partially disordered states is also seen for large $s_p$ in the APQ results. The deviation between the two is most likely due to the different boundary conditions of the respective simulation cells, see Fig.~\ref{fig:coumpound1}. Furthermore, in the QMC simulation results, the depletion of the $\Gamma/J \sim 0.8$ histogram towards $\theta_{2K} = \pm \pi$ is due to the very large transverse field which polarizes the spins to point in the direction of $(h, \Gamma)$. This datapoint is deep into the paramagnetic regime, and when we measure the transverse component of the magnetization in the $Z$ basis, it returns a broad feature with two peaks around $\theta_{2K} = \pm \pi/3$, corresponding to a finite $m_x$ measurement. However, in the APQ simulations the six peaks in the angular histogram do not disappear for large transverse field (small pause point $s_p$), which is likely due to the slow readout quench. Indeed, the disordered configurations that are stable at large transverse field (small $s_p$) appear to be carried through by the slow quench to the dimer star ordered states, for which the system seems to be particularly susceptible to. Altogether the analysis of the histogram of the phase angle at the high-symmetry ${\bf q}=2K$ point reveals that, although no long-range dimer star ordered state sets in, we indeed observe consistent correlations in the transverse part of the ice manifold. It remains unknown whether these correlations would lead to long-range order at much lower temperature or simply remain correlated features atop the spin ice manifold.

\subsection{Evidence of the crossover} \label{sec:crossover}

\begin{figure}[t]
\includegraphics[width=\columnwidth]{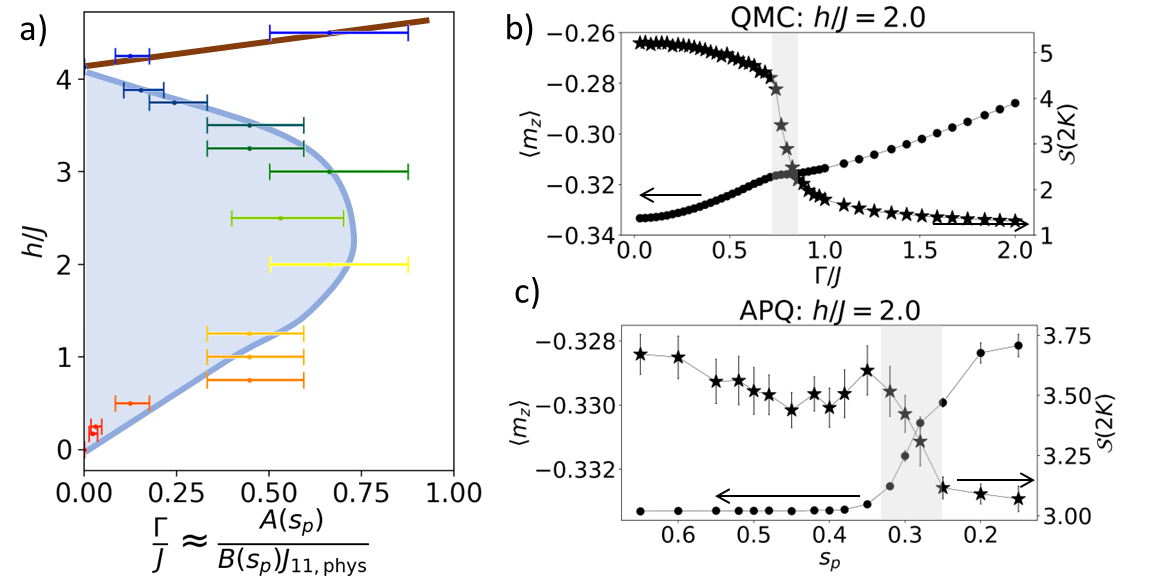} 
\caption{(a) Tracking of the inflection point $s_c$ of the magnetization $\avg{m_z}$, as seen in Fig.~\ref{fig:coumpound2} - we see a rather compelling similarity with the theoretical phase diagram of Fig.~\ref{fig:coumpound1} (a). The blue area are line is meant to be a guide to the eye. The $x$ axis represents extracted values of the ratio $\Gamma/J$, which depends on the values of $A$, $B$ for the pause point $s_p = s_c$ and the bare coupler $J_{11, \rm phys}$ on the device. The magnetization is not the order parameter of any dimer-ordered state. However, similar trends are seen in the structure factor at the $\mathbf{q} = 2K$ point, in both the QMC results (b) and the APQ results (c). We show in both (b-c) the magnetization on the left vertical axis and the structure factor $\mathcal{S}(2K)$ on the right vertical axis. The gray region shows the crossover region that takes the system out of the "ice-rule" region due to the transverse field $\Gamma/J$ and into the paramagnetic state, and this is where we have tracked our inflection points $s_c$ of (a). The parameters of the QMC simulation are $L=9$ and $\beta J = 10$.}
\label{fig:phasediagram}
\end{figure}

We have tracked the inflection point of the magnetization $\avg{m_z}$, which occurs for a given pause point $s_c$. From this $s_c$, we have extracted values of the ratio $\Gamma/J = A(s_c)/(B(s_c) J_{11, \rm phys})$, where $A,B$ are the control functions and $J_{11, \rm phys}$ is the bare value of the magnetic exchange interaction in the coupled Josephson junction device. In Fig.~\ref{fig:phasediagram}, we show the evolution of this tracked inflection point as a function of $h/J$. Blue lines are added as a guide to the eye; it clearly shows common trends with the proposed phase diagram of Fig.~\ref{fig:coumpound1} (a). However, we must be clear: $\avg{m_z} = -\frac13$ by itself is \textbf{not} an order parameter for the possible $\sqrt{3} \times \sqrt{3}$ dimer ordered state \cite{moessner2000two}. As shown in Fig.~\ref{fig:phasediagram} (b-c), the inflection point of the magnetization occurs in the same region as the inflection point of $\mathcal{S}(2K)$, for both the QMC and the APQ data. This is compelling data that there indeed is a crossover from some lower degeneracy manifold of states to a quantum paramagnet as the transverse field is increased (or, experimentally, as the pause point is reduced). Although we are not in position to confirm the existence of a long-range ordered dimer state, the necessary ingredients for its existence, namely the magnetization plateau, the ice rules and the $\mathcal{S}(2K)$ peak, all disappear rather abruptly as a function of transverse field. It is then not inconceivable that, should a low-temperature dimer order exist, it would also be brought to disorder at the same $\Gamma$ and $s_p$ values.

At this point we may venture an explanation as to the fate of the order-by-disorder scenario proposed in Ref.~\cite{moessner2000two}, in which a  true long-range ordered, three-fold degenerate state with $\sqrt{3} \times \sqrt{3}$ unit cell emerges, rather than merely an enhancement of six-fold degenerate clock-ordered states in the transverse fragment of the Kagome ice phase, which is what we observed in both QMC simulations and APQ experiments. Referring to studies of related quantum Kagome ice models where the transverse-field term is replaced by other quantum tunneling terms \cite{damle2006spin,carrasquilla2015two,wu2019tunneling}, it is found that the degeneracy of the Kagome ice manifold could be either lifted \cite{damle2006spin} or not lifted \cite{carrasquilla2015two,wu2019tunneling} by quantum fluctuations. As said earlier, treating the transverse-field term as a perturbation acting on the spin-ice manifold, the lowest order off-diagonal term that could lift the degeneracy is a ring exchange term of order $(\Gamma/h)^6$, which would drive the system into the three-fold degenerate (dimer) ordered state with a $\sqrt{3} \times \sqrt{3}$ unit cell \cite{moessner2000two}. However, as pointed out in \cite{wu2019tunneling}, to the same order of perturbation theory there are also diagonal terms that give a negative energy shift to other states in the spin-ice manifold. Overall, this may lead to a restoration of the classical spin-ice manifold as a quasi-degeneracy \cite{wu2019tunneling}. We note that, in the effective dimer model proposed by Ref.~\cite{moessner2000two}, the dimer flip term goes like $K_{\rm dimer} \sim \Gamma^6/J^5$. In the context of our simulations on D-Wave's annealer, the region or $s_p \in [0.172, 0.318]$ would result in $\log_{10} (K_{\rm dimer}/J) \in [-2,2]$. Considering that the APQ temperature is on the order of $T_{\rm eff}/J \sim 0.4$, this would indicate that only a very narrow range of $s_p$ could result in simulation points where $K_{\rm dimer} \sim T_{\rm eff}$, where the dimer term is on the order of the temperature. As all accessible $s_p$ points were scanned, no indication of the putative clock phase were seen, but it is possible that such phase only reveals itself at temperatures beyond our QMC or APQ protocols, thus we cannot exclude the possibility of a three-fold degenerate $\sqrt{3} \times \sqrt{3}$ ordered state at much lower temperature or larger transverse field. 

Our inability to come to fully quantitative agreement with QMC simulations comes from three current constraints on the device. Firstly, as mentioned before, the readout quench is not fast enough to faithfully collapse a complex entangled state onto the $Z$ basis. Fast readout quenches were recently achieved by the D-Wave team \cite{king2022coherent}, though such features remain unavailable to cloud access users. Secondly, the prototype allows for the embedding of only 231 sites of the Kagome lattice, leading to a cylindrical lattice with a few unit cells width. Fortunately, this point will soon be tackled as the Advantage2 device becomes available within the year. This device will have thousands of qubits in a $Z_{15}$ Zephyr graph (as opposed to $Z_4$), meaning that Kagome cylinders of up to $3000$ sites would be realizable using our embedding. This would additionally allow to carry out a full finite-size scaling analysis. Third, improvements in calibration and device manufacturing will make it so that higher bare $J_{ij, \rm phys}$ couplers will be achieved on the D-Wave devices, thus permitting even lower temperature quantum simulations.

\section{Conclusion and outlook} 
\label{sec:conclusion}

We presented the results of the quantum simulation of the antiferromagnetic transverse field Ising model on the Kagome lattice on a programmable quantum annealer. After embedding a cylinder of the Kagome lattice onto the Zephyr graph of the latest D-Wave prototype, we used a specific type of annealing schedule called anneal-pause-quench (APQ) to generate a set of states representative of the ground states of the Hamiltonian in the presence of both a longitudinal field and a transverse field. We were then able to characterize the magnetic properties, including the spin structure factor, of the obtained states, and compare those to the results of quantum Monte-Carlo simulations. We uncovered good qualitative agreement between the features of the magnetization, frustration and structure factor across the range of fields that we simulated the model at. From this analysis, we were able to construct a partial phase diagram, reconciling the pioneering studies of Moessner, Sondhi and Chandra \cite{moessner2000two} with the concept of magnetic moment fragmentation in kagome spin ice \cite{museur2023neutron}, thus shedding new light on the properties of this fundamental model of magnetic frustration.

Our key observation is that the Kagome spin ice regime, with its characteristic 
features in the structure factor due to the ice rules, namely pinch points and diffuse scattering peaks, 
is stable to quantum fluctuations in an extended $-\frac{1}{3}$ magnetization plateau before crossing over to the field-polarized trivial state. 
At the ordering vector ${\bf q}=2K$ of the diffuse scattering peak, a histogram of the phase angle of the complex order parameter reveals that a set of six-fold degenerate states is favoured over other spin configurations. The underlying clock anisotropy \cite{isakov2003interplay} can be attributed to the effect of thermal and quantum fluctuations in the transverse fragment of the magnetization. 

Finally, our study in the absence of a longitudinal field showed conclusive evidence that indeed, as predicted theoretically \cite{moessner2001ising}, the classical spin liquid state that exists at $h = \Gamma = 0$ is not brought to order under the presence of a finite $\Gamma \neq 0$, thus showcasing "disorder-by-disorder" on a quantum computer. This was observed in the transverse field (resp. $s_p$) independent features (magnetization, frustration, structure factor) obtained using the APQ schedule at $h=0$.

The quantum annealing device used in this study poses a number of technical challenges for faithful quantum simulation: 1) the effective temperature of our quantum simulation is too high, as seen from the thermal broadening of the pinch points in Fig.~\ref{fig:coumpound3}(b,e) (see also \cite{zhou2021experimental, zhou2022probing, hearth2022quantum} for similar conclusions); 2) the APQ schedule is not yet correctly implemented, with the quench to the $Z$ basis taking too long so as to result in significant canting of the spins, thereby destroying fragile states in the $X$ basis \cite{king2018observation, king2021quantum, izquierdo2021testing}. These two conclusions about the state of the hardware are consistent with the literature. Nevertheless, we were able to successfully reproduce qualitative features of our model, as benchmarked with QMC, even though quantitative agreement still eludes us. We note that the runtime necessary for the gathering of the APQ and QMC results differs significantly - all APQ calibration, datapoints and errorbars were obtained using less than 2 hours of real-time device access, while the QMC simulations ran for far longer time. It is thus interesting that, should the challenges presented in this paragraph be solved, one could obtain faithful quantum simulation results for the Kagome system in a short time compared to QMC approaches.

We are hopeful that soon much larger AFI-K systems will be able to be studied on the Advantage2 machine, once it is released. By simply recycling our embedding on this device, cylindrical lattices with up to $3000$ sites will be embedded. This, coupled with improved control on the rapidity of the quench (and therefore better implementation of the APQ method) and the next generation of chip fabrication with lower effective temperature and higher coherence, will undoubtedly help sharpen the presented results and charter the path ahead.

Looking even more into the distance, we are hopeful that soon other types of coupling terms between qubits, such as the XY coupling $\sigma_i^x \sigma_j^x + \sigma_i^y \sigma_j^y$, will be implemented in future generations of superconducting quantum annealers. These terms can either be implemented in an effective Hamiltonian evolution, using Floquet dynamics \cite{ciavarella2022floquet, ciavarella2023state, scholl2022microwave}, or directly within the hardware, as was recently demonstrated in Ref.~\cite{ozfidan2020demonstration} for two flux qubits. Indeed, one could then simulate XXZ or XY models on the Kagome lattice, which are thought to leads to exotic $\mathbb{Z}_2$ spin liquid ground states \cite{block2020kagome, lauchli2019s}. In particular, the Heisenberg Kagome antiferromagnet $H = J \sum_{ij} \vec{S}_i \cdot \vec{S}_j$ has been the subject of a quarter of a century of exhaustive study of its ground state. Whether it remains disordered down to $T=0$, and if so, what type of non-trivial spin liquid exists, are all open questions that a future efficient quantum simulation could help untangle. Furthermore, if indeed a non-trivial quantum spin liquid could be realized on the annealer, this could provide a new route to quantum error correction, as non-local qubits can be implemented in the stable ground states of these entangled many-body states. This was recently achieved in neutral atoms \cite{semeghini2021probing, samajdar2021quantum, verresen2022unifying}, where the realization of an underlying network of qubits on a Kagome lattice was essential to the realization of toric-code-like states. These are exciting directions, and the results presented here, notably the novel insight into the low-temperature phase diagram and the observed absence of long-range order, should serve as another clear example of the near-term usefulness of quantum annealers to study frustrated spin systems.

\acknowledgments

VDT would like to thank Premala Chandra and Rhine Samajdar for illuminating discussions. The work of AR and VDT was supported from a grant from the Simons Foundation (825876, TDN). SH was supported by the Simons Collaboration on the Many Electron Problem (319274FY19).
Computing resources of the Center for Materials Theory at the Department of Physics and Astronomy at Rutgers University are acknowledged.









\section*{Appendices}

\appendix

\section{Embedding} \label{appendix1}

\begin{figure*}[!htb]
\includegraphics[width=0.7\textwidth]{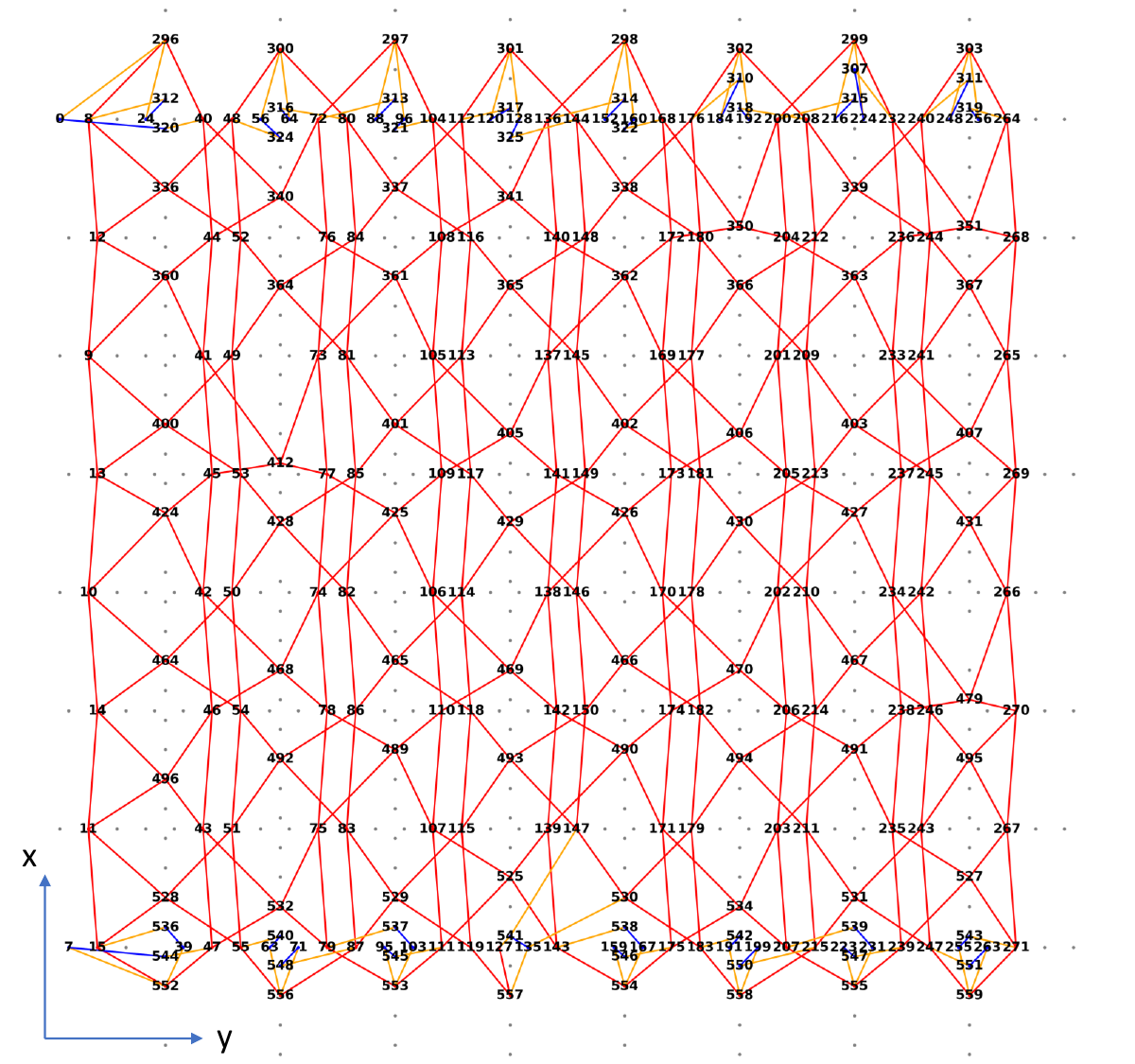} 
\caption{Our embedding of a 231 site Kagome lattice with periodic boundary conditions (in the $y$ direction of the graph) into the D-Wave $\rm Advantage2\_prototype1.1$ QPU with the Zephyr $Z_4$ native graph. Strong ferromagnetic bonds to form large effective qubits are in blue, while orange bonds are called $J_{12}$ couplings (between sites represented by single qubits and chains). The rest of the couplings are named $J_{11}$ and are the bare coupling on the Kagome lattice.}
\label{fig:embedding}
\end{figure*}

In Fig.~\ref{fig:embedding}, we show our full embedding in the prototype machine. The QPU used is the D-Wave $\rm Advantage2\_prototype1.1$ quantum annealer housed in Burnaby, BC, Canada, operating at $12 mK$, accessed remotely through the Leap platform. The nature of alternate embeddings (see qubits 412 versus 405 and 406) in the bulk is explored in this section, while the role of ferromagnetic chains (in blue) is explored in the following section.

The D-Wave quantum annealer has a few missing qubits $\{i\}$ and missing couplers $\{J_{ij}\}$ that are absent of the working graph (here, Zephyr) because of fabrication issues. These defects are static and do not vary while we run our simulations on the QPU through cloud access. Therefore, we can locate the defects and alter our embedding in order to bypass there defects. 

\begin{figure}[h]
\includegraphics[width=\columnwidth]{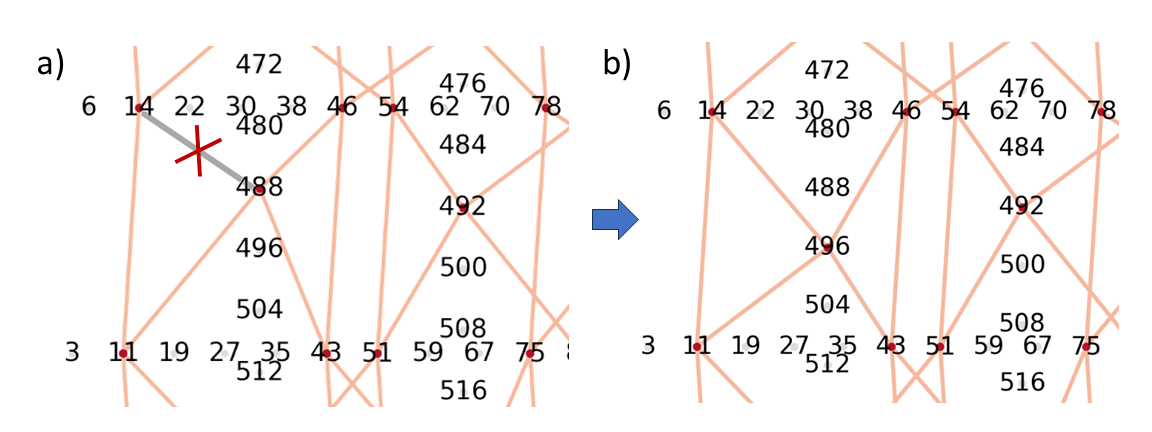} 
\caption{Replacing a missing coupler by altering the local embedding. In (a), the initial embedding requires the coupling between qubits $(14, 488)$, which is allowed on the Zephyr graph but absent in the accessed machine. By moving qubit $488$ to $496$ (b), we can obtain all the necessary finite couplers to be active.}
\label{fig:supp1}
\end{figure}

We found that none of the qubits needed for our embedding were missing, and that we only had to face missing couplers in the bulk of our Kagome lattice. Since our embedding on the Zephyr graph is rather sparse (only 231 qubits are used out of 576 available), we can simply replace the use of a qubit $i$ with missing coupler $J_{ij}=0$ with an alternate qubit $i'$ such that $J_{i'j} = J$. When doing so, we have to make sure that qubit $i'$ is also connected to all other qubits $\{j\}$ that the initial qubit $i$ was connected to. A schematic of this is presented in Fig.~\ref{fig:supp1}

Therefore, all bulk sites of the Kagome lattice are identically represented by a single qubit of the Advantage2 device. Some sites at the edges of our lattice are, however, represented using chains of strongly ferromagnetically coupled qubits (called 2-chains). Since the graph connectivity is lessened at the edge and we required that all sites be part of a complete triangular unit cell of the Kagome lattice (therefore not having dangling sites), we use this gadget to embed more sites on the device. To preserve uniformity on the lattice, we are required to tune the couplers and local fields on the 2-chains, which we present in the next section.

\section{Chain fixing using anneal offsets} \label{appendix2}

Some sites on our lattice may be represented by a chain of strongly ferromagnetically coupled qubits (called n-chains). This is due to the limited connectivity of the native Zephyr graph as physically implemented. Indeed, in the realistic devices available, some couplers $J_{ij}$ or qubits $\sigma_i$ are absent due to manufacture defects, while the available device only implements a finite size Zephyr graph (namely, the prototype we had access implements the $Z_4$ Zephyr graph). This means that the edges of the graph are sparse in couplers, and that we may need to implement some qubits as 2-chains to enhance their connectivity. As we saw in the previous section, we fix missing couplers in the bulk by shifting the embedding locally, so that all bulk sites of our Kagome lattice are represented by single qubits in the annealer. We cannot do the same for the two open edges of our cylindrical lattice, and therefore must resort to 2-chains.

The ferromagnetic bond in 2-chains is implemented through the D-Wave machine's extended J range, which gives us access to a strong FM coupling value of $K=-2$. We limit the other exchange interactions to $|J| \leq 1$ so that the chain interaction remains the strongest. Examinations of our results shows that chain breaks (where measurement in the Z basis of the chain gives a non-aligned result) occur extremely rarely - such configurations are discarded. The local Hamiltonian of a site $i$ represented by a 2-chain arrangement of qubits $(i,1)$ and $(i,2)$ is given by 

\begin{equation}
    H_{2ch} = K\sigma^z_{i,1} \sigma^z_{i,2} -h[\sigma^z_{i,1} + \sigma^z_{i,2}] - \Gamma [\sigma^x_{i,1} + \sigma^x_{i,2}]
\end{equation}

We consider a situation where this ferromagnetic interaction dominates $K \gg \Gamma$, and we project the 2 qubit Hilbert space onto the low-energy manifold of an effective qubit $\ket{\sigma_i^z}$ given by $\ket{\uparrow} = \ket{\uparrow \uparrow}$ and  $\ket{\downarrow} = \ket{\downarrow \downarrow}$, where the second term is the state of both qubits $\ket{\sigma_{i,1}^z \sigma_{i,2}^z}$. Operators $\tilde{\sigma}^{\mu}$ apply on this new effective spin at the site. After performing a Schrieffer-Wolff transformation \cite{schrieffer1966relation} and shifting the energy levels by a constant, we get the effective Hamiltonian

\begin{equation}
    H_{2ch}^{\rm eff} = -h_{\rm eff}\tilde{\sigma}^z_{i} - \Gamma_{\rm eff}\tilde{\sigma}^x_{i}
\end{equation}

with 

\begin{equation}
    \begin{split}
    h_{\rm eff} &= 2 h \\
    \Gamma_{\rm eff} &= 2\frac{\Gamma^2}{|K|}  
    \end{split}
    \label{eq:schriefferwolff}
\end{equation}

We then find that the transverse field (and therefore, the quantum dynamics) on the 2-chains will be different from the sites of our lattice represented by a single qubit, in accordance to previous work on cotunneling of strongly coupled qubits \cite{lanting2010cotunneling}. In this section, we present a way to circumvent this using anneal offsets, that is parameters tunable via cloud access of D-Wave's devices, such that the lattice parameters $J$, $\Gamma$ and $h$ are isotropic on the lattice, no matter if the site is represented as a single qubit or through a 2-chain. Extension to longer chains could be done in the future, depending on the range of accessible anneal offsets, which we now introduce. Parts of the protocol presented in this section have been presented in different references over the year \cite{lanting2010cotunneling, andriyash2016boosting}; for clarity of future investigations, we present it fully here.

\begin{figure}[h]
\includegraphics[width=\columnwidth]{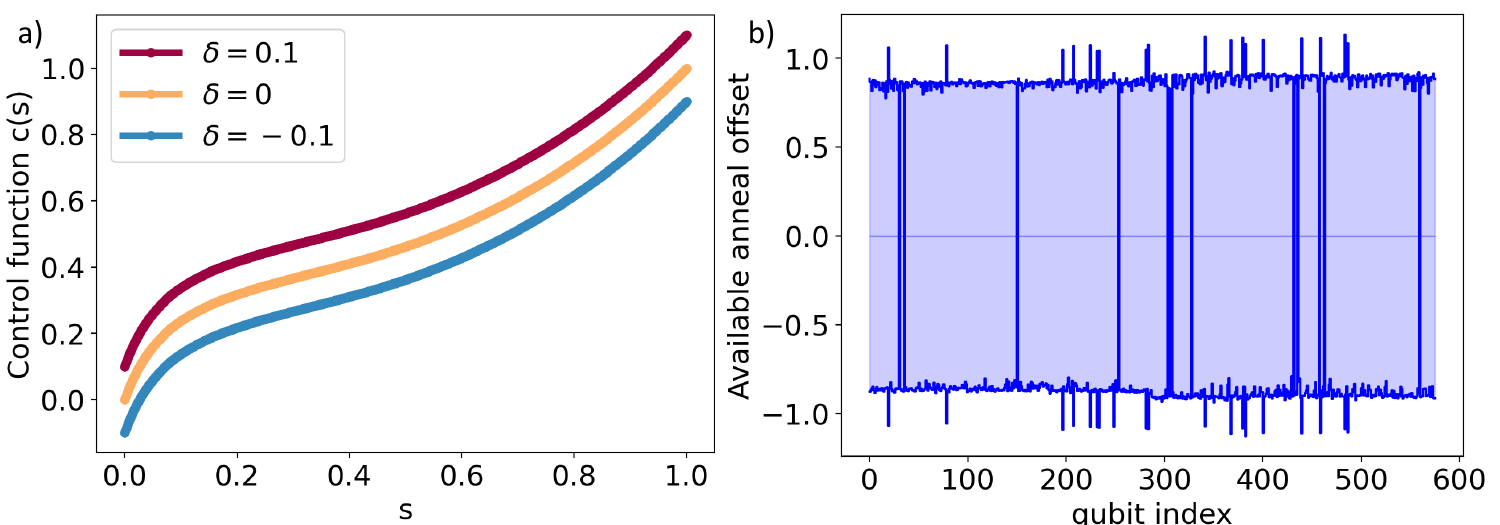} 
\caption{(a) The control bias $c$ (which itself controls both $A(s)$ and $B(s)$) as a function of the anneal fraction $s$. We plot the bare schedule ($\delta = 0$) as well as an advance schedule $\delta = 0.1$ and a retarded schedule $\delta = -0.1$. (b) The full range of anneal offsets $\delta$ that are accessible via cloud access on the D-Wave $\rm Advantage2\_prototype1.1$ device.}
\label{fig:supp2}
\end{figure}

The full Hamiltonian implemented on the D-Wave device is the following:

\begin{equation}
\begin{split}
        &\mathcal{H}_{\rm DWave} (s) = - \frac{A(s)}{2} \Big[ \sum_i \sigma_i^x \Big] \\
    &+ \frac{B(s)}{2} \Big[ \sum_{\avg{i,j}} J_{ij, \rm phys} \sigma^z_i \sigma^z_j + \sum_i h_{i,\rm phys} \sigma^z_i \Big]
\end{split}
\end{equation}

The values of $A(s)$, $B(s)$ are predetermined for a given QPU, as shown in the main text in Fig.~\ref{fig:coumpound1} (c). When we perform an anneal-pause-quench (APQ) schedule, as illustrated in the main text in Fig.~\ref{fig:coumpound1} (b), we anneal to a specific $s=s_p$ where we pause for $100 \mu s$ before quenching as fast as possible (highest slope available on the cloud access is $1/\mu s$) to $s=1$, where it is possible to measure in the $Z$ basis. The hope of this protocol is that the system equilibrates under the full Hamiltonian $\mathcal{H}_{\rm DWave} (s_p)$. For our TFIM Kagome lattice experiments, this means that the chosen $s_p$ corresponds to a chosen $A(s_p)/B(s_p)$ ratio which is related to $\Gamma/J$ in the spin model we are embedding.

The functions $A(s)$ and $B(s)$ are both controlled by a joint function $c(s)$, which is an external room temperature current source that sets $\Phi_{CCJJ}(s)$, the external flux applied to all qubit compound Josephson-junction structures to change the potential energy shape of the rf-SQUID qubit in D-Wave's devices. One has

\begin{equation}
    c(s) = \frac{\Phi_{CCJJ}(s) - \Phi_{CCJJ}^{\rm initial}}{\Phi_{CCJJ}^{\rm final} - \Phi_{CCJJ}^{\rm initial}}
\end{equation}

This single $c(s)$ curve, shown in Fig.~\ref{fig:supp2} (a), is a priori identical for all qubits, prescribing a single annealing path. This is ideal if (1) one is only interested in anneals to $s=1$ and not in quantum dynamics at a given $s_p$ or if (2) all sites of the system simulated are represented by chains of equal length. In the second situation, the effective transverse field on the sites of the lattice is less than that of the bare transverse field on the qubits, but it is uniform so it poses less of a worry. In our case, the mixing of single qubits and 2-chains means we may need to alter this curve \textit{per qubit}. Fortunately, the anneal offset parameter $\delta_i$ provides that effect. A delayed ($\delta >0$) and advanced ($\delta <0$) control curve is shown in Fig.~\ref{fig:supp2} (c), while the range of $\delta_i$ is shown in Fig.~\ref{fig:supp2} (b). Absent qubits from the Zephyr graph due to manufacturing issues also have absent offset, but largely all qubits have a range that hovers around $|\delta_i|< 0.8$. 

When using the anneal offset, one is individually delaying or advancing the dynamics of certain qubits, so that $c_i(s_p)$ may vary for given qubits $i$ - we can now use this in our advantage to make it so that the effective transverse field on 2-chains is identical to the bare transverse field parameter $A(s_p)$. This entire process then permits us to rewrite $A(s) \rightarrow A_i(s)$ and $B(s) \rightarrow B_i(s)$ as they become site dependent using the anneal offsets $\delta_i$. This impacts the effective coupling $\tilde{J}_{ij}$, which becomes 

\begin{equation}
\begin{split}
    \tilde{J}_{ij} &= J_{ij, \rm phys} \sqrt{B_i (s) B_j(s)} \\
    &\equiv J_{ij, \rm phys} \sqrt{B (s + \delta_{s,i}) B(s + \delta_{s,j})} \label{eq:delayJ} 
\end{split}
\end{equation}

We consider our Kagome embedding at a given pause point $s_p$, and attempt to fix the couplings and single site fields to be uniform through the system for a defect-free quantum simulation. At this point, we have the following bare parameters for the bulk of the lattice:

\begin{equation}
    \begin{split}
    h &= B(s_p)h_{1,\rm phys}/2 \\
    \Gamma &= A(s_p)/2 \\
    J &= B(s_p)J_{11,\rm phys}/2
    \end{split}
\end{equation}
where $h_{1,\rm phys}$ is the user specified longitudinal field on single qubit sites, while $J_{11,\rm phys}$ is the nearest neighbor coupling constant between two "normal" sites, i.e. represented by single qubits in the embedding. In addition, we have the effective longitudinal and transverse fields for the 2-chains, as obtained by the Schrieffer-Wolff transformation of Eq.~\ref{eq:schriefferwolff}:

\begin{equation}
    \begin{split}
    h_{2,\rm eff} &= B(s_p + \delta_s) h_{2,\rm phys} \\
    \Gamma_{2, \rm eff} &= \frac{A(s_p + \delta_s)^2}{2 B(s_p + \delta_s)} 
    \end{split}
\end{equation}
where $h_{2,\rm phys}$ is the user specified longitudinal field on qubits part of a 2-chain, while $B(s_p + \delta_s)$ and $A(s_p + \delta_s)$ are the modified $A$ and $B$ functions after the use of anneal offsets $\delta$. For qubits belonging to a 2-chain, we shift the control function $c(s_p) \rightarrow c_2(s_p) = c(s_p) + \delta \equiv c(s_p + \delta_s)$ where we associate a new shift $\delta_s \neq \delta$ for simpler notation. In effect, these qubits are either advanced or retarded such that they evolve at a different $\tilde{s}_p = s_p + \delta_s$ with different $A$ and $B$ values. Our conditions for uniformity on the lattice then becomes that

\begin{equation}
\begin{split}
    h &\equiv h_{2,\rm eff} \\
    \Gamma &\equiv \Gamma_{2, \rm eff}
     \label{eq:conditionhGamma}
\end{split}
\end{equation}

The most constraining equation here is the one on the transverse field. We can see that solving it leads to the condition

\begin{equation}
    A(s_p) = \frac{A(s_p + \delta_s)^2}{B(s_p + \delta_s)} 
    \label{eq:constraintGamma}
\end{equation}

\begin{figure}[h]
\includegraphics[width=\columnwidth]{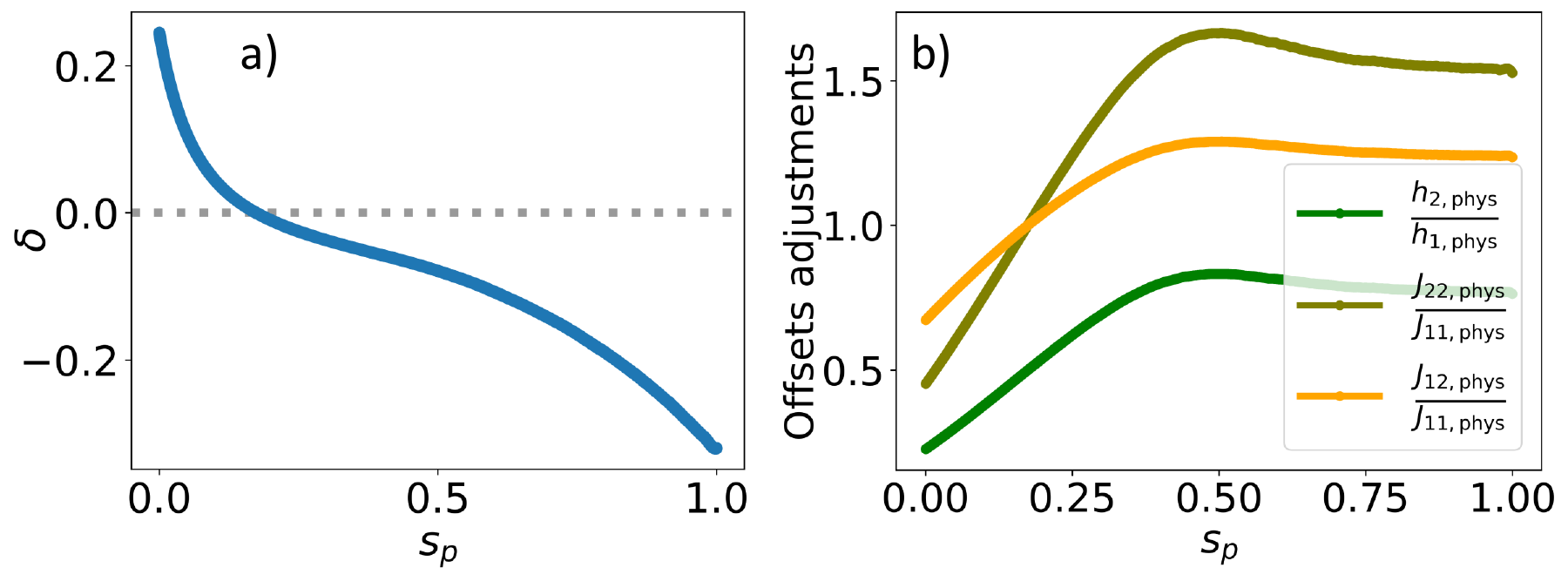} 
\caption{(a) Anneal offset $\delta$ needed on the 2-chains to satisfy Eq.~\ref{eq:constraintGamma} as a function of the pause point $s_p$. Note that is always stays within the range available (see Fig.~\ref{fig:supp1} (d)). (b) The values of $J_{12,\rm phys}$, $h_{2, \rm phys}$ and $J_{22,\rm phys}$ that must be set with respect to the bare ones to preserve uniformity once the anneal offsets are applied.}
\label{fig:supp3}
\end{figure}

Having access to the values of $A$, $B$ and $c$ as a function of $s$, accessible online at \href{https://docs.dwavesys.com/docs/latest/doc\_physical\_properties.html}{this link}, we can find a shift $\delta_{s,i}$ for given 2-chain at site $i$ that leads to the effective evolution at a distinct $\tilde{s}_{p,i} = s_p + \delta_{s,i}$. For the $\rm Advantage2\_prototype1.1$ QPU, we obtain the curve shown at Fig.~\ref{fig:supp3} (a). 

We note however that, unfortunately for us, the available data only at \href{https://docs.dwavesys.com/docs/latest/doc\_physical\_properties.html}{this link} for the $A(s)$ and $B(s)$ values is not perfect. The company itself specifies that these values can vary by up to $30\%$, which is not ideal for the purpose of our method. We opted to proceed along, though we encourage D-Wave to provide an updated real-time table of these values for cloud access, so that processes like ours may be completed more reliably.

Having now satisfied the condition on the transverse field, we must deal with the consequences of altered $B(s_p + \delta)$ values for the longitudinal field and the coupling values $J_{ij}$. The equation on the longitudinal fields is solved readily, and we find

\begin{equation}
    h_{2,\rm phys} = h_{1,\rm phys} \frac{B(s_p)}{2B(s_p + \delta)}
\end{equation}

Therefore, setting $h_{2,\rm phys}$ to this value for any given $s_p$ leads to uniform $h_{\rm eff}$. 

The final difficulty resides in the AF $J$ coupling between the single qubit sites and the 2-chain sites. The $J$ value has to be corrected because of the delay/advance on $B(s_p + \delta)$ on the 2-chain, as shown in Eq.~\ref{eq:delayJ}. The $J$ value between two single qubit sites is $J = B(s_p)J_{11,\rm phys}/2$, where $J_{11,\rm phys}$ is set by the user. We have two other types of spin couplings: $J_{12}$ and $J_{22}$, which respectively couple a single qubit to a 2-chain, and a 2-chain with a 2-chain. We need to set $J_{22, \rm phys}$ and $J_{12, \rm phys}$ so that $J \equiv J_{12} \equiv J_{22}$. Let us first consider $J_{22}$, the simplest to write down; we have

\begin{align}
    J_{22} &= J_{22, \rm phys} B(s_p + \delta_s)/2 \nonumber\\
    &\equiv J = B(s_p) J_{11, \rm phys}/2 \\
    \Rightarrow& \qquad J_{22, \rm phys} = J_{11, \rm phys} \frac{B(s_p)}{B(s_p + \delta_s)}
\end{align}

The situation is a bit more murky for the $J_{12}$ couplings, where, per Eq.~\ref{eq:delayJ}, we have

\begin{align}
     J_{12} &= J_{12,\rm phys} \sqrt{B(s_p)B(s_p + \delta_s)}/2 \nonumber \\
     &\equiv J = J_{11,\rm phys}B(s_p)/2 \\
     \Rightarrow& \qquad J_{12, \rm phys} = J_{11,\rm phys} \sqrt{\frac{B(s_p)}{B(s_p + \delta_s)}}    
\end{align}

We find that all of these physical input parameters are within the range of realizable ones on the cloud access. The values of $J_{12, \rm phys}$, $J_{22, \rm phys}$ and $h_{2,\rm phys}$ are shown in Fig.~\ref{fig:supp3} (b). One sees that, in order to maintain $|K| > |J_{ij}|$ for the strong ferromagnetic couplings, we need to set $J_{11, \rm phys} \leq 1$ so that itself $\max \{J_{12, \rm phys}, J_{22, \rm phys} \} \leq 1$ and we preserve the separation of scales. We find in general that this procedure leads to an extremely rare event of chain breaks, and we immediately reject a configuration containing chain breaks. Otherwise, when qubits in a 2-chain remain align at the measurement, we use a majority rule for the output on the Kagome lattice.

Note that this technique, which uniformizes the interactions $J$ and single site terms $h$ and $\Gamma$ across the lattice, can also be used to replace missing couplers $J_{ij}$ when an alternate embedding is unavailable. Should one have $J_{ij}=0$ because of manufacturing issues, one can then implement the following two 2-chains, via this Hamiltonian:

\begin{equation}
    H_{\rm stitch} = K \sigma_{i,1}^z \sigma_{i,2}^z + J \sigma_{i,2}^z \sigma_{j,1}^z + K \sigma_{j,1}^z \sigma_{j,2}^z
\end{equation}

Both 2-chains then need to have rescaled $h$ and $J$ values following the equations above. We have tested this stitching procedure, and it leads to similar results than the alternate embedding presented in section I.

\section{Shimming the flux bias values for improved statistics} \label{appendix3}

In this section, we briefly review the process outlined in the Tutorial paper on shimming \cite{chern2023tutorial}, and also presented in Refs.~\cite{king2018observation, kairys2020simulating, nishimura2020griffiths}. In real devices like the one accessed for this study, many sources can leads to less than ideal performance. Non-uniform fabrication can lead to small stray fluctuations in the magnetic environment, leading to some qubits being more biased towards one state - in effect, a random $\tilde{h}_i \sigma^z_i$ field on each qubit. Furthermore, a finite nearby coupler $J_{ij} \neq 0$ can lead to crosstalk and, again, a bias field $\tilde{h}_i$. For this reason, careful calibration of the device is achieved through the use of "shimming", where features of the Hamiltonian are tuned so as to reproduce as faithfully as possible the statistical behavior of an ideal and uniform system. 

In the case of our quantum simulation of the Kagome lattice at finite transverse field using the APQ schedule, we have to adapt the shimming process slightly. The first part is standard: we use an iterative gradient descent method to update the flux-bias offsets (FBO) $\Phi_i$ for each used qubit such that the \textit{local} magnetization converges to the \textit{global} magnetization. All code needed to operate the shimming process and gather the measurements is available online~\cite{githubref}.  Any given iteration $n$ has $N_{\rm reads}$ shots of APQ schedule, with a readout thermalization time of $100\; \mu s$. Then, the observed magnetization per qubit $m_i = \avg{\sigma_i^z}$ is obtained, as well as the average global magnetization $\bar{m} = \frac{1}{N_{\mathcal{B}}} \sum_{i \in \mathcal{B}} \avg{\sigma_i^z}$. Note that not all qubits are part of the sum here - we only use qubits in the \textbf{bulk} of our lattice (the set $\mathcal{B}$ with $N_{\mathcal{B}}$ sites), i.e. sites part of an elementary triangle that does not touch the boundaries of the system. This helps in two ways: sites at the edge already have a preliminary condition on them of a halved longitudinal field due to their reduced number of connections, and the adjustment of flux bias offset on the edges will then further help in assuring that the local magnetization on those sites is as close as possible to that in the bulk. 

\begin{figure}
\includegraphics[width=\columnwidth]{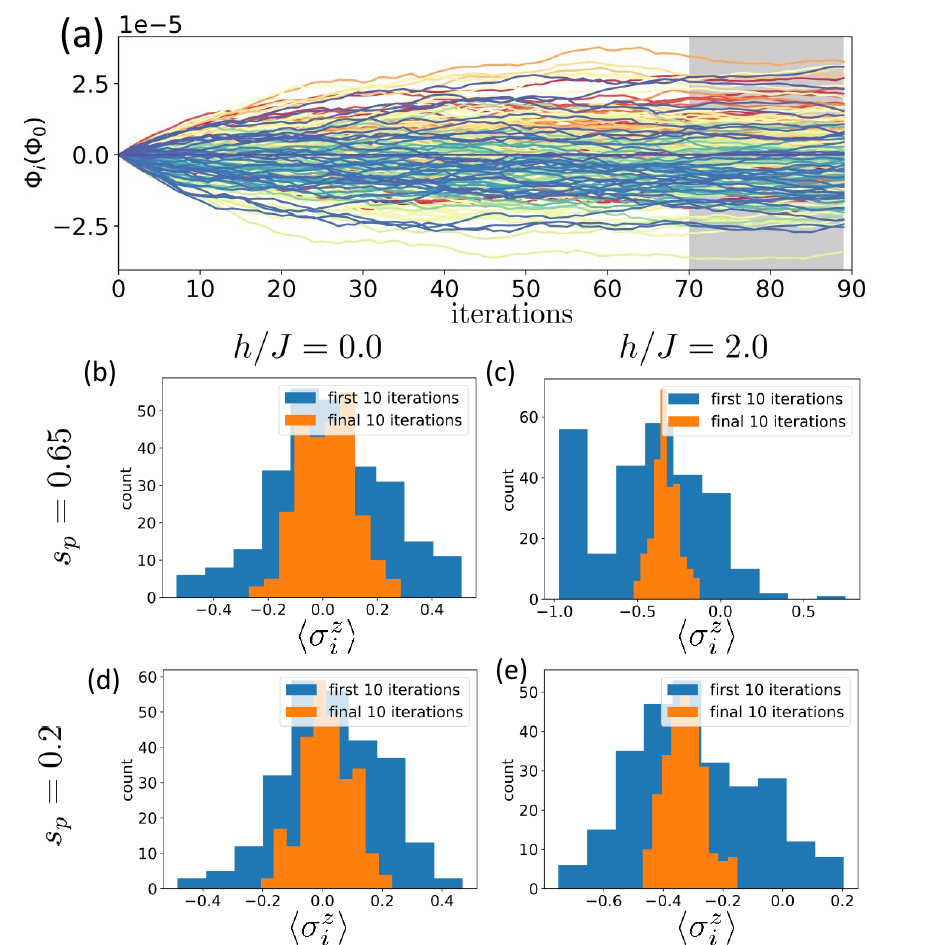} 
\caption{(a) Individual flux bias per qubit versus iteration. In the region of measurement, which is shaded in grey, each iteration consists of $N_{\rm reads} = 1000$ measurement shots, while in the shimming iterations that precede it, each iteraction is comprised of $N_{\rm reads} = 100$ shots. In (b-e), we show the difference in the distribution of individual magnetization per site $\langle \sigma_i^z \rangle$ for the first 10 iterations (in blue) and the final 10 iterations in the measurement section (in orange). In (b-c) $s_p =0.65$ results for different $h/J$. In (b), we are in the paramagnetic phase where $\avg{m} = 0$, while in (c) we are in the dimer phase where $\avg{m} \simeq \frac{1}{3}$. Note the sharpening of the distribution after shimming. (d-e) $s_p =0.2$ results for different $h/J$, which are both in the paramagnetic phase but with different structure: whereas in (d) we have $\avg{m} = 0$, in (e) we have $\avg{m} = \frac{1}{3} + \eta$ with $\eta \neq 0$ departing from the dimer state. Again, shimming sharpens those distributions and more accurately simulates a uniform lattice.}
\label{fig:shim}
\end{figure}

Mathematically, this process then becomes, for the FBOs at iteration $n$

\begin{equation}
 \Phi_i^{(n)} = \Phi_i^{(n+1)} - \alpha (m_i - \bar{m})
\end{equation}

The shimming constant $\alpha$ is taken to be $\alpha = 3 \times 10^{-6}$, after larger and smaller values were tested; larger values lead to chaotic $\Phi_i$ values while smaller values take too many iterations to converge \cite{chern2023tutorial}. As opposed to other shimming process where all $J_{ij}$ and $h_i$ are set to zero during the calibration, in this routine the model parameters are as is - we are merely calibrating the device to simulate the input model as ideally as possible. In practice, we perform this shimming in two steps:

\begin{enumerate}
    \item The \textit{shimming steps}: There are $N_{\rm reps} = 70$ iterations of the shimming where we perform $N_{\rm reads} = 100$ using the APQ schedule; at then end of these steps the FBOs are mostly stable. 
    \item The \textit{measure steps}: There are $N_{\rm reps} = 20$ iterations of the shimming where we perform $N_{\rm reads} = 1000$ using the APQ schedule; while the FBOs can still fluctuate in these steps, we collect the result of all shots and use those to obtain the value of the observables presented in the text.  
\end{enumerate}

The result of one such iterative process for the FBOs leads to a "spaghetti" plot as shown in Fig.~\ref{fig:shim} (a). Each line corresponds to an individual qubit. While many qubits that are unused are not updated (see horizontal line at $\Phi_i = 0$), all used qubits seem to settle around a fixed value after few iterations. The gray area on the right then corresponds to the measure steps.

This process helps tremendously to settle the statistics and improve our results. In Fig.~\ref{fig:shim} (b-e) we show a histogram of the average magnetization $m_i$ for each qubit, as averages over the first 10 iterations of the shimming versus the last 10 iterations. As can be seen, for both $h/J = 0.0$ and $h/J = 2.0$ and extreme cases of $s_p$, the shimming process helps center the distribution of qubits closer to $\bar{m}$, signifying a more uniform behavior of the simulated model.

Note that, in its most general setting, one also does shimming iterations on the couplings $J_{ij}$, updating those couplings by calculating a \textit{frustration probability} $f_{ij} = (1 + \text{sign}(J_{ij}) \avg{\sigma_i^z \sigma_j^z})/2$ such that, for each iteration, $J_{i,j}^{(n)} = J_{i,j}^{(n-1)}(1 + \alpha_J(f_{ij} - \bar{f}))$. The frustration becomes then uniform and this compensates for the case where couplings are not ideally realized. We opted to not do this routine, as it was too costly in term of QPU access time, and because we generally found that the shimming of the flux bias already largely helped the results.  

\section{Details on the Quantum Monte-Carlo Simulations}

In this section, we present additional data stemming from quantum Monte-Carlo simulations of the antiferromagnetic Kagome TFIM, as well as details with the algorithm implementation.

\subsection{Additional data}


\begin{figure}
\includegraphics[width=0.9\linewidth]{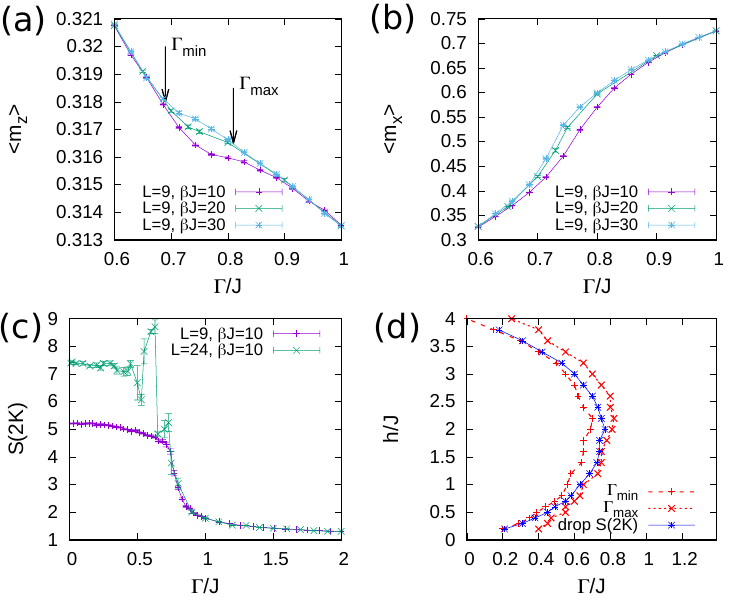}
 \caption{Effects of finite temperature in the crossover region are visible in the longitudinal (a) and transverse magnetization (b), which can be used to define a crossover region between $\Gamma_{\text{min}}$ and $\Gamma_{\text{max}}$ (d). The kink-like feature in the magnetization coincides with the drop in the structure factor $S(2K)$, see (c). The Zeeman field in (a-c) is $h/J=-2$.}
 \label{fig:crossover_boundary}
\end{figure}
The largest inverse temperature of the QMC data presented in the main text is $\beta J=10$.
Curves of $\langle m_z \rangle(\Gamma)$, transverse magnetization $\langle m_x \rangle(\Gamma)$ and energy (not shown) taken at larger inverse temperatures $\beta J= 20, 30$ for $L=9$ indicate that in the crossover region an inverse temperature of $\beta J=10$ is not below the finite size gap and temperature effects are still present (Fig.~\ref{fig:crossover_boundary}(a,b)). 
As illustrated in Fig.~\ref{fig:crossover_boundary}(a), one could use the points where magnetization curves for $\beta J=20$ and $\beta J=30$ start to deviate, $\Gamma_{\text{min}}$, and merge again, $\Gamma_{\text{max}}$, as a criterion to delineate a crossover region where the gap of the system is significantly reduced. 
The location of the sudden drop of $S({\bf q}=2K)$ falls within this region. The resulting crossover phase diagram for $L=9$ shown in Fig.~\ref{fig:crossover_boundary}(d) is in agreement with the experimentally determined phase diagram in Fig.~\ref{fig:phasediagram} of the main text, where the crossover is located based on the inflection point of the magnetization. 
It should be pointed out that there is only an indirect connection between the structure factor $S({\bf q}=2K)$ and the uniform magnetization $\langle m_z \rangle$ in that a drop in the former happens to be accompanied by a kink-like feature in the latter. 
 In the crossover region the transverse magnetization $\langle m_x \rangle$ increases superlinearly with transverse field while the longitudinal magnetization $\langle m_z \rangle$ stays almost constant. In order for the $\langle m_x \rangle$ to rise rapidly while $\langle m_z \rangle$ stays at $\frac{1}{3}$, magnetic fluctuations may involve strings of flipped spins. This observation points towards some form of topological metamagnetic transition \cite{pili2022topological,baez3d2017}, which requires further investigation.

\subsection{QMC method}

\begin{figure*}
\includegraphics[width=0.9\textwidth]{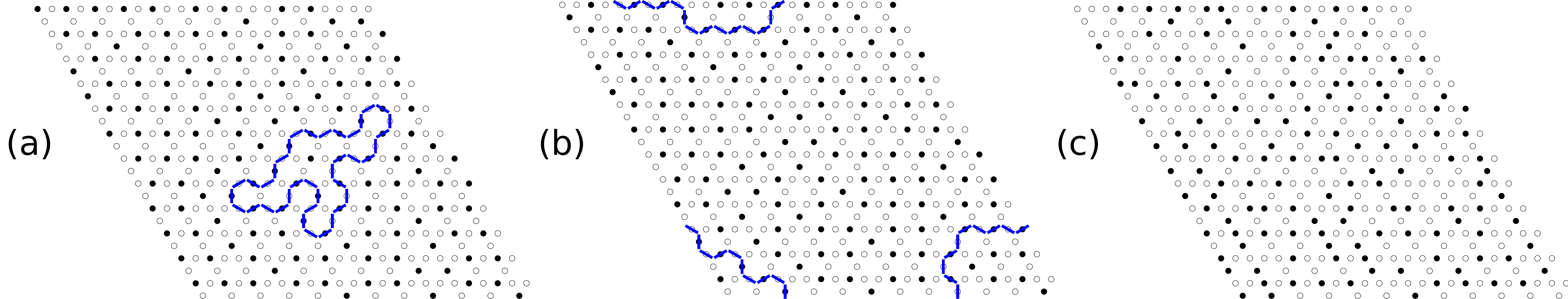}
 \caption{(a,b) Snapshot at a randomly chosen imaginary time slice of configurations in the spin ice regime ($h/J=2, \Gamma/J=0.2, T/J=0.1$) with kagome loops
 which, when reversed, do (b) or do not (a) change the winding number sector of the configuration.
 (c) Configuration in the paramagnetic regime ($h/J=2, \Gamma / J = 1$). Open circles denote spin down.}
 \label{fig:loops}
\end{figure*}

We performed standard path integral quantum Monte Carlo simulations with a finite Trotter discretization of imaginary time $\delta \tau$ given by
$\Gamma \delta \tau = 0.02$ with $N_{\tau} = \frac{\beta}{\delta \tau}$ imaginary time slices. 
To avoid dynamical freezing due to strong ferromagnetic couplings in imaginary time the 
``line update'' \cite{Nakamura2008efficient} has been used, which consists in building Wolff clusters in imaginary time. 
A reason for formulating the algorithm in discrete imaginary time is that efficient loop updates for classical ice models \cite{Barkema1998ice} can be extended to the quantum model in the form of a ``membrane update'' \cite{Henry2014squareice}. 
This approach has been implemented for the case of quantum kagome ice in Ref.~\cite{YaoWang2020}.
The kagome loop update is also straightforwardly applicable to the spin ice regime in the $m=-\frac{1}{3}$ magnetization plateau ($h>0$). Since closed loops always contain an even number of alternating spins (see Fig.~\ref{fig:loops}), the Zeeman energy cancels out and flipping a loop is an iso-energetic move that does not leave the $m=-\frac{1}{3}$ plateau.
Long loops, where the loop head is required to close on its starting position (rather than any loop segment as for ``short loops``), are necessary \cite{Barkema1998ice} for ergodic sampling of all winding number sectors (see Fig.\ref{fig:loops}).
\begin{figure}
 \includegraphics[width=0.9\linewidth]{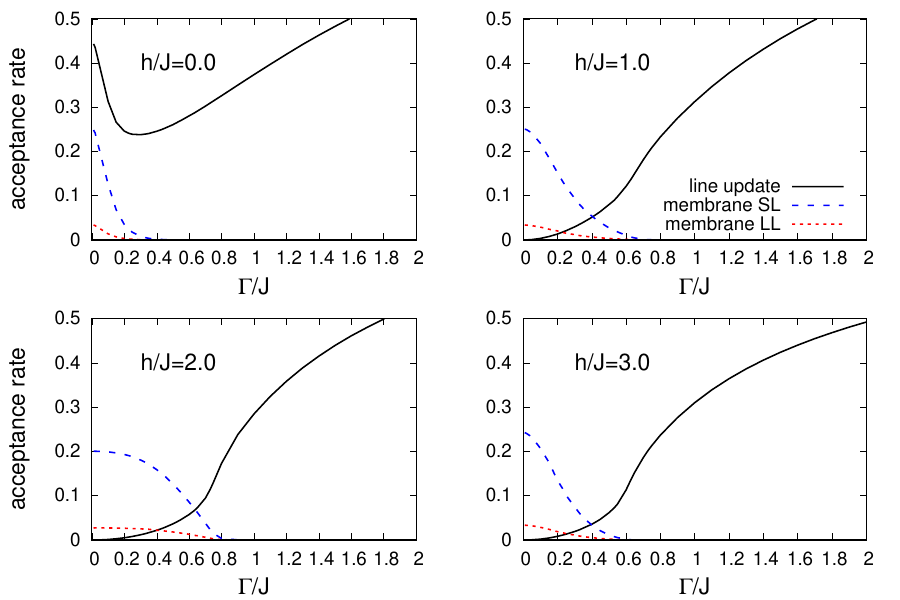}
 \caption{Acceptance rate of different cluster updates employed in the path integral QMC simulations: The ``line update'' of Wolff clusters in imaginary time and the ``membrane update'' with  short loops (SL) and long loops (LL) grown in imaginary time. Parameters: $L=24$, $\beta J=10$.}
 \label{fig:acc_rate}
\end{figure}
Fig.~\ref{fig:acc_rate} shows the acceptance rate for the different cluster updates used. The small acceptance rate of the line update for small $\Gamma / J$ illustrates the necessity of flipping spatially extended clusters in order to sample states from the spin-ice manifold. As to be expected, the parameter regime where the acceptance rate of the membrane update is non-vanishing coincides with the spin-ice regime. 

Each Monte Carlo step (MCS) consists of 10 lattice sweeps of line updates followed by 4 membrane updates each consisting of the attempted construction of $L$ long kagome loops and $3L^2 / 6$ short kagome loops. $10^4$ thermalization MCS and $5\times 10^4$ measurement MCS have been used.
 When the acceptance rate as determined during the thermalization phase becomes too small, the membrane update is switched off.  
 Error bars are determined using the binning analysis or the jackknife method.

\bibliography{biblio}

\begin{thebibliography}{83}%
\makeatletter
\providecommand \@ifxundefined [1]{%
 \@ifx{#1\undefined}
}%
\providecommand \@ifnum [1]{%
 \ifnum #1\expandafter \@firstoftwo
 \else \expandafter \@secondoftwo
 \fi
}%
\providecommand \@ifx [1]{%
 \ifx #1\expandafter \@firstoftwo
 \else \expandafter \@secondoftwo
 \fi
}%
\providecommand \natexlab [1]{#1}%
\providecommand \enquote  [1]{``#1''}%
\providecommand \bibnamefont  [1]{#1}%
\providecommand \bibfnamefont [1]{#1}%
\providecommand \citenamefont [1]{#1}%
\providecommand \href@noop [0]{\@secondoftwo}%
\providecommand \href [0]{\begingroup \@sanitize@url \@href}%
\providecommand \@href[1]{\@@startlink{#1}\@@href}%
\providecommand \@@href[1]{\endgroup#1\@@endlink}%
\providecommand \@sanitize@url [0]{\catcode `\\12\catcode `\$12\catcode
  `\&12\catcode `\#12\catcode `\^12\catcode `\_12\catcode `\%12\relax}%
\providecommand \@@startlink[1]{}%
\providecommand \@@endlink[0]{}%
\providecommand \url  [0]{\begingroup\@sanitize@url \@url }%
\providecommand \@url [1]{\endgroup\@href {#1}{\urlprefix }}%
\providecommand \urlprefix  [0]{URL }%
\providecommand \Eprint [0]{\href }%
\providecommand \doibase [0]{https://doi.org/}%
\providecommand \selectlanguage [0]{\@gobble}%
\providecommand \bibinfo  [0]{\@secondoftwo}%
\providecommand \bibfield  [0]{\@secondoftwo}%
\providecommand \translation [1]{[#1]}%
\providecommand \BibitemOpen [0]{}%
\providecommand \bibitemStop [0]{}%
\providecommand \bibitemNoStop [0]{.\EOS\space}%
\providecommand \EOS [0]{\spacefactor3000\relax}%
\providecommand \BibitemShut  [1]{\csname bibitem#1\endcsname}%
\let\auto@bib@innerbib\@empty
\bibitem [{\citenamefont {Moessner}\ \emph {et~al.}(2000)\citenamefont
  {Moessner}, \citenamefont {Sondhi},\ and\ \citenamefont
  {Chandra}}]{moessner2000two}%
  \BibitemOpen
  \bibfield  {author} {\bibinfo {author} {\bibfnamefont {R.}~\bibnamefont
  {Moessner}}, \bibinfo {author} {\bibfnamefont {S.~L.}\ \bibnamefont
  {Sondhi}},\ and\ \bibinfo {author} {\bibfnamefont {P.}~\bibnamefont
  {Chandra}},\ }\bibfield  {title} {\bibinfo {title} {Two-dimensional periodic
  frustrated ising models in a transverse field},\ }\href@noop {} {\bibfield
  {journal} {\bibinfo  {journal} {Physical review letters}\ }\textbf {\bibinfo
  {volume} {84}},\ \bibinfo {pages} {4457} (\bibinfo {year}
  {2000})}\BibitemShut {NoStop}%
\bibitem [{\citenamefont {Chandra}\ \emph {et~al.}(1990)\citenamefont
  {Chandra}, \citenamefont {Coleman},\ and\ \citenamefont
  {Larkin}}]{chandra1990ising}%
  \BibitemOpen
  \bibfield  {author} {\bibinfo {author} {\bibfnamefont {P.}~\bibnamefont
  {Chandra}}, \bibinfo {author} {\bibfnamefont {P.}~\bibnamefont {Coleman}},\
  and\ \bibinfo {author} {\bibfnamefont {A.~I.}\ \bibnamefont {Larkin}},\
  }\bibfield  {title} {\bibinfo {title} {Ising transition in frustrated
  heisenberg models},\ }\href@noop {} {\bibfield  {journal} {\bibinfo
  {journal} {Physical review letters}\ }\textbf {\bibinfo {volume} {64}},\
  \bibinfo {pages} {88} (\bibinfo {year} {1990})}\BibitemShut {NoStop}%
\bibitem [{\citenamefont {Shastry}\ and\ \citenamefont
  {Sutherland}(1981)}]{shastry1981exact}%
  \BibitemOpen
  \bibfield  {author} {\bibinfo {author} {\bibfnamefont {B.~S.}\ \bibnamefont
  {Shastry}}\ and\ \bibinfo {author} {\bibfnamefont {B.}~\bibnamefont
  {Sutherland}},\ }\bibfield  {title} {\bibinfo {title} {Exact ground state of
  a quantum mechanical antiferromagnet},\ }\href@noop {} {\bibfield  {journal}
  {\bibinfo  {journal} {Physica B+ C}\ }\textbf {\bibinfo {volume} {108}},\
  \bibinfo {pages} {1069} (\bibinfo {year} {1981})}\BibitemShut {NoStop}%
\bibitem [{\citenamefont {Kitaev}(2006)}]{kitaev2006anyons}%
  \BibitemOpen
  \bibfield  {author} {\bibinfo {author} {\bibfnamefont {A.}~\bibnamefont
  {Kitaev}},\ }\bibfield  {title} {\bibinfo {title} {Anyons in an exactly
  solved model and beyond},\ }\href@noop {} {\bibfield  {journal} {\bibinfo
  {journal} {Annals of Physics}\ }\textbf {\bibinfo {volume} {321}},\ \bibinfo
  {pages} {2} (\bibinfo {year} {2006})}\BibitemShut {NoStop}%
\bibitem [{\citenamefont {Anderson}(1973)}]{anderson1973resonating}%
  \BibitemOpen
  \bibfield  {author} {\bibinfo {author} {\bibfnamefont {P.~W.}\ \bibnamefont
  {Anderson}},\ }\bibfield  {title} {\bibinfo {title} {Resonating valence
  bonds: A new kind of insulator?},\ }\href@noop {} {\bibfield  {journal}
  {\bibinfo  {journal} {Materials Research Bulletin}\ }\textbf {\bibinfo
  {volume} {8}},\ \bibinfo {pages} {153} (\bibinfo {year} {1973})}\BibitemShut
  {NoStop}%
\bibitem [{\citenamefont {Balents}(2010)}]{balents2010spin}%
  \BibitemOpen
  \bibfield  {author} {\bibinfo {author} {\bibfnamefont {L.}~\bibnamefont
  {Balents}},\ }\bibfield  {title} {\bibinfo {title} {Spin liquids in
  frustrated magnets},\ }\href@noop {} {\bibfield  {journal} {\bibinfo
  {journal} {nature}\ }\textbf {\bibinfo {volume} {464}},\ \bibinfo {pages}
  {199} (\bibinfo {year} {2010})}\BibitemShut {NoStop}%
\bibitem [{\citenamefont {Iqbal}\ \emph {et~al.}(2011)\citenamefont {Iqbal},
  \citenamefont {Becca},\ and\ \citenamefont {Poilblanc}}]{Iqbal_2011}%
  \BibitemOpen
  \bibfield  {author} {\bibinfo {author} {\bibfnamefont {Y.}~\bibnamefont
  {Iqbal}}, \bibinfo {author} {\bibfnamefont {F.}~\bibnamefont {Becca}},\ and\
  \bibinfo {author} {\bibfnamefont {D.}~\bibnamefont {Poilblanc}},\ }\bibfield
  {title} {\bibinfo {title} {Valence-bond crystal in the extended kagome
  spin-$1/2$ quantum heisenberg antiferromagnet: A variational monte carlo
  approach},\ }\bibfield  {journal} {\bibinfo  {journal} {Physical Review B}\
  }\textbf {\bibinfo {volume} {83}},\ \href
  {https://doi.org/10.1103/physrevb.83.100404} {10.1103/physrevb.83.100404}
  (\bibinfo {year} {2011})\BibitemShut {NoStop}%
\bibitem [{\citenamefont {Kolley}\ \emph {et~al.}(2015)\citenamefont {Kolley},
  \citenamefont {Depenbrock}, \citenamefont {McCulloch}, \citenamefont
  {Schollwock},\ and\ \citenamefont {Alba}}]{Kolley_2015}%
  \BibitemOpen
  \bibfield  {author} {\bibinfo {author} {\bibfnamefont {F.}~\bibnamefont
  {Kolley}}, \bibinfo {author} {\bibfnamefont {S.}~\bibnamefont {Depenbrock}},
  \bibinfo {author} {\bibfnamefont {I.~P.}\ \bibnamefont {McCulloch}}, \bibinfo
  {author} {\bibfnamefont {U.}~\bibnamefont {Schollwock}},\ and\ \bibinfo
  {author} {\bibfnamefont {V.}~\bibnamefont {Alba}},\ }\bibfield  {title}
  {\bibinfo {title} {Phase diagram of the $j_1 - j_2$ heisenberg model on the
  kagome lattice},\ }\bibfield  {journal} {\bibinfo  {journal} {Physical Review
  B}\ }\textbf {\bibinfo {volume} {91}},\ \href
  {https://doi.org/10.1103/physrevb.91.104418} {10.1103/physrevb.91.104418}
  (\bibinfo {year} {2015})\BibitemShut {NoStop}%
\bibitem [{\citenamefont {He}\ \emph {et~al.}(2017)\citenamefont {He},
  \citenamefont {Zaletel}, \citenamefont {Oshikawa},\ and\ \citenamefont
  {Pollmann}}]{He_2017}%
  \BibitemOpen
  \bibfield  {author} {\bibinfo {author} {\bibfnamefont {Y.-C.}\ \bibnamefont
  {He}}, \bibinfo {author} {\bibfnamefont {M.~P.}\ \bibnamefont {Zaletel}},
  \bibinfo {author} {\bibfnamefont {M.}~\bibnamefont {Oshikawa}},\ and\
  \bibinfo {author} {\bibfnamefont {F.}~\bibnamefont {Pollmann}},\ }\bibfield
  {title} {\bibinfo {title} {Signatures of dirac cones in a dmrg study of the
  kagome heisenberg model},\ }\bibfield  {journal} {\bibinfo  {journal}
  {Physical Review X}\ }\textbf {\bibinfo {volume} {7}},\ \href
  {https://doi.org/10.1103/physrevx.7.031020} {10.1103/physrevx.7.031020}
  (\bibinfo {year} {2017})\BibitemShut {NoStop}%
\bibitem [{\citenamefont {Ebadi}\ \emph {et~al.}(2021)\citenamefont {Ebadi},
  \citenamefont {Wang}, \citenamefont {Levine}, \citenamefont {Keesling},
  \citenamefont {Semeghini}, \citenamefont {Omran}, \citenamefont {Bluvstein},
  \citenamefont {Samajdar}, \citenamefont {Pichler}, \citenamefont {Ho} \emph
  {et~al.}}]{ebadi2021quantum}%
  \BibitemOpen
  \bibfield  {author} {\bibinfo {author} {\bibfnamefont {S.}~\bibnamefont
  {Ebadi}}, \bibinfo {author} {\bibfnamefont {T.~T.}\ \bibnamefont {Wang}},
  \bibinfo {author} {\bibfnamefont {H.}~\bibnamefont {Levine}}, \bibinfo
  {author} {\bibfnamefont {A.}~\bibnamefont {Keesling}}, \bibinfo {author}
  {\bibfnamefont {G.}~\bibnamefont {Semeghini}}, \bibinfo {author}
  {\bibfnamefont {A.}~\bibnamefont {Omran}}, \bibinfo {author} {\bibfnamefont
  {D.}~\bibnamefont {Bluvstein}}, \bibinfo {author} {\bibfnamefont
  {R.}~\bibnamefont {Samajdar}}, \bibinfo {author} {\bibfnamefont
  {H.}~\bibnamefont {Pichler}}, \bibinfo {author} {\bibfnamefont {W.~W.}\
  \bibnamefont {Ho}}, \emph {et~al.},\ }\bibfield  {title} {\bibinfo {title}
  {Quantum phases of matter on a 256-atom programmable quantum simulator},\
  }\href@noop {} {\bibfield  {journal} {\bibinfo  {journal} {Nature}\ }\textbf
  {\bibinfo {volume} {595}},\ \bibinfo {pages} {227} (\bibinfo {year}
  {2021})}\BibitemShut {NoStop}%
\bibitem [{\citenamefont {Semeghini}\ \emph {et~al.}(2021)\citenamefont
  {Semeghini}, \citenamefont {Levine}, \citenamefont {Keesling}, \citenamefont
  {Ebadi}, \citenamefont {Wang}, \citenamefont {Bluvstein}, \citenamefont
  {Verresen}, \citenamefont {Pichler}, \citenamefont {Kalinowski},
  \citenamefont {Samajdar} \emph {et~al.}}]{semeghini2021probing}%
  \BibitemOpen
  \bibfield  {author} {\bibinfo {author} {\bibfnamefont {G.}~\bibnamefont
  {Semeghini}}, \bibinfo {author} {\bibfnamefont {H.}~\bibnamefont {Levine}},
  \bibinfo {author} {\bibfnamefont {A.}~\bibnamefont {Keesling}}, \bibinfo
  {author} {\bibfnamefont {S.}~\bibnamefont {Ebadi}}, \bibinfo {author}
  {\bibfnamefont {T.~T.}\ \bibnamefont {Wang}}, \bibinfo {author}
  {\bibfnamefont {D.}~\bibnamefont {Bluvstein}}, \bibinfo {author}
  {\bibfnamefont {R.}~\bibnamefont {Verresen}}, \bibinfo {author}
  {\bibfnamefont {H.}~\bibnamefont {Pichler}}, \bibinfo {author} {\bibfnamefont
  {M.}~\bibnamefont {Kalinowski}}, \bibinfo {author} {\bibfnamefont
  {R.}~\bibnamefont {Samajdar}}, \emph {et~al.},\ }\bibfield  {title} {\bibinfo
  {title} {Probing topological spin liquids on a programmable quantum
  simulator},\ }\href@noop {} {\bibfield  {journal} {\bibinfo  {journal}
  {Science}\ }\textbf {\bibinfo {volume} {374}},\ \bibinfo {pages} {1242}
  (\bibinfo {year} {2021})}\BibitemShut {NoStop}%
\bibitem [{\citenamefont {Kattem{\"o}lle}\ and\ \citenamefont {van
  Wezel}(2022)}]{kattemolle2022variational}%
  \BibitemOpen
  \bibfield  {author} {\bibinfo {author} {\bibfnamefont {J.}~\bibnamefont
  {Kattem{\"o}lle}}\ and\ \bibinfo {author} {\bibfnamefont {J.}~\bibnamefont
  {van Wezel}},\ }\bibfield  {title} {\bibinfo {title} {Variational quantum
  eigensolver for the heisenberg antiferromagnet on the kagome lattice},\
  }\href@noop {} {\bibfield  {journal} {\bibinfo  {journal} {Physical Review
  B}\ }\textbf {\bibinfo {volume} {106}},\ \bibinfo {pages} {214429} (\bibinfo
  {year} {2022})}\BibitemShut {NoStop}%
\bibitem [{\citenamefont {Iqbal}\ \emph {et~al.}(2024)\citenamefont {Iqbal},
  \citenamefont {Tantivasadakarn}, \citenamefont {Verresen}, \citenamefont
  {Campbell}, \citenamefont {Dreiling}, \citenamefont {Figgatt}, \citenamefont
  {Gaebler}, \citenamefont {Johansen}, \citenamefont {Mills}, \citenamefont
  {Moses}, \citenamefont {Pino}, \citenamefont {Ransford}, \citenamefont
  {Rowe}, \citenamefont {Siegfried}, \citenamefont {Stutz}, \citenamefont
  {Foss-Feig}, \citenamefont {Vishwanath},\ and\ \citenamefont
  {Dreyer}}]{iqbal2023creation}%
  \BibitemOpen
  \bibfield  {author} {\bibinfo {author} {\bibfnamefont {M.}~\bibnamefont
  {Iqbal}}, \bibinfo {author} {\bibfnamefont {N.}~\bibnamefont
  {Tantivasadakarn}}, \bibinfo {author} {\bibfnamefont {R.}~\bibnamefont
  {Verresen}}, \bibinfo {author} {\bibfnamefont {S.~L.}\ \bibnamefont
  {Campbell}}, \bibinfo {author} {\bibfnamefont {J.~M.}\ \bibnamefont
  {Dreiling}}, \bibinfo {author} {\bibfnamefont {C.}~\bibnamefont {Figgatt}},
  \bibinfo {author} {\bibfnamefont {J.~P.}\ \bibnamefont {Gaebler}}, \bibinfo
  {author} {\bibfnamefont {J.}~\bibnamefont {Johansen}}, \bibinfo {author}
  {\bibfnamefont {M.}~\bibnamefont {Mills}}, \bibinfo {author} {\bibfnamefont
  {S.~A.}\ \bibnamefont {Moses}}, \bibinfo {author} {\bibfnamefont {J.~M.}\
  \bibnamefont {Pino}}, \bibinfo {author} {\bibfnamefont {A.}~\bibnamefont
  {Ransford}}, \bibinfo {author} {\bibfnamefont {M.}~\bibnamefont {Rowe}},
  \bibinfo {author} {\bibfnamefont {P.}~\bibnamefont {Siegfried}}, \bibinfo
  {author} {\bibfnamefont {R.~P.}\ \bibnamefont {Stutz}}, \bibinfo {author}
  {\bibfnamefont {M.}~\bibnamefont {Foss-Feig}}, \bibinfo {author}
  {\bibfnamefont {A.}~\bibnamefont {Vishwanath}},\ and\ \bibinfo {author}
  {\bibfnamefont {H.}~\bibnamefont {Dreyer}},\ }\bibfield  {title} {\bibinfo
  {title} {Non-abelian topological order and anyons on a trapped-ion
  processor},\ }\href {https://doi.org/10.1038/s41586-023-06934-4} {\bibfield
  {journal} {\bibinfo  {journal} {Nature}\ }\textbf {\bibinfo {volume} {626}},\
  \bibinfo {pages} {505–511} (\bibinfo {year} {2024})}\BibitemShut {NoStop}%
\bibitem [{\citenamefont {{Google Quantum AI}}\ and\ \citenamefont
  {Collaborators}(2023)}]{google2023non}%
  \BibitemOpen
  \bibfield  {author} {\bibinfo {author} {\bibnamefont {{Google Quantum AI}}}\
  and\ \bibinfo {author} {\bibnamefont {Collaborators}},\ }\bibfield  {title}
  {\bibinfo {title} {Non-abelian braiding of graph vertices in a
  superconducting processor},\ }\href@noop {} {\bibfield  {journal} {\bibinfo
  {journal} {Nature}\ }\textbf {\bibinfo {volume} {618}},\ \bibinfo {pages}
  {264–269} (\bibinfo {year} {2023})}\BibitemShut {NoStop}%
\bibitem [{\citenamefont {Powalski}\ \emph {et~al.}(2013)\citenamefont
  {Powalski}, \citenamefont {Coester}, \citenamefont {Moessner},\ and\
  \citenamefont {Schmidt}}]{powalski2013disorder}%
  \BibitemOpen
  \bibfield  {author} {\bibinfo {author} {\bibfnamefont {M.}~\bibnamefont
  {Powalski}}, \bibinfo {author} {\bibfnamefont {K.}~\bibnamefont {Coester}},
  \bibinfo {author} {\bibfnamefont {R.}~\bibnamefont {Moessner}},\ and\
  \bibinfo {author} {\bibfnamefont {K.~P.}\ \bibnamefont {Schmidt}},\
  }\bibfield  {title} {\bibinfo {title} {Disorder by disorder and flat bands in
  the kagome transverse field ising model},\ }\href@noop {} {\bibfield
  {journal} {\bibinfo  {journal} {Physical Review B}\ }\textbf {\bibinfo
  {volume} {87}},\ \bibinfo {pages} {054404} (\bibinfo {year}
  {2013})}\BibitemShut {NoStop}%
\bibitem [{\citenamefont {Villain}\ \emph {et~al.}(1980)\citenamefont
  {Villain}, \citenamefont {Bidaux}, \citenamefont {Carton},\ and\
  \citenamefont {Conte}}]{villain1980order}%
  \BibitemOpen
  \bibfield  {author} {\bibinfo {author} {\bibfnamefont {J.}~\bibnamefont
  {Villain}}, \bibinfo {author} {\bibfnamefont {R.}~\bibnamefont {Bidaux}},
  \bibinfo {author} {\bibfnamefont {J.-P.}\ \bibnamefont {Carton}},\ and\
  \bibinfo {author} {\bibfnamefont {R.}~\bibnamefont {Conte}},\ }\bibfield
  {title} {\bibinfo {title} {Order as an effect of disorder},\ }\href@noop {}
  {\bibfield  {journal} {\bibinfo  {journal} {Journal de Physique}\ }\textbf
  {\bibinfo {volume} {41}},\ \bibinfo {pages} {1263} (\bibinfo {year}
  {1980})}\BibitemShut {NoStop}%
\bibitem [{\citenamefont {Moessner}\ and\ \citenamefont
  {Sondhi}(2001{\natexlab{a}})}]{moessner2001ising}%
  \BibitemOpen
  \bibfield  {author} {\bibinfo {author} {\bibfnamefont {R.}~\bibnamefont
  {Moessner}}\ and\ \bibinfo {author} {\bibfnamefont {S.~L.}\ \bibnamefont
  {Sondhi}},\ }\bibfield  {title} {\bibinfo {title} {Ising models of quantum
  frustration},\ }\href@noop {} {\bibfield  {journal} {\bibinfo  {journal}
  {Physical Review B}\ }\textbf {\bibinfo {volume} {63}},\ \bibinfo {pages}
  {224401} (\bibinfo {year} {2001}{\natexlab{a}})}\BibitemShut {NoStop}%
\bibitem [{\citenamefont {Moessner}\ \emph
  {et~al.}(2001{\natexlab{a}})\citenamefont {Moessner}, \citenamefont
  {Sondhi},\ and\ \citenamefont {Chandra}}]{moessner2001phase}%
  \BibitemOpen
  \bibfield  {author} {\bibinfo {author} {\bibfnamefont {R.}~\bibnamefont
  {Moessner}}, \bibinfo {author} {\bibfnamefont {S.~L.}\ \bibnamefont
  {Sondhi}},\ and\ \bibinfo {author} {\bibfnamefont {P.}~\bibnamefont
  {Chandra}},\ }\bibfield  {title} {\bibinfo {title} {Phase diagram of the
  hexagonal lattice quantum dimer model},\ }\href@noop {} {\bibfield  {journal}
  {\bibinfo  {journal} {Physical Review B}\ }\textbf {\bibinfo {volume} {64}},\
  \bibinfo {pages} {144416} (\bibinfo {year} {2001}{\natexlab{a}})}\BibitemShut
  {NoStop}%
\bibitem [{\citenamefont {Moessner}\ and\ \citenamefont
  {Sondhi}(2001{\natexlab{b}})}]{moessner2001resonating}%
  \BibitemOpen
  \bibfield  {author} {\bibinfo {author} {\bibfnamefont {R.}~\bibnamefont
  {Moessner}}\ and\ \bibinfo {author} {\bibfnamefont {S.~L.}\ \bibnamefont
  {Sondhi}},\ }\bibfield  {title} {\bibinfo {title} {Resonating valence bond
  phase in the triangular lattice quantum dimer model},\ }\href@noop {}
  {\bibfield  {journal} {\bibinfo  {journal} {Physical Review Letters}\
  }\textbf {\bibinfo {volume} {86}},\ \bibinfo {pages} {1881} (\bibinfo {year}
  {2001}{\natexlab{b}})}\BibitemShut {NoStop}%
\bibitem [{\citenamefont {Moessner}\ \emph
  {et~al.}(2001{\natexlab{b}})\citenamefont {Moessner}, \citenamefont
  {Sondhi},\ and\ \citenamefont {Fradkin}}]{moessner2001short}%
  \BibitemOpen
  \bibfield  {author} {\bibinfo {author} {\bibfnamefont {R.}~\bibnamefont
  {Moessner}}, \bibinfo {author} {\bibfnamefont {S.~L.}\ \bibnamefont
  {Sondhi}},\ and\ \bibinfo {author} {\bibfnamefont {E.}~\bibnamefont
  {Fradkin}},\ }\bibfield  {title} {\bibinfo {title} {Short-ranged resonating
  valence bond physics, quantum dimer models, and ising gauge theories},\
  }\href@noop {} {\bibfield  {journal} {\bibinfo  {journal} {Physical Review
  B}\ }\textbf {\bibinfo {volume} {65}},\ \bibinfo {pages} {024504} (\bibinfo
  {year} {2001}{\natexlab{b}})}\BibitemShut {NoStop}%
\bibitem [{\citenamefont {King}\ \emph {et~al.}(2018)\citenamefont {King},
  \citenamefont {Carrasquilla}, \citenamefont {Raymond}, \citenamefont
  {Ozfidan}, \citenamefont {Andriyash}, \citenamefont {Berkley}, \citenamefont
  {Reis}, \citenamefont {Lanting}, \citenamefont {Harris}, \citenamefont
  {Altomare} \emph {et~al.}}]{king2018observation}%
  \BibitemOpen
  \bibfield  {author} {\bibinfo {author} {\bibfnamefont {A.~D.}\ \bibnamefont
  {King}}, \bibinfo {author} {\bibfnamefont {J.}~\bibnamefont {Carrasquilla}},
  \bibinfo {author} {\bibfnamefont {J.}~\bibnamefont {Raymond}}, \bibinfo
  {author} {\bibfnamefont {I.}~\bibnamefont {Ozfidan}}, \bibinfo {author}
  {\bibfnamefont {E.}~\bibnamefont {Andriyash}}, \bibinfo {author}
  {\bibfnamefont {A.}~\bibnamefont {Berkley}}, \bibinfo {author} {\bibfnamefont
  {M.}~\bibnamefont {Reis}}, \bibinfo {author} {\bibfnamefont {T.}~\bibnamefont
  {Lanting}}, \bibinfo {author} {\bibfnamefont {R.}~\bibnamefont {Harris}},
  \bibinfo {author} {\bibfnamefont {F.}~\bibnamefont {Altomare}}, \emph
  {et~al.},\ }\bibfield  {title} {\bibinfo {title} {Observation of topological
  phenomena in a programmable lattice of 1,800 qubits},\ }\href@noop {}
  {\bibfield  {journal} {\bibinfo  {journal} {Nature}\ }\textbf {\bibinfo
  {volume} {560}},\ \bibinfo {pages} {456} (\bibinfo {year}
  {2018})}\BibitemShut {NoStop}%
\bibitem [{\citenamefont {Isakov}\ and\ \citenamefont
  {Moessner}(2003)}]{isakov2003interplay}%
  \BibitemOpen
  \bibfield  {author} {\bibinfo {author} {\bibfnamefont {S.~V.}\ \bibnamefont
  {Isakov}}\ and\ \bibinfo {author} {\bibfnamefont {R.}~\bibnamefont
  {Moessner}},\ }\bibfield  {title} {\bibinfo {title} {Interplay of quantum and
  thermal fluctuations in a frustrated magnet},\ }\href@noop {} {\bibfield
  {journal} {\bibinfo  {journal} {Physical Review B}\ }\textbf {\bibinfo
  {volume} {68}},\ \bibinfo {pages} {104409} (\bibinfo {year}
  {2003})}\BibitemShut {NoStop}%
\bibitem [{\citenamefont {Wang}\ \emph {et~al.}(2017)\citenamefont {Wang},
  \citenamefont {Qi}, \citenamefont {Chen},\ and\ \citenamefont
  {Meng}}]{wang2017caution}%
  \BibitemOpen
  \bibfield  {author} {\bibinfo {author} {\bibfnamefont {Y.-C.}\ \bibnamefont
  {Wang}}, \bibinfo {author} {\bibfnamefont {Y.}~\bibnamefont {Qi}}, \bibinfo
  {author} {\bibfnamefont {S.}~\bibnamefont {Chen}},\ and\ \bibinfo {author}
  {\bibfnamefont {Z.~Y.}\ \bibnamefont {Meng}},\ }\bibfield  {title} {\bibinfo
  {title} {Caution on emergent continuous symmetry: A monte carlo investigation
  of the transverse-field frustrated ising model on the triangular and
  honeycomb lattices},\ }\href@noop {} {\bibfield  {journal} {\bibinfo
  {journal} {Physical Review B}\ }\textbf {\bibinfo {volume} {96}},\ \bibinfo
  {pages} {115160} (\bibinfo {year} {2017})}\BibitemShut {NoStop}%
\bibitem [{\citenamefont {Lopez-Bezanilla}\ and\ \citenamefont
  {Nisoli}(2023)}]{lopez2023field}%
  \BibitemOpen
  \bibfield  {author} {\bibinfo {author} {\bibfnamefont {A.}~\bibnamefont
  {Lopez-Bezanilla}}\ and\ \bibinfo {author} {\bibfnamefont {C.}~\bibnamefont
  {Nisoli}},\ }\bibfield  {title} {\bibinfo {title} {Field-induced magnetic
  phases in a qubit penrose quasicrystal},\ }\href@noop {} {\bibfield
  {journal} {\bibinfo  {journal} {Science Advances}\ }\textbf {\bibinfo
  {volume} {9}},\ \bibinfo {pages} {eadf6631} (\bibinfo {year}
  {2023})}\BibitemShut {NoStop}%
\bibitem [{\citenamefont {Lopez-Bezanilla}\ \emph {et~al.}(2024)\citenamefont
  {Lopez-Bezanilla}, \citenamefont {King}, \citenamefont {Nisoli},\ and\
  \citenamefont {Saxena}}]{lopez2024quantum}%
  \BibitemOpen
  \bibfield  {author} {\bibinfo {author} {\bibfnamefont {A.}~\bibnamefont
  {Lopez-Bezanilla}}, \bibinfo {author} {\bibfnamefont {A.~D.}\ \bibnamefont
  {King}}, \bibinfo {author} {\bibfnamefont {C.}~\bibnamefont {Nisoli}},\ and\
  \bibinfo {author} {\bibfnamefont {A.}~\bibnamefont {Saxena}},\ }\bibfield
  {title} {\bibinfo {title} {Quantum fluctuations drive nonmonotonic
  correlations in a qubit lattice},\ }\href@noop {} {\bibfield  {journal}
  {\bibinfo  {journal} {Nature Communications}\ }\textbf {\bibinfo {volume}
  {15}},\ \bibinfo {pages} {589} (\bibinfo {year} {2024})}\BibitemShut
  {NoStop}%
\bibitem [{\citenamefont {Fazekas}\ and\ \citenamefont
  {Anderson}(1974)}]{fazekas1974ground}%
  \BibitemOpen
  \bibfield  {author} {\bibinfo {author} {\bibfnamefont {P.}~\bibnamefont
  {Fazekas}}\ and\ \bibinfo {author} {\bibfnamefont {P.~W.}\ \bibnamefont
  {Anderson}},\ }\bibfield  {title} {\bibinfo {title} {On the ground state
  properties of the anisotropic triangular antiferromagnet},\ }\href@noop {}
  {\bibfield  {journal} {\bibinfo  {journal} {Philosophical Magazine}\ }\textbf
  {\bibinfo {volume} {30}},\ \bibinfo {pages} {423} (\bibinfo {year}
  {1974})}\BibitemShut {NoStop}%
\bibitem [{\citenamefont {Lopez-Bezanilla}\ \emph {et~al.}(2023)\citenamefont
  {Lopez-Bezanilla}, \citenamefont {Raymond}, \citenamefont {Boothby},
  \citenamefont {Carrasquilla}, \citenamefont {Nisoli},\ and\ \citenamefont
  {King}}]{lopez2023kagome}%
  \BibitemOpen
  \bibfield  {author} {\bibinfo {author} {\bibfnamefont {A.}~\bibnamefont
  {Lopez-Bezanilla}}, \bibinfo {author} {\bibfnamefont {J.}~\bibnamefont
  {Raymond}}, \bibinfo {author} {\bibfnamefont {K.}~\bibnamefont {Boothby}},
  \bibinfo {author} {\bibfnamefont {J.}~\bibnamefont {Carrasquilla}}, \bibinfo
  {author} {\bibfnamefont {C.}~\bibnamefont {Nisoli}},\ and\ \bibinfo {author}
  {\bibfnamefont {A.~D.}\ \bibnamefont {King}},\ }\bibfield  {title} {\bibinfo
  {title} {Kagome qubit ice},\ }\href@noop {} {\bibfield  {journal} {\bibinfo
  {journal} {Nature Communications}\ }\textbf {\bibinfo {volume} {14}},\
  \bibinfo {pages} {1105} (\bibinfo {year} {2023})}\BibitemShut {NoStop}%
\bibitem [{\citenamefont {Nikoli{\'c}}\ and\ \citenamefont
  {Senthil}(2005)}]{nikolic2005theory}%
  \BibitemOpen
  \bibfield  {author} {\bibinfo {author} {\bibfnamefont {P.}~\bibnamefont
  {Nikoli{\'c}}}\ and\ \bibinfo {author} {\bibfnamefont {T.}~\bibnamefont
  {Senthil}},\ }\bibfield  {title} {\bibinfo {title} {Theory of the kagome
  lattice ising antiferromagnet in weak transverse fields},\ }\href@noop {}
  {\bibfield  {journal} {\bibinfo  {journal} {Physical Review B}\ }\textbf
  {\bibinfo {volume} {71}},\ \bibinfo {pages} {024401} (\bibinfo {year}
  {2005})}\BibitemShut {NoStop}%
\bibitem [{\citenamefont {Nikoli{\'c}}(2005)}]{nikolic2005disordered}%
  \BibitemOpen
  \bibfield  {author} {\bibinfo {author} {\bibfnamefont {P.}~\bibnamefont
  {Nikoli{\'c}}},\ }\bibfield  {title} {\bibinfo {title} {Disordered, spin
  liquid, and valence-bond ordered phases of kagome lattice quantum ising
  models with transverse field and x x z dynamics},\ }\href@noop {} {\bibfield
  {journal} {\bibinfo  {journal} {Physical Review B}\ }\textbf {\bibinfo
  {volume} {72}},\ \bibinfo {pages} {064423} (\bibinfo {year}
  {2005})}\BibitemShut {NoStop}%
\bibitem [{Note1()}]{Note1}%
  \BibitemOpen
  \bibinfo {note} {This is obtained by comparing the energy of a polarized
  configuration, where each bond carries an energy of $J$ while each site an
  energy of $-h$. There are two bonds per site, resulting in an energy per site
  of $2J - h$. Comparing this with the energy of a $M/N = \pm \protect \frac
  {1}{3}$ state, where the energy per site is, on average, $-\protect \frac
  {2J}{3} - \protect \frac {h}{3}$, one obtains the value of $h/J = 4.0$ as the
  field beyond which the fully polarized state is favored.}\BibitemShut {Stop}%
\bibitem [{\citenamefont {Wu}\ \emph {et~al.}(2019)\citenamefont {Wu},
  \citenamefont {Huang},\ and\ \citenamefont {Kao}}]{wu2019tunneling}%
  \BibitemOpen
  \bibfield  {author} {\bibinfo {author} {\bibfnamefont {K.-H.}\ \bibnamefont
  {Wu}}, \bibinfo {author} {\bibfnamefont {Y.-P.}\ \bibnamefont {Huang}},\ and\
  \bibinfo {author} {\bibfnamefont {Y.-J.}\ \bibnamefont {Kao}},\ }\bibfield
  {title} {\bibinfo {title} {Tunneling-induced restoration of classical
  degeneracy in quantum kagome ice},\ }\href@noop {} {\bibfield  {journal}
  {\bibinfo  {journal} {Physical Review B}\ }\textbf {\bibinfo {volume} {99}},\
  \bibinfo {pages} {134440} (\bibinfo {year} {2019})}\BibitemShut {NoStop}%
\bibitem [{\citenamefont {Museur}\ \emph {et~al.}(2023)\citenamefont {Museur},
  \citenamefont {Lhotel},\ and\ \citenamefont
  {Holdsworth}}]{museur2023neutron}%
  \BibitemOpen
  \bibfield  {author} {\bibinfo {author} {\bibfnamefont {F.}~\bibnamefont
  {Museur}}, \bibinfo {author} {\bibfnamefont {E.}~\bibnamefont {Lhotel}},\
  and\ \bibinfo {author} {\bibfnamefont {P.~C.~W.}\ \bibnamefont
  {Holdsworth}},\ }\bibfield  {title} {\bibinfo {title} {Neutron scattering
  from fragmented frustrated magnets},\ }\href
  {https://doi.org/10.1103/PhysRevB.107.214425} {\bibfield  {journal} {\bibinfo
   {journal} {Phys. Rev. B}\ }\textbf {\bibinfo {volume} {107}},\ \bibinfo
  {pages} {214425} (\bibinfo {year} {2023})}\BibitemShut {NoStop}%
\bibitem [{\citenamefont {Farhi}\ \emph {et~al.}(2001)\citenamefont {Farhi},
  \citenamefont {Goldstone}, \citenamefont {Gutmann}, \citenamefont {Lapan},
  \citenamefont {Lundgren},\ and\ \citenamefont {Preda}}]{farhi2001quantum}%
  \BibitemOpen
  \bibfield  {author} {\bibinfo {author} {\bibfnamefont {E.}~\bibnamefont
  {Farhi}}, \bibinfo {author} {\bibfnamefont {J.}~\bibnamefont {Goldstone}},
  \bibinfo {author} {\bibfnamefont {S.}~\bibnamefont {Gutmann}}, \bibinfo
  {author} {\bibfnamefont {J.}~\bibnamefont {Lapan}}, \bibinfo {author}
  {\bibfnamefont {A.}~\bibnamefont {Lundgren}},\ and\ \bibinfo {author}
  {\bibfnamefont {D.}~\bibnamefont {Preda}},\ }\bibfield  {title} {\bibinfo
  {title} {A quantum adiabatic evolution algorithm applied to random instances
  of an np-complete problem},\ }\href@noop {} {\bibfield  {journal} {\bibinfo
  {journal} {Science}\ }\textbf {\bibinfo {volume} {292}},\ \bibinfo {pages}
  {472} (\bibinfo {year} {2001})}\BibitemShut {NoStop}%
\bibitem [{\citenamefont {Albash}\ \emph {et~al.}(2015)\citenamefont {Albash},
  \citenamefont {R{\o}nnow}, \citenamefont {Troyer},\ and\ \citenamefont
  {Lidar}}]{albash2015reexamining}%
  \BibitemOpen
  \bibfield  {author} {\bibinfo {author} {\bibfnamefont {T.}~\bibnamefont
  {Albash}}, \bibinfo {author} {\bibfnamefont {T.~F.}\ \bibnamefont
  {R{\o}nnow}}, \bibinfo {author} {\bibfnamefont {M.}~\bibnamefont {Troyer}},\
  and\ \bibinfo {author} {\bibfnamefont {D.~A.}\ \bibnamefont {Lidar}},\
  }\bibfield  {title} {\bibinfo {title} {Reexamining classical and quantum
  models for the d-wave one processor},\ }\href@noop {} {\bibfield  {journal}
  {\bibinfo  {journal} {The European Physical Journal Special Topics}\ }\textbf
  {\bibinfo {volume} {224}},\ \bibinfo {pages} {111} (\bibinfo {year}
  {2015})}\BibitemShut {NoStop}%
\bibitem [{\citenamefont {Das}\ and\ \citenamefont
  {Chakrabarti}(2008)}]{das2008colloquium}%
  \BibitemOpen
  \bibfield  {author} {\bibinfo {author} {\bibfnamefont {A.}~\bibnamefont
  {Das}}\ and\ \bibinfo {author} {\bibfnamefont {B.~K.}\ \bibnamefont
  {Chakrabarti}},\ }\bibfield  {title} {\bibinfo {title} {Colloquium: Quantum
  annealing and analog quantum computation},\ }\href@noop {} {\bibfield
  {journal} {\bibinfo  {journal} {Reviews of Modern Physics}\ }\textbf
  {\bibinfo {volume} {80}},\ \bibinfo {pages} {1061} (\bibinfo {year}
  {2008})}\BibitemShut {NoStop}%
\bibitem [{\citenamefont {Hauke}\ \emph {et~al.}(2020)\citenamefont {Hauke},
  \citenamefont {Katzgraber}, \citenamefont {Lechner}, \citenamefont
  {Nishimori},\ and\ \citenamefont {Oliver}}]{hauke2020perspectives}%
  \BibitemOpen
  \bibfield  {author} {\bibinfo {author} {\bibfnamefont {P.}~\bibnamefont
  {Hauke}}, \bibinfo {author} {\bibfnamefont {H.~G.}\ \bibnamefont
  {Katzgraber}}, \bibinfo {author} {\bibfnamefont {W.}~\bibnamefont {Lechner}},
  \bibinfo {author} {\bibfnamefont {H.}~\bibnamefont {Nishimori}},\ and\
  \bibinfo {author} {\bibfnamefont {W.~D.}\ \bibnamefont {Oliver}},\ }\bibfield
   {title} {\bibinfo {title} {Perspectives of quantum annealing: Methods and
  implementations},\ }\href@noop {} {\bibfield  {journal} {\bibinfo  {journal}
  {Reports on Progress in Physics}\ }\textbf {\bibinfo {volume} {83}},\
  \bibinfo {pages} {054401} (\bibinfo {year} {2020})}\BibitemShut {NoStop}%
\bibitem [{\citenamefont {Brooke}\ \emph {et~al.}(1999)\citenamefont {Brooke},
  \citenamefont {Bitko}, \citenamefont {Rosenbaum},\ and\ \citenamefont
  {Aeppli}}]{brooke1999quantum}%
  \BibitemOpen
  \bibfield  {author} {\bibinfo {author} {\bibfnamefont {J.}~\bibnamefont
  {Brooke}}, \bibinfo {author} {\bibfnamefont {D.}~\bibnamefont {Bitko}},
  \bibinfo {author} {\bibnamefont {Rosenbaum}},\ and\ \bibinfo {author}
  {\bibfnamefont {G.}~\bibnamefont {Aeppli}},\ }\bibfield  {title} {\bibinfo
  {title} {Quantum annealing of a disordered magnet},\ }\href@noop {}
  {\bibfield  {journal} {\bibinfo  {journal} {Science}\ }\textbf {\bibinfo
  {volume} {284}},\ \bibinfo {pages} {779} (\bibinfo {year}
  {1999})}\BibitemShut {NoStop}%
\bibitem [{\citenamefont {Kadowaki}\ and\ \citenamefont
  {Nishimori}(1998)}]{kadowaki1998quantum}%
  \BibitemOpen
  \bibfield  {author} {\bibinfo {author} {\bibfnamefont {T.}~\bibnamefont
  {Kadowaki}}\ and\ \bibinfo {author} {\bibfnamefont {H.}~\bibnamefont
  {Nishimori}},\ }\bibfield  {title} {\bibinfo {title} {Quantum annealing in
  the transverse ising model},\ }\href@noop {} {\bibfield  {journal} {\bibinfo
  {journal} {Physical Review E}\ }\textbf {\bibinfo {volume} {58}},\ \bibinfo
  {pages} {5355} (\bibinfo {year} {1998})}\BibitemShut {NoStop}%
\bibitem [{\citenamefont {Henriet}\ \emph {et~al.}(2020)\citenamefont
  {Henriet}, \citenamefont {Beguin}, \citenamefont {Signoles}, \citenamefont
  {Lahaye}, \citenamefont {Browaeys}, \citenamefont {Reymond},\ and\
  \citenamefont {Jurczak}}]{henriet2020quantum}%
  \BibitemOpen
  \bibfield  {author} {\bibinfo {author} {\bibfnamefont {L.}~\bibnamefont
  {Henriet}}, \bibinfo {author} {\bibfnamefont {L.}~\bibnamefont {Beguin}},
  \bibinfo {author} {\bibfnamefont {A.}~\bibnamefont {Signoles}}, \bibinfo
  {author} {\bibfnamefont {T.}~\bibnamefont {Lahaye}}, \bibinfo {author}
  {\bibfnamefont {A.}~\bibnamefont {Browaeys}}, \bibinfo {author}
  {\bibfnamefont {G.-O.}\ \bibnamefont {Reymond}},\ and\ \bibinfo {author}
  {\bibfnamefont {C.}~\bibnamefont {Jurczak}},\ }\bibfield  {title} {\bibinfo
  {title} {Quantum computing with neutral atoms},\ }\href@noop {} {\bibfield
  {journal} {\bibinfo  {journal} {Quantum}\ }\textbf {\bibinfo {volume} {4}},\
  \bibinfo {pages} {327} (\bibinfo {year} {2020})}\BibitemShut {NoStop}%
\bibitem [{\citenamefont {King}\ \emph
  {et~al.}(2022{\natexlab{a}})\citenamefont {King}, \citenamefont {Raymond},
  \citenamefont {Lanting}, \citenamefont {Harris}, \citenamefont {Zucca},
  \citenamefont {Altomare}, \citenamefont {Berkley}, \citenamefont {Boothby},
  \citenamefont {Ejtemaee}, \citenamefont {Enderud} \emph
  {et~al.}}]{king2022quantum}%
  \BibitemOpen
  \bibfield  {author} {\bibinfo {author} {\bibfnamefont {A.~D.}\ \bibnamefont
  {King}}, \bibinfo {author} {\bibfnamefont {J.}~\bibnamefont {Raymond}},
  \bibinfo {author} {\bibfnamefont {T.}~\bibnamefont {Lanting}}, \bibinfo
  {author} {\bibfnamefont {R.}~\bibnamefont {Harris}}, \bibinfo {author}
  {\bibfnamefont {A.}~\bibnamefont {Zucca}}, \bibinfo {author} {\bibfnamefont
  {F.}~\bibnamefont {Altomare}}, \bibinfo {author} {\bibfnamefont {A.~J.}\
  \bibnamefont {Berkley}}, \bibinfo {author} {\bibfnamefont {K.}~\bibnamefont
  {Boothby}}, \bibinfo {author} {\bibfnamefont {S.}~\bibnamefont {Ejtemaee}},
  \bibinfo {author} {\bibfnamefont {C.}~\bibnamefont {Enderud}}, \emph
  {et~al.},\ }\bibfield  {title} {\bibinfo {title} {Quantum critical dynamics
  in a 5000-qubit programmable spin glass},\ }\href@noop {} {\bibfield
  {journal} {\bibinfo  {journal} {arXiv preprint arXiv:2207.13800}\ } (\bibinfo
  {year} {2022}{\natexlab{a}})}\BibitemShut {NoStop}%
\bibitem [{\citenamefont {King}\ \emph
  {et~al.}(2022{\natexlab{b}})\citenamefont {King}, \citenamefont {Suzuki},
  \citenamefont {Raymond}, \citenamefont {Zucca}, \citenamefont {Lanting},
  \citenamefont {Altomare}, \citenamefont {Berkley}, \citenamefont {Ejtemaee},
  \citenamefont {Hoskinson}, \citenamefont {Huang} \emph
  {et~al.}}]{king2022coherent}%
  \BibitemOpen
  \bibfield  {author} {\bibinfo {author} {\bibfnamefont {A.~D.}\ \bibnamefont
  {King}}, \bibinfo {author} {\bibfnamefont {S.}~\bibnamefont {Suzuki}},
  \bibinfo {author} {\bibfnamefont {J.}~\bibnamefont {Raymond}}, \bibinfo
  {author} {\bibfnamefont {A.}~\bibnamefont {Zucca}}, \bibinfo {author}
  {\bibfnamefont {T.}~\bibnamefont {Lanting}}, \bibinfo {author} {\bibfnamefont
  {F.}~\bibnamefont {Altomare}}, \bibinfo {author} {\bibfnamefont {A.~J.}\
  \bibnamefont {Berkley}}, \bibinfo {author} {\bibfnamefont {S.}~\bibnamefont
  {Ejtemaee}}, \bibinfo {author} {\bibfnamefont {E.}~\bibnamefont {Hoskinson}},
  \bibinfo {author} {\bibfnamefont {S.}~\bibnamefont {Huang}}, \emph {et~al.},\
  }\bibfield  {title} {\bibinfo {title} {Coherent quantum annealing in a
  programmable 2000-qubit ising chain},\ }\href@noop {} {\bibfield  {journal}
  {\bibinfo  {journal} {arXiv preprint arXiv:2202.05847}\ } (\bibinfo {year}
  {2022}{\natexlab{b}})}\BibitemShut {NoStop}%
\bibitem [{\citenamefont {King}\ \emph
  {et~al.}(2021{\natexlab{a}})\citenamefont {King}, \citenamefont {Batista},
  \citenamefont {Raymond}, \citenamefont {Lanting}, \citenamefont {Ozfidan},
  \citenamefont {Poulin-Lamarre}, \citenamefont {Zhang},\ and\ \citenamefont
  {Amin}}]{king2021quantum}%
  \BibitemOpen
  \bibfield  {author} {\bibinfo {author} {\bibfnamefont {A.~D.}\ \bibnamefont
  {King}}, \bibinfo {author} {\bibfnamefont {C.~D.}\ \bibnamefont {Batista}},
  \bibinfo {author} {\bibfnamefont {J.}~\bibnamefont {Raymond}}, \bibinfo
  {author} {\bibfnamefont {T.}~\bibnamefont {Lanting}}, \bibinfo {author}
  {\bibfnamefont {I.}~\bibnamefont {Ozfidan}}, \bibinfo {author} {\bibfnamefont
  {G.}~\bibnamefont {Poulin-Lamarre}}, \bibinfo {author} {\bibfnamefont
  {H.}~\bibnamefont {Zhang}},\ and\ \bibinfo {author} {\bibfnamefont {M.~H.}\
  \bibnamefont {Amin}},\ }\bibfield  {title} {\bibinfo {title} {Quantum
  annealing simulation of out-of-equilibrium magnetization in a spin-chain
  compound},\ }\href@noop {} {\bibfield  {journal} {\bibinfo  {journal} {PRX
  Quantum}\ }\textbf {\bibinfo {volume} {2}},\ \bibinfo {pages} {030317}
  (\bibinfo {year} {2021}{\natexlab{a}})}\BibitemShut {NoStop}%
\bibitem [{\citenamefont {Park}\ and\ \citenamefont
  {Lee}(2021)}]{park2021phase}%
  \BibitemOpen
  \bibfield  {author} {\bibinfo {author} {\bibfnamefont {H.}~\bibnamefont
  {Park}}\ and\ \bibinfo {author} {\bibfnamefont {H.}~\bibnamefont {Lee}},\
  }\bibfield  {title} {\bibinfo {title} {Phase transition of frustrated ising
  model via d-wave quantum annealing machine},\ }\href@noop {} {\bibfield
  {journal} {\bibinfo  {journal} {arXiv preprint arXiv:2110.05124}\ } (\bibinfo
  {year} {2021})}\BibitemShut {NoStop}%
\bibitem [{\citenamefont {Hearth}\ \emph {et~al.}(2022)\citenamefont {Hearth},
  \citenamefont {Morampudi},\ and\ \citenamefont
  {Laumann}}]{hearth2022quantum}%
  \BibitemOpen
  \bibfield  {author} {\bibinfo {author} {\bibfnamefont {S.~N.}\ \bibnamefont
  {Hearth}}, \bibinfo {author} {\bibfnamefont {S.~C.}\ \bibnamefont
  {Morampudi}},\ and\ \bibinfo {author} {\bibfnamefont {C.~R.}\ \bibnamefont
  {Laumann}},\ }\bibfield  {title} {\bibinfo {title} {Quantum orders in the
  frustrated ising model on the bathroom tile lattice},\ }\href@noop {}
  {\bibfield  {journal} {\bibinfo  {journal} {Physical Review B}\ }\textbf
  {\bibinfo {volume} {105}},\ \bibinfo {pages} {195101} (\bibinfo {year}
  {2022})}\BibitemShut {NoStop}%
\bibitem [{\citenamefont {Zhou}\ \emph {et~al.}(2021)\citenamefont {Zhou},
  \citenamefont {Green}, \citenamefont {Dahl},\ and\ \citenamefont
  {Chamon}}]{zhou2021experimental}%
  \BibitemOpen
  \bibfield  {author} {\bibinfo {author} {\bibfnamefont {S.}~\bibnamefont
  {Zhou}}, \bibinfo {author} {\bibfnamefont {D.}~\bibnamefont {Green}},
  \bibinfo {author} {\bibfnamefont {E.~D.}\ \bibnamefont {Dahl}},\ and\
  \bibinfo {author} {\bibfnamefont {C.}~\bibnamefont {Chamon}},\ }\bibfield
  {title} {\bibinfo {title} {Experimental realization of classical z 2 spin
  liquids in a programmable quantum device},\ }\href@noop {} {\bibfield
  {journal} {\bibinfo  {journal} {Physical Review B}\ }\textbf {\bibinfo
  {volume} {104}},\ \bibinfo {pages} {L081107} (\bibinfo {year}
  {2021})}\BibitemShut {NoStop}%
\bibitem [{\citenamefont {Zhou}\ \emph {et~al.}(2022)\citenamefont {Zhou},
  \citenamefont {Zelenayova}, \citenamefont {Hart}, \citenamefont {Chamon},\
  and\ \citenamefont {Castelnovo}}]{zhou2022probing}%
  \BibitemOpen
  \bibfield  {author} {\bibinfo {author} {\bibfnamefont {S.}~\bibnamefont
  {Zhou}}, \bibinfo {author} {\bibfnamefont {M.}~\bibnamefont {Zelenayova}},
  \bibinfo {author} {\bibfnamefont {O.}~\bibnamefont {Hart}}, \bibinfo {author}
  {\bibfnamefont {C.}~\bibnamefont {Chamon}},\ and\ \bibinfo {author}
  {\bibfnamefont {C.}~\bibnamefont {Castelnovo}},\ }\bibfield  {title}
  {\bibinfo {title} {Probing fractional statistics in quantum simulators of
  spin liquid hamiltonians},\ }\href@noop {} {\bibfield  {journal} {\bibinfo
  {journal} {arXiv preprint arXiv:2211.09784}\ } (\bibinfo {year}
  {2022})}\BibitemShut {NoStop}%
\bibitem [{\citenamefont {King}\ \emph
  {et~al.}(2021{\natexlab{b}})\citenamefont {King}, \citenamefont {Nisoli},
  \citenamefont {Dahl}, \citenamefont {Poulin-Lamarre},\ and\ \citenamefont
  {Lopez-Bezanilla}}]{king2021qubit}%
  \BibitemOpen
  \bibfield  {author} {\bibinfo {author} {\bibfnamefont {A.~D.}\ \bibnamefont
  {King}}, \bibinfo {author} {\bibfnamefont {C.}~\bibnamefont {Nisoli}},
  \bibinfo {author} {\bibfnamefont {E.~D.}\ \bibnamefont {Dahl}}, \bibinfo
  {author} {\bibfnamefont {G.}~\bibnamefont {Poulin-Lamarre}},\ and\ \bibinfo
  {author} {\bibfnamefont {A.}~\bibnamefont {Lopez-Bezanilla}},\ }\bibfield
  {title} {\bibinfo {title} {Qubit spin ice},\ }\href@noop {} {\bibfield
  {journal} {\bibinfo  {journal} {Science}\ }\textbf {\bibinfo {volume}
  {373}},\ \bibinfo {pages} {576} (\bibinfo {year}
  {2021}{\natexlab{b}})}\BibitemShut {NoStop}%
\bibitem [{\citenamefont {Harris}\ \emph {et~al.}(2018)\citenamefont {Harris},
  \citenamefont {Sato}, \citenamefont {Berkley}, \citenamefont {Reis},
  \citenamefont {Altomare}, \citenamefont {Amin}, \citenamefont {Boothby},
  \citenamefont {Bunyk}, \citenamefont {Deng}, \citenamefont {Enderud} \emph
  {et~al.}}]{harris2018phase}%
  \BibitemOpen
  \bibfield  {author} {\bibinfo {author} {\bibfnamefont {R.}~\bibnamefont
  {Harris}}, \bibinfo {author} {\bibfnamefont {Y.}~\bibnamefont {Sato}},
  \bibinfo {author} {\bibfnamefont {A.}~\bibnamefont {Berkley}}, \bibinfo
  {author} {\bibfnamefont {M.}~\bibnamefont {Reis}}, \bibinfo {author}
  {\bibfnamefont {F.}~\bibnamefont {Altomare}}, \bibinfo {author}
  {\bibfnamefont {M.}~\bibnamefont {Amin}}, \bibinfo {author} {\bibfnamefont
  {K.}~\bibnamefont {Boothby}}, \bibinfo {author} {\bibfnamefont
  {P.}~\bibnamefont {Bunyk}}, \bibinfo {author} {\bibfnamefont
  {C.}~\bibnamefont {Deng}}, \bibinfo {author} {\bibfnamefont {C.}~\bibnamefont
  {Enderud}}, \emph {et~al.},\ }\bibfield  {title} {\bibinfo {title} {Phase
  transitions in a programmable quantum spin glass simulator},\ }\href@noop {}
  {\bibfield  {journal} {\bibinfo  {journal} {Science}\ }\textbf {\bibinfo
  {volume} {361}},\ \bibinfo {pages} {162} (\bibinfo {year}
  {2018})}\BibitemShut {NoStop}%
\bibitem [{\citenamefont {King}\ \emph
  {et~al.}(2021{\natexlab{c}})\citenamefont {King}, \citenamefont {Raymond},
  \citenamefont {Lanting}, \citenamefont {Isakov}, \citenamefont {Mohseni},
  \citenamefont {Poulin-Lamarre}, \citenamefont {Ejtemaee}, \citenamefont
  {Bernoudy}, \citenamefont {Ozfidan}, \citenamefont {Smirnov} \emph
  {et~al.}}]{king2021scaling}%
  \BibitemOpen
  \bibfield  {author} {\bibinfo {author} {\bibfnamefont {A.~D.}\ \bibnamefont
  {King}}, \bibinfo {author} {\bibfnamefont {J.}~\bibnamefont {Raymond}},
  \bibinfo {author} {\bibfnamefont {T.}~\bibnamefont {Lanting}}, \bibinfo
  {author} {\bibfnamefont {S.~V.}\ \bibnamefont {Isakov}}, \bibinfo {author}
  {\bibfnamefont {M.}~\bibnamefont {Mohseni}}, \bibinfo {author} {\bibfnamefont
  {G.}~\bibnamefont {Poulin-Lamarre}}, \bibinfo {author} {\bibfnamefont
  {S.}~\bibnamefont {Ejtemaee}}, \bibinfo {author} {\bibfnamefont
  {W.}~\bibnamefont {Bernoudy}}, \bibinfo {author} {\bibfnamefont
  {I.}~\bibnamefont {Ozfidan}}, \bibinfo {author} {\bibfnamefont {A.~Y.}\
  \bibnamefont {Smirnov}}, \emph {et~al.},\ }\bibfield  {title} {\bibinfo
  {title} {Scaling advantage over path-integral monte carlo in quantum
  simulation of geometrically frustrated magnets},\ }\href@noop {} {\bibfield
  {journal} {\bibinfo  {journal} {Nature communications}\ }\textbf {\bibinfo
  {volume} {12}},\ \bibinfo {pages} {1} (\bibinfo {year}
  {2021}{\natexlab{c}})}\BibitemShut {NoStop}%
\bibitem [{\citenamefont {Nishimura}\ \emph {et~al.}(2020)\citenamefont
  {Nishimura}, \citenamefont {Nishimori},\ and\ \citenamefont
  {Katzgraber}}]{nishimura2020griffiths}%
  \BibitemOpen
  \bibfield  {author} {\bibinfo {author} {\bibfnamefont {K.}~\bibnamefont
  {Nishimura}}, \bibinfo {author} {\bibfnamefont {H.}~\bibnamefont
  {Nishimori}},\ and\ \bibinfo {author} {\bibfnamefont {H.~G.}\ \bibnamefont
  {Katzgraber}},\ }\bibfield  {title} {\bibinfo {title} {Griffiths-mccoy
  singularity on the diluted chimera graph: Monte carlo simulations and
  experiments on quantum hardware},\ }\href@noop {} {\bibfield  {journal}
  {\bibinfo  {journal} {Physical Review A}\ }\textbf {\bibinfo {volume}
  {102}},\ \bibinfo {pages} {042403} (\bibinfo {year} {2020})}\BibitemShut
  {NoStop}%
\bibitem [{\citenamefont {Kairys}\ \emph {et~al.}(2020)\citenamefont {Kairys},
  \citenamefont {King}, \citenamefont {Ozfidan}, \citenamefont {Boothby},
  \citenamefont {Raymond}, \citenamefont {Banerjee},\ and\ \citenamefont
  {Humble}}]{kairys2020simulating}%
  \BibitemOpen
  \bibfield  {author} {\bibinfo {author} {\bibfnamefont {P.}~\bibnamefont
  {Kairys}}, \bibinfo {author} {\bibfnamefont {A.~D.}\ \bibnamefont {King}},
  \bibinfo {author} {\bibfnamefont {I.}~\bibnamefont {Ozfidan}}, \bibinfo
  {author} {\bibfnamefont {K.}~\bibnamefont {Boothby}}, \bibinfo {author}
  {\bibfnamefont {J.}~\bibnamefont {Raymond}}, \bibinfo {author} {\bibfnamefont
  {A.}~\bibnamefont {Banerjee}},\ and\ \bibinfo {author} {\bibfnamefont
  {T.~S.}\ \bibnamefont {Humble}},\ }\bibfield  {title} {\bibinfo {title}
  {Simulating the shastry-sutherland ising model using quantum annealing},\
  }\href@noop {} {\bibfield  {journal} {\bibinfo  {journal} {PRX Quantum}\
  }\textbf {\bibinfo {volume} {1}},\ \bibinfo {pages} {020320} (\bibinfo {year}
  {2020})}\BibitemShut {NoStop}%
\bibitem [{\citenamefont {Boothby}\ \emph {et~al.}(2020)\citenamefont
  {Boothby}, \citenamefont {Bunyk}, \citenamefont {Raymond},\ and\
  \citenamefont {Roy}}]{boothby2020next}%
  \BibitemOpen
  \bibfield  {author} {\bibinfo {author} {\bibfnamefont {K.}~\bibnamefont
  {Boothby}}, \bibinfo {author} {\bibfnamefont {P.}~\bibnamefont {Bunyk}},
  \bibinfo {author} {\bibfnamefont {J.}~\bibnamefont {Raymond}},\ and\ \bibinfo
  {author} {\bibfnamefont {A.}~\bibnamefont {Roy}},\ }\bibfield  {title}
  {\bibinfo {title} {Next-generation topology of d-wave quantum processors},\
  }\href@noop {} {\bibfield  {journal} {\bibinfo  {journal} {arXiv preprint
  arXiv:2003.00133}\ } (\bibinfo {year} {2020})}\BibitemShut {NoStop}%
\bibitem [{\citenamefont {Boothby}\ \emph
  {et~al.}(2021{\natexlab{a}})\citenamefont {Boothby}, \citenamefont {Enderud},
  \citenamefont {Lanting}, \citenamefont {Molavi}, \citenamefont {Tsai},
  \citenamefont {Volkmann}, \citenamefont {Altomare}, \citenamefont {Amin},
  \citenamefont {Babcock}, \citenamefont {Berkley} \emph
  {et~al.}}]{boothby2021architectural}%
  \BibitemOpen
  \bibfield  {author} {\bibinfo {author} {\bibfnamefont {K.}~\bibnamefont
  {Boothby}}, \bibinfo {author} {\bibfnamefont {C.}~\bibnamefont {Enderud}},
  \bibinfo {author} {\bibfnamefont {T.}~\bibnamefont {Lanting}}, \bibinfo
  {author} {\bibfnamefont {R.}~\bibnamefont {Molavi}}, \bibinfo {author}
  {\bibfnamefont {N.}~\bibnamefont {Tsai}}, \bibinfo {author} {\bibfnamefont
  {M.~H.}\ \bibnamefont {Volkmann}}, \bibinfo {author} {\bibfnamefont
  {F.}~\bibnamefont {Altomare}}, \bibinfo {author} {\bibfnamefont {M.~H.}\
  \bibnamefont {Amin}}, \bibinfo {author} {\bibfnamefont {M.}~\bibnamefont
  {Babcock}}, \bibinfo {author} {\bibfnamefont {A.~J.}\ \bibnamefont
  {Berkley}}, \emph {et~al.},\ }\bibfield  {title} {\bibinfo {title}
  {Architectural considerations in the design of a third-generation
  superconducting quantum annealing processor},\ }\href@noop {} {\bibfield
  {journal} {\bibinfo  {journal} {arXiv preprint arXiv:2108.02322}\ } (\bibinfo
  {year} {2021}{\natexlab{a}})}\BibitemShut {NoStop}%
\bibitem [{\citenamefont {Boothby}\ \emph
  {et~al.}(2021{\natexlab{b}})\citenamefont {Boothby}, \citenamefont {King},\
  and\ \citenamefont {Raymond}}]{boothby2021zephyr}%
  \BibitemOpen
  \bibfield  {author} {\bibinfo {author} {\bibfnamefont {K.}~\bibnamefont
  {Boothby}}, \bibinfo {author} {\bibfnamefont {A.~D.}\ \bibnamefont {King}},\
  and\ \bibinfo {author} {\bibfnamefont {J.}~\bibnamefont {Raymond}},\
  }\href@noop {} {\bibinfo {title} {Zephyr topology of d-wave quantum
  processors}} (\bibinfo {year} {2021}{\natexlab{b}})\BibitemShut {NoStop}%
\bibitem [{\citenamefont {Yarkoni}\ \emph {et~al.}(2019)\citenamefont
  {Yarkoni}, \citenamefont {Wang}, \citenamefont {Plaat},\ and\ \citenamefont
  {B{\"a}ck}}]{yarkoni2019boosting}%
  \BibitemOpen
  \bibfield  {author} {\bibinfo {author} {\bibfnamefont {S.}~\bibnamefont
  {Yarkoni}}, \bibinfo {author} {\bibfnamefont {H.}~\bibnamefont {Wang}},
  \bibinfo {author} {\bibfnamefont {A.}~\bibnamefont {Plaat}},\ and\ \bibinfo
  {author} {\bibfnamefont {T.}~\bibnamefont {B{\"a}ck}},\ }\bibfield  {title}
  {\bibinfo {title} {Boosting quantum annealing performance using evolution
  strategies for annealing offsets tuning},\ }in\ \href@noop {} {\emph
  {\bibinfo {booktitle} {Quantum Technology and Optimization Problems: First
  International Workshop, QTOP 2019, Munich, Germany, March 18, 2019,
  Proceedings 1}}}\ (\bibinfo {organization} {Springer},\ \bibinfo {year}
  {2019})\ pp.\ \bibinfo {pages} {157--168}\BibitemShut {NoStop}%
\bibitem [{\citenamefont {Chern}\ \emph {et~al.}(2023)\citenamefont {Chern},
  \citenamefont {Boothby}, \citenamefont {Raymond}, \citenamefont {Farr{\'e}},\
  and\ \citenamefont {King}}]{chern2023tutorial}%
  \BibitemOpen
  \bibfield  {author} {\bibinfo {author} {\bibfnamefont {K.}~\bibnamefont
  {Chern}}, \bibinfo {author} {\bibfnamefont {K.}~\bibnamefont {Boothby}},
  \bibinfo {author} {\bibfnamefont {J.}~\bibnamefont {Raymond}}, \bibinfo
  {author} {\bibfnamefont {P.}~\bibnamefont {Farr{\'e}}},\ and\ \bibinfo
  {author} {\bibfnamefont {A.~D.}\ \bibnamefont {King}},\ }\bibfield  {title}
  {\bibinfo {title} {Tutorial: Calibration refinement in quantum annealing},\
  }\href@noop {} {\bibfield  {journal} {\bibinfo  {journal} {arXiv preprint
  arXiv:2304.10352}\ } (\bibinfo {year} {2023})}\BibitemShut {NoStop}%
\bibitem [{\citenamefont {Izquierdo}\ \emph {et~al.}(2021)\citenamefont
  {Izquierdo}, \citenamefont {Hen},\ and\ \citenamefont
  {Albash}}]{izquierdo2021testing}%
  \BibitemOpen
  \bibfield  {author} {\bibinfo {author} {\bibfnamefont {Z.~G.}\ \bibnamefont
  {Izquierdo}}, \bibinfo {author} {\bibfnamefont {I.}~\bibnamefont {Hen}},\
  and\ \bibinfo {author} {\bibfnamefont {T.}~\bibnamefont {Albash}},\
  }\bibfield  {title} {\bibinfo {title} {Testing a quantum annealer as a
  quantum thermal sampler},\ }\href@noop {} {\bibfield  {journal} {\bibinfo
  {journal} {ACM Transactions on Quantum Computing}\ }\textbf {\bibinfo
  {volume} {2}},\ \bibinfo {pages} {1} (\bibinfo {year} {2021})}\BibitemShut
  {NoStop}%
\bibitem [{\citenamefont {Moessner}\ and\ \citenamefont
  {Sondhi}(2003)}]{moessner2003theory}%
  \BibitemOpen
  \bibfield  {author} {\bibinfo {author} {\bibfnamefont {R.}~\bibnamefont
  {Moessner}}\ and\ \bibinfo {author} {\bibfnamefont {S.~L.}\ \bibnamefont
  {Sondhi}},\ }\bibfield  {title} {\bibinfo {title} {Theory of the [111]
  magnetization plateau in spin ice},\ }\href
  {https://doi.org/10.1103/PhysRevB.68.064411} {\bibfield  {journal} {\bibinfo
  {journal} {Phys. Rev. B}\ }\textbf {\bibinfo {volume} {68}},\ \bibinfo
  {pages} {064411} (\bibinfo {year} {2003})}\BibitemShut {NoStop}%
\bibitem [{\citenamefont {Turrini}\ \emph {et~al.}(2022)\citenamefont
  {Turrini}, \citenamefont {Harman-Clarke}, \citenamefont {Haeseler},
  \citenamefont {Fennell}, \citenamefont {Wood}, \citenamefont {Henelius},
  \citenamefont {Bramwell},\ and\ \citenamefont
  {Holdsworth}}]{turrini2022tunable}%
  \BibitemOpen
  \bibfield  {author} {\bibinfo {author} {\bibfnamefont {A.~A.}\ \bibnamefont
  {Turrini}}, \bibinfo {author} {\bibfnamefont {A.}~\bibnamefont
  {Harman-Clarke}}, \bibinfo {author} {\bibfnamefont {G.}~\bibnamefont
  {Haeseler}}, \bibinfo {author} {\bibfnamefont {T.}~\bibnamefont {Fennell}},
  \bibinfo {author} {\bibfnamefont {I.~G.}\ \bibnamefont {Wood}}, \bibinfo
  {author} {\bibfnamefont {P.}~\bibnamefont {Henelius}}, \bibinfo {author}
  {\bibfnamefont {S.~T.}\ \bibnamefont {Bramwell}},\ and\ \bibinfo {author}
  {\bibfnamefont {P.~C.~W.}\ \bibnamefont {Holdsworth}},\ }\bibfield  {title}
  {\bibinfo {title} {Tunable critical correlations in kagome ice},\ }\href
  {https://doi.org/10.1103/PhysRevB.105.094403} {\bibfield  {journal} {\bibinfo
   {journal} {Phys. Rev. B}\ }\textbf {\bibinfo {volume} {105}},\ \bibinfo
  {pages} {094403} (\bibinfo {year} {2022})}\BibitemShut {NoStop}%
\bibitem [{\citenamefont {Youngblood}\ \emph {et~al.}(1980)\citenamefont
  {Youngblood}, \citenamefont {Axe},\ and\ \citenamefont
  {McCoy}}]{youngblood1980correlations}%
  \BibitemOpen
  \bibfield  {author} {\bibinfo {author} {\bibfnamefont {R.}~\bibnamefont
  {Youngblood}}, \bibinfo {author} {\bibfnamefont {J.~D.}\ \bibnamefont
  {Axe}},\ and\ \bibinfo {author} {\bibfnamefont {B.~M.}\ \bibnamefont
  {McCoy}},\ }\bibfield  {title} {\bibinfo {title} {Correlations in ice-rule
  ferroelectrics},\ }\href {https://doi.org/10.1103/PhysRevB.21.5212}
  {\bibfield  {journal} {\bibinfo  {journal} {Phys. Rev. B}\ }\textbf {\bibinfo
  {volume} {21}},\ \bibinfo {pages} {5212} (\bibinfo {year}
  {1980})}\BibitemShut {NoStop}%
\bibitem [{\citenamefont {Garanin}\ and\ \citenamefont
  {Canals}(1999)}]{garanin1999classical}%
  \BibitemOpen
  \bibfield  {author} {\bibinfo {author} {\bibfnamefont {D.}~\bibnamefont
  {Garanin}}\ and\ \bibinfo {author} {\bibfnamefont {B.}~\bibnamefont
  {Canals}},\ }\bibfield  {title} {\bibinfo {title} {Classical spin liquid:
  Exact solution for the infinite-component antiferromagnetic model on the
  kagom{\'e} lattice},\ }\href@noop {} {\bibfield  {journal} {\bibinfo
  {journal} {Physical Review B}\ }\textbf {\bibinfo {volume} {59}},\ \bibinfo
  {pages} {443} (\bibinfo {year} {1999})}\BibitemShut {NoStop}%
\bibitem [{\citenamefont {Brooks-Bartlett}\ \emph {et~al.}(2014)\citenamefont
  {Brooks-Bartlett}, \citenamefont {Banks}, \citenamefont {Jaubert},
  \citenamefont {Harman-Clarke},\ and\ \citenamefont
  {Holdsworth}}]{brooks-bartlett2014fragmentation}%
  \BibitemOpen
  \bibfield  {author} {\bibinfo {author} {\bibfnamefont {M.~E.}\ \bibnamefont
  {Brooks-Bartlett}}, \bibinfo {author} {\bibfnamefont {S.~T.}\ \bibnamefont
  {Banks}}, \bibinfo {author} {\bibfnamefont {L.~D.~C.}\ \bibnamefont
  {Jaubert}}, \bibinfo {author} {\bibfnamefont {A.}~\bibnamefont
  {Harman-Clarke}},\ and\ \bibinfo {author} {\bibfnamefont {P.~C.~W.}\
  \bibnamefont {Holdsworth}},\ }\bibfield  {title} {\bibinfo {title}
  {Magnetic-moment fragmentation and monopole crystallization},\ }\href
  {https://doi.org/10.1103/PhysRevX.4.011007} {\bibfield  {journal} {\bibinfo
  {journal} {Phys. Rev. X}\ }\textbf {\bibinfo {volume} {4}},\ \bibinfo {pages}
  {011007} (\bibinfo {year} {2014})}\BibitemShut {NoStop}%
\bibitem [{\citenamefont {Rougemaille}\ and\ \citenamefont
  {Canals}(2019)}]{rougemaille2019cooperative}%
  \BibitemOpen
  \bibfield  {author} {\bibinfo {author} {\bibfnamefont {N.}~\bibnamefont
  {Rougemaille}}\ and\ \bibinfo {author} {\bibfnamefont {B.}~\bibnamefont
  {Canals}},\ }\bibfield  {title} {\bibinfo {title} {Cooperative magnetic
  phenomena in artificial spin systems: spin liquids, coulomb phase and
  fragmentation of magnetism--a colloquium},\ }\href@noop {} {\bibfield
  {journal} {\bibinfo  {journal} {The European Physical Journal B}\ }\textbf
  {\bibinfo {volume} {92}},\ \bibinfo {pages} {1} (\bibinfo {year}
  {2019})}\BibitemShut {NoStop}%
\bibitem [{\citenamefont {Damle}\ and\ \citenamefont
  {Senthil}(2006)}]{damle2006spin}%
  \BibitemOpen
  \bibfield  {author} {\bibinfo {author} {\bibfnamefont {K.}~\bibnamefont
  {Damle}}\ and\ \bibinfo {author} {\bibfnamefont {T.}~\bibnamefont
  {Senthil}},\ }\bibfield  {title} {\bibinfo {title} {Spin nematics and
  magnetization plateau transition in anisotropic kagome magnets},\ }\href
  {https://doi.org/10.1103/PhysRevLett.97.067202} {\bibfield  {journal}
  {\bibinfo  {journal} {Phys. Rev. Lett.}\ }\textbf {\bibinfo {volume} {97}},\
  \bibinfo {pages} {067202} (\bibinfo {year} {2006})}\BibitemShut {NoStop}%
\bibitem [{\citenamefont {Carrasquilla}\ \emph {et~al.}(2015)\citenamefont
  {Carrasquilla}, \citenamefont {Hao},\ and\ \citenamefont
  {Melko}}]{carrasquilla2015two}%
  \BibitemOpen
  \bibfield  {author} {\bibinfo {author} {\bibfnamefont {J.}~\bibnamefont
  {Carrasquilla}}, \bibinfo {author} {\bibfnamefont {Z.}~\bibnamefont {Hao}},\
  and\ \bibinfo {author} {\bibfnamefont {R.~G.}\ \bibnamefont {Melko}},\
  }\bibfield  {title} {\bibinfo {title} {A two-dimensional spin liquid in
  quantum kagome ice},\ }\href@noop {} {\bibfield  {journal} {\bibinfo
  {journal} {Nature communications}\ }\textbf {\bibinfo {volume} {6}},\
  \bibinfo {pages} {7421} (\bibinfo {year} {2015})}\BibitemShut {NoStop}%
\bibitem [{\citenamefont {Ciavarella}\ \emph
  {et~al.}(2023{\natexlab{a}})\citenamefont {Ciavarella}, \citenamefont
  {Caspar}, \citenamefont {Illa},\ and\ \citenamefont
  {Savage}}]{ciavarella2022floquet}%
  \BibitemOpen
  \bibfield  {author} {\bibinfo {author} {\bibfnamefont {A.~N.}\ \bibnamefont
  {Ciavarella}}, \bibinfo {author} {\bibfnamefont {S.}~\bibnamefont {Caspar}},
  \bibinfo {author} {\bibfnamefont {M.}~\bibnamefont {Illa}},\ and\ \bibinfo
  {author} {\bibfnamefont {M.~J.}\ \bibnamefont {Savage}},\ }\bibfield  {title}
  {\bibinfo {title} {State preparation in the heisenberg model through
  adiabatic spiraling},\ }\href {https://doi.org/10.22331/q-2023-04-06-970}
  {\bibfield  {journal} {\bibinfo  {journal} {Quantum}\ }\textbf {\bibinfo
  {volume} {7}},\ \bibinfo {pages} {970} (\bibinfo {year}
  {2023}{\natexlab{a}})}\BibitemShut {NoStop}%
\bibitem [{\citenamefont {Ciavarella}\ \emph
  {et~al.}(2023{\natexlab{b}})\citenamefont {Ciavarella}, \citenamefont
  {Caspar}, \citenamefont {Illa},\ and\ \citenamefont
  {Savage}}]{ciavarella2023state}%
  \BibitemOpen
  \bibfield  {author} {\bibinfo {author} {\bibfnamefont {A.~N.}\ \bibnamefont
  {Ciavarella}}, \bibinfo {author} {\bibfnamefont {S.}~\bibnamefont {Caspar}},
  \bibinfo {author} {\bibfnamefont {M.}~\bibnamefont {Illa}},\ and\ \bibinfo
  {author} {\bibfnamefont {M.~J.}\ \bibnamefont {Savage}},\ }\bibfield  {title}
  {\bibinfo {title} {State preparation in the heisenberg model through
  adiabatic spiraling},\ }\href@noop {} {\bibfield  {journal} {\bibinfo
  {journal} {Quantum}\ }\textbf {\bibinfo {volume} {7}},\ \bibinfo {pages}
  {970} (\bibinfo {year} {2023}{\natexlab{b}})}\BibitemShut {NoStop}%
\bibitem [{\citenamefont {Scholl}\ \emph {et~al.}(2022)\citenamefont {Scholl},
  \citenamefont {Williams}, \citenamefont {Bornet}, \citenamefont {Wallner},
  \citenamefont {Barredo}, \citenamefont {Henriet}, \citenamefont {Signoles},
  \citenamefont {Hainaut}, \citenamefont {Franz}, \citenamefont {Geier} \emph
  {et~al.}}]{scholl2022microwave}%
  \BibitemOpen
  \bibfield  {author} {\bibinfo {author} {\bibfnamefont {P.}~\bibnamefont
  {Scholl}}, \bibinfo {author} {\bibfnamefont {H.~J.}\ \bibnamefont
  {Williams}}, \bibinfo {author} {\bibfnamefont {G.}~\bibnamefont {Bornet}},
  \bibinfo {author} {\bibfnamefont {F.}~\bibnamefont {Wallner}}, \bibinfo
  {author} {\bibfnamefont {D.}~\bibnamefont {Barredo}}, \bibinfo {author}
  {\bibfnamefont {L.}~\bibnamefont {Henriet}}, \bibinfo {author} {\bibfnamefont
  {A.}~\bibnamefont {Signoles}}, \bibinfo {author} {\bibfnamefont
  {C.}~\bibnamefont {Hainaut}}, \bibinfo {author} {\bibfnamefont
  {T.}~\bibnamefont {Franz}}, \bibinfo {author} {\bibfnamefont
  {S.}~\bibnamefont {Geier}}, \emph {et~al.},\ }\bibfield  {title} {\bibinfo
  {title} {Microwave engineering of programmable x x z hamiltonians in arrays
  of rydberg atoms},\ }\href@noop {} {\bibfield  {journal} {\bibinfo  {journal}
  {PRX Quantum}\ }\textbf {\bibinfo {volume} {3}},\ \bibinfo {pages} {020303}
  (\bibinfo {year} {2022})}\BibitemShut {NoStop}%
\bibitem [{\citenamefont {Ozfidan}\ \emph {et~al.}(2020)\citenamefont
  {Ozfidan}, \citenamefont {Deng}, \citenamefont {Smirnov}, \citenamefont
  {Lanting}, \citenamefont {Harris}, \citenamefont {Swenson}, \citenamefont
  {Whittaker}, \citenamefont {Altomare}, \citenamefont {Babcock}, \citenamefont
  {Baron}, \citenamefont {Berkley}, \citenamefont {Boothby}, \citenamefont
  {Christiani}, \citenamefont {Bunyk}, \citenamefont {Enderud}, \citenamefont
  {Evert}, \citenamefont {Hager}, \citenamefont {Hajda}, \citenamefont
  {Hilton}, \citenamefont {Huang}, \citenamefont {Hoskinson}, \citenamefont
  {Johnson}, \citenamefont {Jooya}, \citenamefont {Ladizinsky}, \citenamefont
  {Ladizinsky}, \citenamefont {Li}, \citenamefont {MacDonald}, \citenamefont
  {Marsden}, \citenamefont {Marsden}, \citenamefont {Medina}, \citenamefont
  {Molavi}, \citenamefont {Neufeld}, \citenamefont {Nissen}, \citenamefont
  {Norouzpour}, \citenamefont {Oh}, \citenamefont {Pavlov}, \citenamefont
  {Perminov}, \citenamefont {Poulin-Lamarre}, \citenamefont {Reis},
  \citenamefont {Prescott}, \citenamefont {Rich}, \citenamefont {Sato},
  \citenamefont {Sterling}, \citenamefont {Tsai}, \citenamefont {Volkmann},
  \citenamefont {Wilkinson}, \citenamefont {Yao},\ and\ \citenamefont
  {Amin}}]{ozfidan2020demonstration}%
  \BibitemOpen
  \bibfield  {author} {\bibinfo {author} {\bibfnamefont {I.}~\bibnamefont
  {Ozfidan}}, \bibinfo {author} {\bibfnamefont {C.}~\bibnamefont {Deng}},
  \bibinfo {author} {\bibfnamefont {A.}~\bibnamefont {Smirnov}}, \bibinfo
  {author} {\bibfnamefont {T.}~\bibnamefont {Lanting}}, \bibinfo {author}
  {\bibfnamefont {R.}~\bibnamefont {Harris}}, \bibinfo {author} {\bibfnamefont
  {L.}~\bibnamefont {Swenson}}, \bibinfo {author} {\bibfnamefont
  {J.}~\bibnamefont {Whittaker}}, \bibinfo {author} {\bibfnamefont
  {F.}~\bibnamefont {Altomare}}, \bibinfo {author} {\bibfnamefont
  {M.}~\bibnamefont {Babcock}}, \bibinfo {author} {\bibfnamefont
  {C.}~\bibnamefont {Baron}}, \bibinfo {author} {\bibfnamefont
  {A.}~\bibnamefont {Berkley}}, \bibinfo {author} {\bibfnamefont
  {K.}~\bibnamefont {Boothby}}, \bibinfo {author} {\bibfnamefont
  {H.}~\bibnamefont {Christiani}}, \bibinfo {author} {\bibfnamefont
  {P.}~\bibnamefont {Bunyk}}, \bibinfo {author} {\bibfnamefont
  {C.}~\bibnamefont {Enderud}}, \bibinfo {author} {\bibfnamefont
  {B.}~\bibnamefont {Evert}}, \bibinfo {author} {\bibfnamefont
  {M.}~\bibnamefont {Hager}}, \bibinfo {author} {\bibfnamefont
  {A.}~\bibnamefont {Hajda}}, \bibinfo {author} {\bibfnamefont
  {J.}~\bibnamefont {Hilton}}, \bibinfo {author} {\bibfnamefont
  {S.}~\bibnamefont {Huang}}, \bibinfo {author} {\bibfnamefont
  {E.}~\bibnamefont {Hoskinson}}, \bibinfo {author} {\bibfnamefont
  {M.}~\bibnamefont {Johnson}}, \bibinfo {author} {\bibfnamefont
  {K.}~\bibnamefont {Jooya}}, \bibinfo {author} {\bibfnamefont
  {E.}~\bibnamefont {Ladizinsky}}, \bibinfo {author} {\bibfnamefont
  {N.}~\bibnamefont {Ladizinsky}}, \bibinfo {author} {\bibfnamefont
  {R.}~\bibnamefont {Li}}, \bibinfo {author} {\bibfnamefont {A.}~\bibnamefont
  {MacDonald}}, \bibinfo {author} {\bibfnamefont {D.}~\bibnamefont {Marsden}},
  \bibinfo {author} {\bibfnamefont {G.}~\bibnamefont {Marsden}}, \bibinfo
  {author} {\bibfnamefont {T.}~\bibnamefont {Medina}}, \bibinfo {author}
  {\bibfnamefont {R.}~\bibnamefont {Molavi}}, \bibinfo {author} {\bibfnamefont
  {R.}~\bibnamefont {Neufeld}}, \bibinfo {author} {\bibfnamefont
  {M.}~\bibnamefont {Nissen}}, \bibinfo {author} {\bibfnamefont
  {M.}~\bibnamefont {Norouzpour}}, \bibinfo {author} {\bibfnamefont
  {T.}~\bibnamefont {Oh}}, \bibinfo {author} {\bibfnamefont {I.}~\bibnamefont
  {Pavlov}}, \bibinfo {author} {\bibfnamefont {I.}~\bibnamefont {Perminov}},
  \bibinfo {author} {\bibfnamefont {G.}~\bibnamefont {Poulin-Lamarre}},
  \bibinfo {author} {\bibfnamefont {M.}~\bibnamefont {Reis}}, \bibinfo {author}
  {\bibfnamefont {T.}~\bibnamefont {Prescott}}, \bibinfo {author}
  {\bibfnamefont {C.}~\bibnamefont {Rich}}, \bibinfo {author} {\bibfnamefont
  {Y.}~\bibnamefont {Sato}}, \bibinfo {author} {\bibfnamefont {G.}~\bibnamefont
  {Sterling}}, \bibinfo {author} {\bibfnamefont {N.}~\bibnamefont {Tsai}},
  \bibinfo {author} {\bibfnamefont {M.}~\bibnamefont {Volkmann}}, \bibinfo
  {author} {\bibfnamefont {W.}~\bibnamefont {Wilkinson}}, \bibinfo {author}
  {\bibfnamefont {J.}~\bibnamefont {Yao}},\ and\ \bibinfo {author}
  {\bibfnamefont {M.}~\bibnamefont {Amin}},\ }\bibfield  {title} {\bibinfo
  {title} {Demonstration of a nonstoquastic hamiltonian in coupled
  superconducting flux qubits},\ }\bibfield  {journal} {\bibinfo  {journal}
  {Physical Review Applied}\ }\textbf {\bibinfo {volume} {13}},\ \href
  {https://doi.org/10.1103/physrevapplied.13.034037}
  {10.1103/physrevapplied.13.034037} (\bibinfo {year} {2020})\BibitemShut
  {NoStop}%
\bibitem [{\citenamefont {Block}\ \emph {et~al.}(2020)\citenamefont {Block},
  \citenamefont {D'Emidio},\ and\ \citenamefont {Kaul}}]{block2020kagome}%
  \BibitemOpen
  \bibfield  {author} {\bibinfo {author} {\bibfnamefont {M.~S.}\ \bibnamefont
  {Block}}, \bibinfo {author} {\bibfnamefont {J.}~\bibnamefont {D'Emidio}},\
  and\ \bibinfo {author} {\bibfnamefont {R.~K.}\ \bibnamefont {Kaul}},\
  }\bibfield  {title} {\bibinfo {title} {Kagome model for a z 2 quantum spin
  liquid},\ }\href@noop {} {\bibfield  {journal} {\bibinfo  {journal} {Physical
  Review B}\ }\textbf {\bibinfo {volume} {101}},\ \bibinfo {pages} {020402(R)}
  (\bibinfo {year} {2020})}\BibitemShut {NoStop}%
\bibitem [{\citenamefont {L{\"a}uchli}\ \emph {et~al.}(2019)\citenamefont
  {L{\"a}uchli}, \citenamefont {Sudan},\ and\ \citenamefont
  {Moessner}}]{lauchli2019s}%
  \BibitemOpen
  \bibfield  {author} {\bibinfo {author} {\bibfnamefont {A.~M.}\ \bibnamefont
  {L{\"a}uchli}}, \bibinfo {author} {\bibfnamefont {J.}~\bibnamefont {Sudan}},\
  and\ \bibinfo {author} {\bibfnamefont {R.}~\bibnamefont {Moessner}},\
  }\bibfield  {title} {\bibinfo {title} {S= 1 2 kagome heisenberg
  antiferromagnet revisited},\ }\href@noop {} {\bibfield  {journal} {\bibinfo
  {journal} {Physical Review B}\ }\textbf {\bibinfo {volume} {100}},\ \bibinfo
  {pages} {155142} (\bibinfo {year} {2019})}\BibitemShut {NoStop}%
\bibitem [{\citenamefont {Samajdar}\ \emph {et~al.}(2021)\citenamefont
  {Samajdar}, \citenamefont {Ho}, \citenamefont {Pichler}, \citenamefont
  {Lukin},\ and\ \citenamefont {Sachdev}}]{samajdar2021quantum}%
  \BibitemOpen
  \bibfield  {author} {\bibinfo {author} {\bibfnamefont {R.}~\bibnamefont
  {Samajdar}}, \bibinfo {author} {\bibfnamefont {W.~W.}\ \bibnamefont {Ho}},
  \bibinfo {author} {\bibfnamefont {H.}~\bibnamefont {Pichler}}, \bibinfo
  {author} {\bibfnamefont {M.~D.}\ \bibnamefont {Lukin}},\ and\ \bibinfo
  {author} {\bibfnamefont {S.}~\bibnamefont {Sachdev}},\ }\bibfield  {title}
  {\bibinfo {title} {Quantum phases of rydberg atoms on a kagome lattice},\
  }\href@noop {} {\bibfield  {journal} {\bibinfo  {journal} {Proceedings of the
  National Academy of Sciences}\ }\textbf {\bibinfo {volume} {118}} (\bibinfo
  {year} {2021})}\BibitemShut {NoStop}%
\bibitem [{\citenamefont {Verresen}\ and\ \citenamefont
  {Vishwanath}(2022)}]{verresen2022unifying}%
  \BibitemOpen
  \bibfield  {author} {\bibinfo {author} {\bibfnamefont {R.}~\bibnamefont
  {Verresen}}\ and\ \bibinfo {author} {\bibfnamefont {A.}~\bibnamefont
  {Vishwanath}},\ }\bibfield  {title} {\bibinfo {title} {Unifying kitaev
  magnets, kagome dimer models and ruby rydberg spin liquids},\ }\href@noop {}
  {\bibfield  {journal} {\bibinfo  {journal} {arXiv preprint arXiv:2205.15302}\
  } (\bibinfo {year} {2022})}\BibitemShut {NoStop}%
\bibitem [{\citenamefont {Schrieffer}\ and\ \citenamefont
  {Wolff}(1966)}]{schrieffer1966relation}%
  \BibitemOpen
  \bibfield  {author} {\bibinfo {author} {\bibfnamefont {J.~R.}\ \bibnamefont
  {Schrieffer}}\ and\ \bibinfo {author} {\bibfnamefont {P.~A.}\ \bibnamefont
  {Wolff}},\ }\bibfield  {title} {\bibinfo {title} {Relation between the
  anderson and kondo hamiltonians},\ }\href@noop {} {\bibfield  {journal}
  {\bibinfo  {journal} {Physical Review}\ }\textbf {\bibinfo {volume} {149}},\
  \bibinfo {pages} {491} (\bibinfo {year} {1966})}\BibitemShut {NoStop}%
\bibitem [{\citenamefont {Lanting}\ \emph {et~al.}(2010)\citenamefont
  {Lanting}, \citenamefont {Harris}, \citenamefont {Johansson}, \citenamefont
  {Amin}, \citenamefont {Berkley}, \citenamefont {Gildert}, \citenamefont
  {Johnson}, \citenamefont {Bunyk}, \citenamefont {Tolkacheva}, \citenamefont
  {Ladizinsky} \emph {et~al.}}]{lanting2010cotunneling}%
  \BibitemOpen
  \bibfield  {author} {\bibinfo {author} {\bibfnamefont {T.}~\bibnamefont
  {Lanting}}, \bibinfo {author} {\bibfnamefont {R.}~\bibnamefont {Harris}},
  \bibinfo {author} {\bibfnamefont {J.}~\bibnamefont {Johansson}}, \bibinfo
  {author} {\bibfnamefont {M.}~\bibnamefont {Amin}}, \bibinfo {author}
  {\bibfnamefont {A.}~\bibnamefont {Berkley}}, \bibinfo {author} {\bibfnamefont
  {S.}~\bibnamefont {Gildert}}, \bibinfo {author} {\bibfnamefont
  {M.}~\bibnamefont {Johnson}}, \bibinfo {author} {\bibfnamefont
  {P.}~\bibnamefont {Bunyk}}, \bibinfo {author} {\bibfnamefont
  {E.}~\bibnamefont {Tolkacheva}}, \bibinfo {author} {\bibfnamefont
  {E.}~\bibnamefont {Ladizinsky}}, \emph {et~al.},\ }\bibfield  {title}
  {\bibinfo {title} {Cotunneling in pairs of coupled flux qubits},\ }\href@noop
  {} {\bibfield  {journal} {\bibinfo  {journal} {Physical Review B}\ }\textbf
  {\bibinfo {volume} {82}},\ \bibinfo {pages} {060512(R)} (\bibinfo {year}
  {2010})}\BibitemShut {NoStop}%
\bibitem [{\citenamefont {Andriyash}\ \emph {et~al.}(2016)\citenamefont
  {Andriyash}, \citenamefont {Bian}, \citenamefont {Chudak}, \citenamefont
  {Drew-Brook}, \citenamefont {King}, \citenamefont {Macready},\ and\
  \citenamefont {Roy}}]{andriyash2016boosting}%
  \BibitemOpen
  \bibfield  {author} {\bibinfo {author} {\bibfnamefont {E.}~\bibnamefont
  {Andriyash}}, \bibinfo {author} {\bibfnamefont {Z.}~\bibnamefont {Bian}},
  \bibinfo {author} {\bibfnamefont {F.}~\bibnamefont {Chudak}}, \bibinfo
  {author} {\bibfnamefont {M.}~\bibnamefont {Drew-Brook}}, \bibinfo {author}
  {\bibfnamefont {A.~D.}\ \bibnamefont {King}}, \bibinfo {author}
  {\bibfnamefont {W.~G.}\ \bibnamefont {Macready}},\ and\ \bibinfo {author}
  {\bibfnamefont {A.}~\bibnamefont {Roy}},\ }\bibfield  {title} {\bibinfo
  {title} {Boosting integer factoring performance via quantum annealing
  offsets},\ }\href@noop {} {\bibfield  {journal} {\bibinfo  {journal} {D-Wave
  Technical Report Series}\ }\textbf {\bibinfo {volume} {14}} (\bibinfo {year}
  {2016})}\BibitemShut {NoStop}%
\bibitem [{\citenamefont {{(2023) All code, scripts and data used in this work
  are included in a GitHub repository.}}()}]{githubref}%
  \BibitemOpen
  \bibfield  {author} {\bibinfo {author} {\bibnamefont {{(2023) All code,
  scripts and data used in this work are included in a GitHub repository.}}},\
  }\href {https://github.com/vdrouint/d-wave-other} {\bibinfo {title}
  {https://github.com/vdrouint/d-wave-other}}\BibitemShut {NoStop}%
\bibitem [{\citenamefont {Pili}\ \emph {et~al.}(2022)\citenamefont {Pili},
  \citenamefont {Steppke}, \citenamefont {Barber}, \citenamefont {Jerzembeck},
  \citenamefont {Hicks}, \citenamefont {Guruciaga}, \citenamefont
  {Prabhakaran}, \citenamefont {Moessner}, \citenamefont {Mackenzie},
  \citenamefont {Grigera},\ and\ \citenamefont {Borzi}}]{pili2022topological}%
  \BibitemOpen
  \bibfield  {author} {\bibinfo {author} {\bibfnamefont {L.}~\bibnamefont
  {Pili}}, \bibinfo {author} {\bibfnamefont {A.}~\bibnamefont {Steppke}},
  \bibinfo {author} {\bibfnamefont {M.~E.}\ \bibnamefont {Barber}}, \bibinfo
  {author} {\bibfnamefont {F.}~\bibnamefont {Jerzembeck}}, \bibinfo {author}
  {\bibfnamefont {C.~W.}\ \bibnamefont {Hicks}}, \bibinfo {author}
  {\bibfnamefont {P.~C.}\ \bibnamefont {Guruciaga}}, \bibinfo {author}
  {\bibfnamefont {D.}~\bibnamefont {Prabhakaran}}, \bibinfo {author}
  {\bibfnamefont {R.}~\bibnamefont {Moessner}}, \bibinfo {author}
  {\bibfnamefont {A.~P.}\ \bibnamefont {Mackenzie}}, \bibinfo {author}
  {\bibfnamefont {S.~A.}\ \bibnamefont {Grigera}},\ and\ \bibinfo {author}
  {\bibfnamefont {R.~A.}\ \bibnamefont {Borzi}},\ }\bibfield  {title} {\bibinfo
  {title} {Topological metamagnetism: Thermodynamics and dynamics of the
  transition in spin ice under uniaxial compression},\ }\href
  {https://doi.org/10.1103/PhysRevB.105.184422} {\bibfield  {journal} {\bibinfo
   {journal} {Phys. Rev. B}\ }\textbf {\bibinfo {volume} {105}},\ \bibinfo
  {pages} {184422} (\bibinfo {year} {2022})}\BibitemShut {NoStop}%
\bibitem [{\citenamefont {Baez}\ and\ \citenamefont
  {Borzi}(2016)}]{baez3d2017}%
  \BibitemOpen
  \bibfield  {author} {\bibinfo {author} {\bibfnamefont {M.~L.}\ \bibnamefont
  {Baez}}\ and\ \bibinfo {author} {\bibfnamefont {R.~A.}\ \bibnamefont
  {Borzi}},\ }\bibfield  {title} {\bibinfo {title} {The 3d kasteleyn transition
  in dipolar spin ice: a numerical study with the conserved monopoles
  algorithm},\ }\href@noop {} {\bibfield  {journal} {\bibinfo  {journal}
  {Journal of Physics: Condensed Matter}\ }\textbf {\bibinfo {volume} {29}},\
  \bibinfo {pages} {055806} (\bibinfo {year} {2016})}\BibitemShut {NoStop}%
\bibitem [{\citenamefont {Nakamura}(2008)}]{Nakamura2008efficient}%
  \BibitemOpen
  \bibfield  {author} {\bibinfo {author} {\bibfnamefont {T.}~\bibnamefont
  {Nakamura}},\ }\bibfield  {title} {\bibinfo {title} {Efficient monte carlo
  algorithm in quasi-one-dimensional ising spin systems},\ }\href
  {https://doi.org/10.1103/PhysRevLett.101.210602} {\bibfield  {journal}
  {\bibinfo  {journal} {Phys. Rev. Lett.}\ }\textbf {\bibinfo {volume} {101}},\
  \bibinfo {pages} {210602} (\bibinfo {year} {2008})}\BibitemShut {NoStop}%
\bibitem [{\citenamefont {Barkema}\ and\ \citenamefont
  {Newman}(1998)}]{Barkema1998ice}%
  \BibitemOpen
  \bibfield  {author} {\bibinfo {author} {\bibfnamefont {G.~T.}\ \bibnamefont
  {Barkema}}\ and\ \bibinfo {author} {\bibfnamefont {M.~E.~J.}\ \bibnamefont
  {Newman}},\ }\bibfield  {title} {\bibinfo {title} {Monte carlo simulation of
  ice models},\ }\href {https://doi.org/10.1103/PhysRevE.57.1155} {\bibfield
  {journal} {\bibinfo  {journal} {Phys. Rev. E}\ }\textbf {\bibinfo {volume}
  {57}},\ \bibinfo {pages} {1155} (\bibinfo {year} {1998})}\BibitemShut
  {NoStop}%
\bibitem [{\citenamefont {Henry}\ and\ \citenamefont
  {Roscilde}(2014)}]{Henry2014squareice}%
  \BibitemOpen
  \bibfield  {author} {\bibinfo {author} {\bibfnamefont {L.-P.}\ \bibnamefont
  {Henry}}\ and\ \bibinfo {author} {\bibfnamefont {T.}~\bibnamefont
  {Roscilde}},\ }\bibfield  {title} {\bibinfo {title} {Order-by-disorder and
  quantum coulomb phase in quantum square ice},\ }\href
  {https://doi.org/10.1103/PhysRevLett.113.027204} {\bibfield  {journal}
  {\bibinfo  {journal} {Phys. Rev. Lett.}\ }\textbf {\bibinfo {volume} {113}},\
  \bibinfo {pages} {027204} (\bibinfo {year} {2014})}\BibitemShut {NoStop}%
\bibitem [{\citenamefont {Wang}\ \emph {et~al.}(2020)\citenamefont {Wang},
  \citenamefont {Humeniuk},\ and\ \citenamefont {Wan}}]{YaoWang2020}%
  \BibitemOpen
  \bibfield  {author} {\bibinfo {author} {\bibfnamefont {Y.}~\bibnamefont
  {Wang}}, \bibinfo {author} {\bibfnamefont {S.}~\bibnamefont {Humeniuk}},\
  and\ \bibinfo {author} {\bibfnamefont {Y.}~\bibnamefont {Wan}},\ }\bibfield
  {title} {\bibinfo {title} {Tuning the two-step melting of magnetic order in a
  dipolar kagome spin ice by quantum fluctuations},\ }\href
  {https://doi.org/10.1103/PhysRevB.101.134414} {\bibfield  {journal} {\bibinfo
   {journal} {Phys. Rev. B}\ }\textbf {\bibinfo {volume} {101}},\ \bibinfo
  {pages} {134414} (\bibinfo {year} {2020})}\BibitemShut {NoStop}%
\end{thebibliography}%

\end{document}